\documentclass[%
 reprint,
superscriptaddress,
 nofootinbib,
 amsmath,amssymb,
 aps,
 prb,
]{revtex4-2}

\usepackage{graphicx} % Include figure files
\graphicspath{{Figures/}}
\usepackage[utf8]{inputenc}
\usepackage{dcolumn} % Align table columns on decimal point
\usepackage{bm}      % bold math
\usepackage[colorlinks=true,linkcolor=blue,citecolor=blue]{hyperref}% add hypertext capabilities
\usepackage{verbatim}
\usepackage{orcidlink}

\usepackage{xcolor}

\usepackage{braket}

\usepackage{siunitx}

\usepackage{xfrac,faktor}

\usepackage[normalem]{ulem}

\usepackage[capitalise]{cleveref}
\Crefname{section}{Sec.}{Sections~}

\usepackage{verbatim}

\usepackage{multirow}

\usepackage{aas_macros}

\usepackage{bbold}

\DeclareUnicodeCharacter{2212}{-}
\DeclareUnicodeCharacter{2265}{>}

\newcommand{\hubble}{H_{0}}

\newcommand{\hubblep}{H}

\DeclareSIUnit{\msun}{\text{M}_{\odot}}

\DeclareSIUnit{\year}{\text{yr}}

\DeclareSIUnit{\radiant}{\text{rad}}

\DeclareSIUnit{\degree}{\text{deg}}

\DeclareSIUnit{\parsec}{\text{pc}}

\newcommand{\mchirp}{\mathcal{M}_c}

\newcommand{\de}[1]{\partial_{#1}}

\newcommand{\snr}{SNR}
\newcommand{\SNR}{\text{SNR}}
\newcommand{\psd}{S_{n}}

\newcommand{\inner}[2]{(#1|#2)}

\newcommand{\lcdm}{\Lambda \text{CDM}}

\newcommand{\Om}{\Omega_{m}}

\newcommand{\Ol}{\Omega_{\Lambda}}

\newcommand{\dl}{d_{L}}

\newcommand{\FM}{\Gamma}

\newcommand{\CM}{\Sigma}

\newcommand{\epsinv}{\varepsilon_{\text{inv}}}

\newcommand\tabvspace{1.2}

\begin{document}
%-----------------------------
% FRONT MATTER

\preprint{APS/123-QED}

% Title and authors
\title{Dark siren cosmology with binary black holes in the era of third-generation gravitational wave detectors}% Force line breaks with \\

\author{Niccol\`o Muttoni \orcidlink{0000-0002-4214-2344}}
\email{niccolo.muttoni@unige.ch}
\affiliation{Laboratoire des 2 Infinis - Toulouse (L2IT-IN2P3), Universit\'e de Toulouse, CNRS, UPS, F-31062 Toulouse Cedex 9, France}
\affiliation{D\'epartement de Physique Th\'eorique and Gravitational Wave Science Center, Universit\'e de Gen\`eve, 24 quai Ernest Ansermet, 1211 Gen\`eve 4, Switzerland}

\author{Danny Laghi \orcidlink{0000-0001-7462-3794}}
\affiliation{Laboratoire des 2 Infinis - Toulouse (L2IT-IN2P3), Universit\'e de Toulouse, CNRS, UPS, F-31062 Toulouse Cedex 9, France}

\author{Nicola Tamanini \orcidlink{0000-0001-8760-5421}}
\affiliation{Laboratoire des 2 Infinis - Toulouse (L2IT-IN2P3), Universit\'e de Toulouse, CNRS, UPS, F-31062 Toulouse Cedex 9, France}

\author{Sylvain Marsat \orcidlink{0000-0001-9449-1071}}
\affiliation{Laboratoire des 2 Infinis - Toulouse (L2IT-IN2P3), Universit\'e de Toulouse, CNRS, UPS, F-31062 Toulouse Cedex 9, France}
\author{David Izquierdo-Villalba \orcidlink{0000-0002-6143-1491}}
\affiliation{Department of Physics G. Occhialini, University of Milano - Bicocca, Piazza della Scienza 3, 20126 Milano, Italy}
\affiliation{INFN, Sezione di Milano-Bicocca, Piazza della Scienza 3, 20126 Milano, Italy}

\date{\today}% It is always \today, today,

% Abstract
\begin{abstract}

Third-generation (3G) gravitational wave detectors, in particular Einstein Telescope (ET) and Cosmic Explorer (CE), will explore unprecedented cosmic volumes in search for compact binary mergers, providing us with tens of thousands of detections per year.
In this study, we simulate and employ binary black holes detected by 3G interferometers as dark sirens, to extract and infer cosmological parameters by cross-matching gravitational wave data with electromagnetic information retrieved from a simulated galaxy catalog.
Considering a standard $\lcdm$ model, we apply a suitable Bayesian framework to obtain joint posterior distributions for the Hubble constant $\hubble$ and the matter energy density parameter $\Om$ in different scenarios. 
Assuming a galaxy catalog complete up to $z=1$ and dark sirens detected with a network signal-to-noise ratio greater than \num{300}, we show that a network made of ET and two CEs can constrain $\hubble$ ($\Om$) to a promising $0.8\%$ ($10.0\%$) at $90\%$ confidence interval within one year of continuous observations. 
Additionally, we find that most of the information on $\hubble$ is contained in local, single-host dark sirens, and that dark sirens at $z>1$ do not substantially improve these estimates.
Our results imply that a subpercent measure of $\hubble$ can confidently be attained by a network of 3G detectors, highlighting the need for characterizing all systematic effects to a higher accuracy.

\end{abstract}


% To print the title, authors and abstract              
\maketitle

%-----------------------------
% MAIN MATTER

\section{Introduction}

The last few decades have seen a revolutionary change in the paradigms underpinning our knowledge of the evolution and dynamics of the Universe as a whole.
The research field of cosmology has gone from ``the quest for two numbers'' to a new rich phenomenological field rooted in precise astronomical observations; see \cite{2022ARNPS..72....1T} for a vivid historical reconstruction.
The resulting era of ``precision cosmology'' delivered a detailed description of the Universe at the largest scales, with a standard cosmological model capable of explaining all current observations, modulo few persisting statistical tensions; see, e.g.,~\cite{Planck:2018nkj,Planck:2018vyg,Riess:2021jrx,Pan-STARRS1:2017jku,DES:2021wwk,DES:2021bvc}.
This spectacular achievement has been possible by the piling up of ever more accurate astronomical observations, the overwhelming majority of which obtained with electromagnetic (EM) telescopes over the whole accessible band of the EM spectrum.

Nevertheless since the first direct detection of gravitational waves (GWs) by the LIGO and Virgo collaborations in 2015~\cite{LIGOScientific:2016aoc}, we now possess a new whole spectrum that can provide a wealth of cosmological information complementary to EM observations.
GWs can be used as standard cosmological rulers~\cite{schutz86} and thus provide a map of the cosmic expansion history at different redshifts.
The luminosity distance of a binary system emitting GWs can in fact be extracted from the detected GW signal without relying on any phenomenological relation or calibration at lower redshifts.
In other words, compact binaries emitting GWs are \textit{absolute} cosmic distance rulers since they do not depend on the so-called \textit{cosmic distance ladder}.
In analogy to supernovae-type Ia, which are \textit{calibrated} cosmic distance rulers commonly called standard candles, GW signals from compact binaries containing black holes (BHs) and neutron stars (NSs) are commonly known as \textit{standard sirens}~\cite{2005ApJ...629...15H,Dalal:2006qt}.

Unfortunately the redshift of the source is not one of the parameters that we can easily obtain from GWs emitted by compact binaries.
For this reason standard sirens cannot be used straightaway to map the expansion of the Universe through the well-known \textit{distance-redshift relation}, contrary to standard candles for which a redshift measurement is usually readily available.
Different methods have been proposed to obtain complementary redshift information to a standard siren.

The simplest and most intuitive of these methods consists in observing an EM counterpart of the GW event to identify its host galaxy~\cite{schutz86}.
In such cases the redshift of the GW source can be estimated by measuring the redshift of the host galaxy, providing in this way a single redshift value for the distance-redshift diagram.
Unfortunately this method applies only to GW events for which an EM counterpart can be observed, which are commonly referred to as ``bright sirens'' in a cosmological context.
So far the LIGO-Virgo-Kagra (LVK) Collaboration observed only one such bright siren, namely the multimessenger binary neutron star (BNS) merger GW170817~\cite{ligobns,MMApaper}.
The coincident measurements of both distance and redshift of this event delivered the first ever cosmological measurement with GWs: a constraint on the Hubble constant of $\hubble = \SI[parse-numbers = false]{70^{+12}_{-8}}{\kilo\meter\per\second\per\mega\parsec}$~\cite{2017Natur.551...85A}.
Future observational runs of the LVK detectors are expected to improve upon this result with the addition of further bright sirens, not only BNSs but also BH-NS binaries for which an EM counterpart is spotted~\cite{chen17,Feeney:2018mkj,Feeney:2020kxk,Vitale:2018wlg}.
The technological limitations of current GW interferometers however cannot guarantee sufficiently numerous detections to achieve a measurement of $\hubble$ better than a few \%, while constraints on other cosmological parameters are well beyond their reach~\cite{Chen:2020zoq}.

If no EM counterpart can be detected, other methodologies are nevertheless used to gather redshift information complementary to a GW binary signal.
The so-called ``dark siren'' method~\citep{schutz86,PhysRevD.86.043011,chen17,fishbach,2020PhRvD.101l2001G,Finke:2021aom,Gray2022,PhysRevD.105.023523,2022arXiv221208694G}, sometimes referred to as the ``statistical method,'' consists in cross-matching the sky localization error volume, sometimes simply called volume error-box, of the GW source with galaxy catalogs collected by EM surveys, either readily available or constructed \textit{ad hoc} along the sky localization cone of the detected GW signal.
Such a method has been proved to work with both simulated~\cite{PhysRevD.86.043011,2020PhRvD.101l2001G,Gray2022} and observational data~\cite{fishbach,2019ApJ...876L...7S,2020ApJ...900L..33P,2021ApJ...909..218A,Finke:2021aom}.
The latest results from all the LVK observational runs combined yield $\hubble = \SI[parse-numbers = false]{68^{+8}_{-6}}{\kilo\meter\per\second\per\mega\parsec}$~\cite{2021arXiv211103604T}; see also~\cite{2021arXiv211106445P}.
Although this represents only a small improvement with respect to the constraint obtained from GW170817 only, the large number of expected GW detections without EM counterparts renders dark sirens a promising and robust method to estimate cosmological parameters from future GW observations.
Similar approaches exploiting the spatial cross-correlation between GW sources and galaxies, have also been proposed and shown to work well once a large amount of GW events will be observed~\citep{PhysRevD.93.083511,Mukherjee:2019wcg, Mukherjee:2020hyn,Bera:2020jhx, Diaz:2021pem,Balaudo:2022znx}.

A third methodology to obtain redshift information for standard sirens is based on the insight and modeling of the population distribution of intrinsic parameters of the GW sources~\citep{1993ApJ...411L...5C,Taylor_2012,Farr_2019,2020arXiv200602211M,mastrogiovanni_2021,Mukherjee:2021rtw,2022JCAP...09..012L,Ezquiaga_2022, Karathanasis:2022rtr,Mancarella:2022cnu,2022PhRvL.129f1102E}, in particular their masses, spins, and merger rate evolution.
Such a method is usually called ``spectral sirens'' due to the use of features in the distribution spectra of GW source parameters, whose parameters are inferred simultaneously with the cosmological parameters.
It has already been applied to real LVK data with the most recent measurement registering a constraint $\hubble = \SI[parse-numbers = false]{68^{+12}_{-7}}{\kilo\meter\per\second\per\mega\parsec}$~\cite{2021arXiv211103604T}.
The pros of this method are that it does not require any EM information, but the cons are that it introduces a dependence on the modeling of the parameter distributions of the underlying astrophysical population of GW sources.

In general the current status of GW cosmology outlined above, with results from standard sirens still at large experimental uncertainties, begs an analogy between EM cosmology in the era of the ``quest for two numbers.''
Nowadays the main objective of GW cosmology consists in the measurement of the Hubble constant, with low chances to access information on other cosmological parameters.
Similarly to the situation for EM cosmology at the end of the last century, the current 2nd generation of GW interferometers do not have in practice the constraining power needed to push observations beyond the quest for $\hubble$.
This scenario will change dramatically when 3rd generation (3G) interferometers will come online: the new era of ``precision GW cosmology'' will begin.

Two possible concepts for 3G interferometers are currently under consideration for construction in the 2030s: the Einstein Telescope (ET) in Europe~\cite{2010CQGra..27s4002P,2011CQGra..28i4013H,2020JCAP...03..050M} and the Cosmic Explorer (CE) in the USA~\cite{Reitze2019Cosmic,Evans:2021gyd}.
They are both aimed at greatly improving the sensitivity around the same frequency band of the LVK detectors, as well as at extending observations at lower frequencies down to a few \si{\hertz}.
The scientific potential of 3G detectors is huge, with expected observations of \num[retain-unity-mantissa = false]{1e5}-\num[retain-unity-mantissa = false]{1e6} GW signals from compact binary coalescences over few years of observations.
ET and CE will deliver new breakthrough observations on multiple subjects encompassing astrophysics, cosmology, and fundamental physics; see~\cite{Kalogera:2021bya} for a summary of their science case.

In terms of cosmology, 3G detectors will exploit the standard siren methodologies described above to attain accurate and precise measurements of the cosmological parameters.
Bright sirens may yield subpercent constraints on $\hubble$ with $\mathcal{O}$(100) observations of multimessenger BNSs, which are expected after few years of operation~\cite{Cai:2016sby,Zhao:2017cbb,Belgacem:2019tbw,deSouza:2021xtg,deSouza:2021xtg,Califano:2022syd,Dhani:2022ulg,Alfradique:2022tox} (similar results are claimed for binary neutron star-black holes (NSBHs)~\cite{Gupta:2022fwd}).
However, the expected EM counterpart signals from these BNSs can only be detected at relatively low redshift ($z \lesssim 0.5$)~\cite{Belgacem:2019tbw}, implying that further cosmological parameters beyond the Hubble constant may not be well measured, except perhaps the equation of state of dark energy~\cite{Sathyaprakash:2009xt,Zhao:2010sz}.

Spectral sirens on the other hand will be able to exploit the full redshift range of observable binary black holes (BBHs), which extends well beyond the reach of BNSs for 3G detectors.
This method will consequently not only deliver stringent constraints on $\hubble$, but it will also provide interesting results at high redshift for both dark energy and further cosmological parameters~\cite{Taylor:2011fs,Taylor:2012db,2022PhRvL.129f1102E,Leandro:2021qlc,Ye:2021klk}.
Moreover, another similar method that will be applicable to 3G detectors thanks to their exquisite precision, consists in the simultaneous inference of both the equation of state of neutron stars and their redshift~\cite{Messenger:2011gi}.
Such an approach provides a redshift for each BNS, without the need of an EM counterpart, but introduces a dependence on the modeling of the equation of state of neutron stars which can introduce systematics if not properly accounted for.
Nevertheless recent estimates provide forecasts on the measurement of the Hubble constant that range from few \% to subpercent levels, showing that further cosmological parameters are within reach of an accurate measurement~\cite{2022PhRvD.106l3529G,2021PhRvD.104h3528C,Dhani:2022ulg,Jin:2022qnj}.

Contrary to bright sirens, dark sirens in the 3G era have not yet been systematically investigated.
Recent exploratory studies, based on a number of over-simplifying assumptions and restricted to low redshift galaxy catalogs ($z<0.3$), claim that constraints on $\hubble$ can reach an extremely optimistic precision of $\mathcal{O}(0.01\%)$, or even better, in 5 years of observations~\cite{Yu:2020vyy,Song:2022siz}.
Other exploratory investigations using either BBHs or NSBHs as ``golden'' dark sirens, namely well-localized events for which one single galaxy is contained in their sky localization volume, show instead that $\mathcal{O}(0.1)\%$ constraints on $\hubble$ can be obtained again in 5 years of observations~\cite{Borhanian:2020vyr,Gupta:2022fwd}.
Less optimistic results have been recently reported in~\cite{Zhu:2023jti}, where a more realistic simulation yields $\mathcal{O}(1)\%$ constraints on $\hubble$ and $\mathcal{O}(10)\%$ constraints on $\Omega_m$ with 300 BBHs detected by ET plus one CE.
Further analyses, under more realistic assumptions and using the complete information from galaxy catalogs, are clearly needed to make clarity on the expected dark siren potential of 3G detectors.

The scope of the present paper consists in producing reliable cosmological dark sirens forecasts with BBH mergers for the 3G era.
Compared to the existing literature, our study extends to higher redshift, it reduces the underlying simplifying assumptions making our simulation more realistic, and it enlarges the cosmological inference to cosmological parameters beyond $\hubble$.
In \cref{sec:discussion} we will compare our results with the ones reported previously.
All these results, especially if folded together with other cosmological expectations from GW large-scale observatories in the 2030s, notably for example from space-borne detectors~\cite{LISACosmologyWorkingGroup:2022jok,Tamanini:2016zlh,Caprini:2016qxs,Cai:2017yww,DelPozzo:2017kme,LISACosmologyWorkingGroup:2019mwx,Speri:2020hwc,2021MNRAS.508.4512L,Muttoni:2021veo,Yang:2021qge}, show that 3G detectors will usher an era of precision GW cosmology, similarly to how EM telescopes and surveys opened an era of precision EM cosmology 20-30 years ago.
The era of the ``quest for one number'', namely $\hubble$, will leave space for a plethora of different cosmological measurements with GWs which will offer an unprecedented and clear picture of the gravitational Universe.

This study is organised as follows. 
In \cref{sec:GWpopulation} we construct a realistic, simulated population of BBH mergers based on the most recent LVK observations.
In \cref{sec:GWPE} we present our approach to detect the GW signals emitted by BBHs and measure their parameters with Fisher information techniques.
In \cref{sec:errorboxes} we describe how we build our galaxy catalogs, produce GW sky-localization error volumes and associate potential host galaxies to GW events.
In \cref{sec:cosmoinference} the details of our Bayesian inference approach to measure the cosmological parameters are outlined.
In \cref{sec:results} we present the results of our analyses, namely the expected constraints on $\lcdm$ for different observational scenarios.
Finally in \cref{sec:discussion} we discuss our findings, their implications and compare them with the literature, while in \cref{sec:conclusion} we conclude.

\section{Simulation of the mock gravitational-wave event catalog}
\label{sec:GWpopulation}

In order to infer the cosmological parameters with 3G detectors, we first need to define an astrophysical population of BBHs. Each source parameter is extracted from some probability density function which are motivated by astrophysical assumptions. In the following we discuss the generation of the parameters that characterize each BBH, that is, the BH component masses and spins, sky position, redshift, inclination, polarization angles, and coalescence time and phase. While most of these parameters are described by relatively trivial distributions, others need to be investigated more carefully.

\subsection{Masses}

The recent observing runs with the Advanced LIGO and Advanced Virgo interferometers (O1 \cite{abbott2019gwtc1}, O2 \cite{abbott2021gwtc2} and O3 \cite{abbott2021gwtc3}) enriched the graveyard of known compact binary mergers with a total of \num{90} events, and the analysis of the population properties of these events~\cite{ligo2021population} shed light on their nature. 
Here we adopt these latest results to extract the masses of the individual components.

For BBHs, the primary BH mass distribution may be described by different fits~\cite{ligo2021population}. Among them, we choose the \texttt{POWER LAW + PEAK} which provides a good description of the overall observations. 
This model features a power law and a Gaussian peak around $\sim \SI{35}{\msun}$, which reflects the pair instability supernovae lower edge. Specifically, the probability distribution reads
\begin{equation}
    p(m_1) \propto \bigl[(1 - \lambda_{\rm peak})\mathcal{B}(m_1) + \lambda_{\rm peak}\mathcal{G}(m_1)\bigr] \mathcal{S}(m_1) \, ,
\label{eqn:p_of_m_1}
\end{equation}
where $\mathcal{B}(m) \propto m^{-\alpha}$ is a power law with spectral index $\alpha = 3.5$, $\mathcal{G}(m) \propto \mathcal{N}(\mu_{\rm BH}, \sigma^2_{\rm BH})$ is a Gaussian with mean $\mu_{\rm BH} = \SI{34}{\msun}$ and width $\sigma_{\rm BH} = \SI{5.69}{\msun}$, $\lambda_{\rm peak} = 0.038$ is a factor that controls the relative frequency of mergers in the power-law-dominated region and the Gaussian one, and finally $\mathcal{S}(m) \in [0, \, 1]$ is a smoothing piece-wise function, defined through
   \begin{multline}
        \mathcal{S}(m) = \\
        \begin{cases}
        0 & \text{if} \; m < m_{\rm min}, \\
        \Bigl(f(m - m_{\rm min}) + 1\Bigr)^{-1} & \text{if} \; m_{\rm min} \le m < m_{\rm min} + \delta_m, \\
        1 & \text{if} \; m \ge m_{\rm min} + \delta_m,
        \end{cases}
    \end{multline}
    with    
    \begin{equation}
        f(m) = \exp\biggl(\frac{\delta_m}{m} + \frac{\delta_m}{m - \delta_m}\biggr) \, , \qquad \delta_m = \SI{4.9}{\msun} \, .
    \end{equation}

We compute the secondary mass of the BBH through the mass ratio. The probability distribution of this parameter is described by the following expression
\begin{equation}
    p(q) \propto q^{\beta}\mathcal{S}(q m_1) \, ,
\end{equation}
with spectral index $\beta = 1.1$. We refer the reader to Ref.~\cite{ligo2021population} for more details about these mass distributions.

\subsection{Spins} \label{sec:spins}

Each binary component is characterized by a spin vector.
While the proper sample of the spins should keep into account all the \num{3} spatial components, we choose to assume only nonprecessing binaries, i.e., systems where the individual object spins are aligned with the total angular momentum.
Our choice reduces the complexity of the simulations, since each object is now described by \num{1} spin component, which we set to be along the $z$ axis. In particular, BH spin magnitudes are extracted from a uniform distribution between $[\num{-0.75}, \, \num{0.75}]$, as assumed in recent works (e.g., \cite{borhanian2022listening}). Thus, we do not include the effects of precession in the population.

\subsection{Angles and coalescence time}

Each binary is described by a set of different angles which includes:
\begin{itemize}
    \item The sky position angles, typically labeled by right ascension (RA) and declination (DEC). These two parameters range respectively between $[0, \, 2\pi]$ and $[-\pi/2, \, \pi/2]$. 
    Assuming an isotropic Universe, we sample the source
    sky positions uniformly on a spherical surface according to
    \begin{equation}
       p(\theta, \, \varphi) d\theta d\varphi  \propto \sin\theta d\theta d\varphi = p(\theta) d\theta \, p(\varphi) d\varphi \, ,
    \end{equation}
    where $\theta = \pi/2 - \text{DEC}$ is the colatitude and $\varphi = \text{RA}$ is the longitude, while
    \begin{equation}
    \begin{split}
        p(\theta) d\theta & \propto \sin\theta d\theta \, , \\
        p(\varphi) d\varphi & \propto d\varphi \, .
    \end{split}
    \end{equation}
    \item The inclination angle $\iota$, defined as the angle between the line of sight and the angular momentum of the binary. It takes values between $[0, \, \pi]$, where the lower (upper) boundary reflects face-on (face-off) binaries, while the midpoint characterizes edge-on binaries. The inclination angle is extracted uniformly in $\cos\iota$. Hence, its probability distribution follows the same of $\theta$, that is
    \begin{equation}
        p(\iota) d\iota \propto \sin\iota d\iota \propto d\cos\iota \, .
    \end{equation}
    This is to avoid a uniform sample in $\iota$ which would overestimate the number of loud (i.e.~face-on and face-off) sources.
    \item The polarization angle $\psi$, which ranges between $[0, \, \pi]$.
    This parameter represents a generic rotation of the GW main axes on the plane perpendicular to the direction of propagation.
    We extract $\psi$ samples from a uniform distribution.
    \item The coalescence phase $\Phi_c$, with values within $[0, \, 2\pi]$. It represents a reference value, conventionally associated to the merger, from which to determine the evolution of the phase of the GW signal. Its values are drawn from a flat distribution. 
\end{itemize}

In the same fashion, we draw uniformly the GPS coalescence time $t_c$ in a \SI{1}{\year} time window, starting from a fixed GPS reference time.

\subsection{Redshift} \label{sec:redshift_distribution}

While current observations still provide valuable information on the distance distribution of compact binary mergers, the horizon of current detectors is not sufficiently large to allow a precise reconstruction of the redshift distribution of GW sources across the cosmic history. For this reason we design a probability density function $p(z)$ for the redshift which is based on plausible astrophysical assumptions.

The merger of a binary system occurs after a time delay $t_d$ since its formation. Time delay, and the redshifts of the merger $z_m$ and formation $z_f$ of the system are related through
\begin{equation}
\begin{split}
    t_d & = \int_{z_m}^{z_f} \frac{dz}{(1 + z)\hubblep(z)} \\
        & = \int_{0}^{z_f} \frac{dz}{(1 + z)\hubblep(z)} - \int_{0}^{z_m} \frac{dz}{(1 + z)\hubblep(z)} \, ,
\end{split}
\label{eqn:time_delay}
\end{equation}
where $\hubblep(z) = \hubble \sqrt{\Om (1+z)^3 + \Ol}$ is the Hubble parameter, $\hubble$ is the Hubble constant, $\Om \equiv \Omega_{m,0}$ is the matter density parameter and $\Ol$ is the dark energy density parameter. \Cref{eqn:time_delay} represents the lookback time difference between $z_f$ and $z_m$: given $t_d$ and $z_f$, one can compute $z_m$ by inverting \cref{eqn:time_delay}. We assume that binary formation and merger are tracked by the star formation rate density $\Psi(z)$ (units \si{\msun\giga\parsec^{-3}\year^{-1}}) with the addition of a prescription for time delay effects. Throughout this work, we adopt the Madau-Fragos $\Psi(z)$, modeled by
\begin{equation}
    \Psi(z) = a \frac{(1+z)^b}{1 + [c(1+z)]^d} \, \si{\msun\giga\parsec^{-3}\year^{-1}} \, ,
\end{equation}
where $a = 0.01$, $b = 2.6$, $c = 1/3.2$ and $d = 6.2$, as reported in \cite{madau2017radiation}. The merger rate density $\dot{n}(z)$ (units \si{\giga\parsec^{-3}\year^{-1}}) is obtained by integrating $\Psi(z_f)$ over all the possible time delays:
\begin{equation}
    \dot{n}(z) \propto \int_{t_d^{\rm min}}^{t_d^{\rm max}} dt_d \, \Psi(z_f(z, t_d)) \, p(t_d) \, .
\label{eqn:merger_rate_density}
\end{equation}
Here $p(t_d)$ is a probability density function associated to the time delay, while $t_d^{\rm min}$ and $t_d^{\rm max}$ are respectively the minimum and maximum time delay of the distribution. For BBH systems we consider a minimum time delay $t_d^{\rm min} = \SI{10}{\mega\year}$, while we fix the maximum to $t_d^{\rm max} = \SI{10}{\giga\year}$ as in \cite{borhanian2022listening}.
Furthermore, we assume $p(t_d) \propto t_d^{-1}$, which becomes
\begin{equation}
    p(t_d) = \biggl(\ln \biggl(\frac{t_d^{\rm max}}{t_d^{\rm min}}\biggr) t_d\biggr)^{-1} \, ,
\end{equation}
once rescaled. In the same fashion, \cref{eqn:merger_rate_density} must be normalized. To compute the normalization factor (units \si{\msun^{-1}}), we require that $\dot{n}(0) = \dot{n}_0$, i.e.,~that the merger rate density evaluated at $z=0$ must be equal to the state-of-the-art local merger rate density of BBHs. In particular, we adopt $\dot{n}_{0} = \SI{23.9}{\per\cubic\giga\parsec\per\year}$ as reported in \cite{ligo2021population}.

Next, we obtain the merger rate per unit redshift bin $dR/dz$ through
\begin{equation}
    \frac{dR}{dz}(z) = \dot{n}(z) \frac{dV}{dz}(z) = \dot{n}(z) \frac{4\pi c \, d_c(z)^2 }{\hubblep(z)} \, ,
\label{eqn:source_frame_merger_rate}
\end{equation}
where $dV/dz$ is the differential comoving volume element, $c$ is the speed of light and $d_c(z) = c \int_{0}^{z} d\tilde{z}/\hubblep(\tilde{z})$ represents the comoving distance.
However, \cref{eqn:source_frame_merger_rate} is a source-frame quantity. The expansion of the Universe affects the source-frame merger rate with a time dilation factor $dt/dt^{\rm obs} = (1 + z)^{-1}$, therefore the observer-frame merger rate $d\mathcal{R}/dz$ reads
\begin{equation}
    \frac{d\mathcal{R}}{dz}(z) = \frac{1}{(1 + z)}\frac{dR}{dz}(z) \, .
\label{eqn:observer_frame_merger_rate}
\end{equation}

By integrating the observed merger rate, \cref{eqn:observer_frame_merger_rate},
one can recover the total number of mergers per unit time in the integration domain, i.e.,~the cosmic merger rate $\mathcal{R}$. Specifically:
\begin{equation}
    \mathcal{R} = \int_{z_{\rm min}}^{z_{\rm max}} dz \frac{d\mathcal{R}}{dz}(z) \, ,
\end{equation}
where $[z_{\rm min}, \, z_{\rm max}]$ is the relevant redshift interval. Our study focuses on GW sources within $z_{\rm min} = \num{0}$ and $z_{\rm max} = \num{50}$, and the associated BBH cosmic merger rate is $\mathcal{R} = \SI{49056}{\per\year}$, which is in broad agreement with other recent works (see e.g.~\cite{borhanian2022listening,iacovelli2022forecasting}).
The redshift probability density function $p(z)$ is then given by
\begin{equation}
    p(z) = \frac{1}{\mathcal{R}} \frac{d\mathcal{R}}{dz}(z) \, ,
\end{equation}
which represents our sampling distribution (see the dashed-black histogram in the left plot of \cref{fig:total_and_detected_pops} for a representative sample).

We convert the sampled redshifts in to luminosity distances through the redshift-luminosity distance relation
\begin{equation}\label{eqn:luminosity_distance_flat}
    d(z, \Omega) = \frac{c}{\hubble} (1+z) \int_0^z \frac{d\tilde{z}}{\sqrt{\Om(1+\tilde{z})^3 + 1 - \Om}} \,,
\end{equation}
assuming a flat $\lcdm$ cosmology with values coming from the Planck first-year data~\citep{PlanckCollaboration2014}: $h = \hubble/\SI{100}{\kilo\meter\per\second\per\mega\parsec} = 0.673$ and $\Om = 0.315$.
These values define our fiducial cosmology, on which is based the galaxy catalog that we use (see~\cref{subsec:gal_cats}).

%%%%%%%%%%%%%%%%%%%%%%%%%%%%%%%%%%%%%%%%%%%%%%%%%%%%%%%%%%%%%%%%%%%%%%%%%%%%%%%%%%%%%%%%%%%%%%%%%%%%%%%%%%%%%%%%%%%%%%%%%%%%%%%%%%%%%%%%%%%%%%%%%%%%%%%%%%%%%%%%%%%%%%%%%%%%%%%%%%%%%%%%%%%%%%%%%%%%%%%%%%%%%%%%%%%%%%%%%%%%%%%%%%%%%%%%%%%%%%%%%%%%%%%%%%%%%%%%%%%%%%%%%%%%%%%%

\section{Gravitational-wave detection and Fisher analysis}
\label{sec:GWPE}

\begin{figure}
    \centering
    \includegraphics[width=0.47\textwidth]{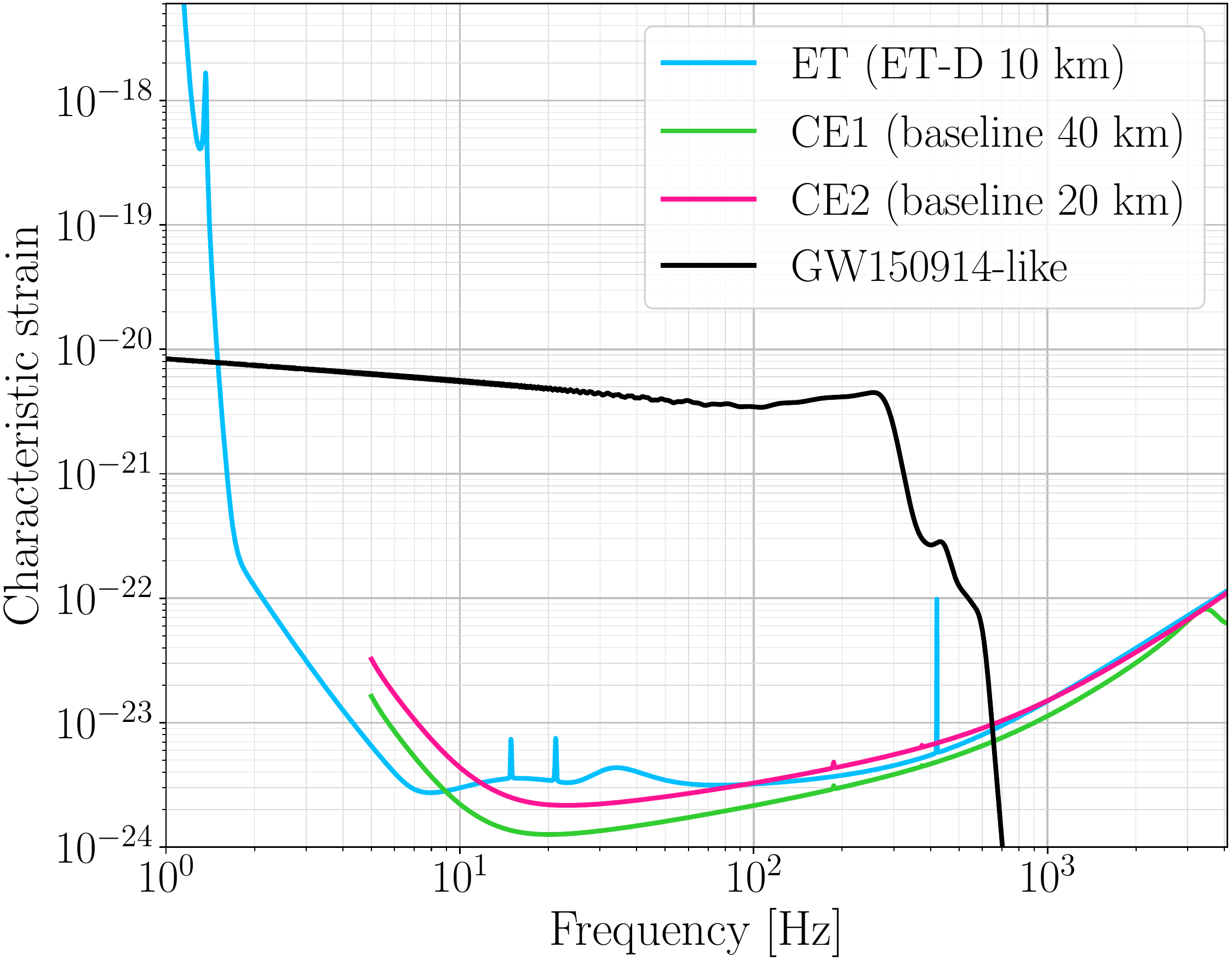}
    \caption{Sensitivity curve of the detectors considered in this study, together with a GW150914-like signal, colors as in legend. The characteristic strain is a dimensionless quantity related to the detector's power spectral density through $\sqrt{f \psd(f)}$. In the case of a GW signal, instead, it is defined as $2 f |\tilde{h}(f)|$. These two expressions, which implicitly enter in \cref{eqn:snr}, are useful to visualize the loudness of a GW signal when plotted together. The ET-D curve is rescaled, as reported in \cite{2011CQGra..28i4013H}, to take into account for the triangular geometry. Each detector observes a different strain, due to the modulation of its respective antenna pattern function. Here we display only the unprojected signal with the \texttt{IMRPhenomXHM} waveform model.}
    \label{fig:sensitivities}
\end{figure}

The future of ground-based GW astronomy will be led by \num{3}G interferometers. This work aims at testing the capabilities of ET and CE(s) by considering different combinations of such detectors. 
In this study, ET is assumed to have 
a triangular-shaped configuration of \num{3} independent detectors co-located in Italy (E1, E2, E3), while CE consists of \num{2} independent L-shaped detectors, placed respectively in the United States (CE1) and in Australia (CE2). We adopt the latest sensitivity curves available, specifically we consider the \SI{10}{\kilo\meter}-arm ET-D noise curve model \cite{2011CQGra..28i4013H} and the baseline \num{40}-\num{20} \si{\kilo\meter} arms CE \cite{Evans:2021gyd} curves, displayed in \cref{fig:sensitivities}. We further limit our work to the assumption that all the detectors are always operative during the whole observation time (i.e.,~full duty cycles are considered), which we set to be \SI{1}{\year}. In \cref{tab:detectors} we summarize the specifics of the individual observatories. In this study we concentrate on three particular networks of 3G detectors: a single ET, ET and CE1 (ET+CE1), and ET with the two CEs (ET+CE1+CE2). We do not consider the network made up by CE1 and CE2 (CE1+CE2) as we want to focus on ET and its potential in a network.

\subsection{Signal modeling and injection settings}

\begin{table*}[t]
\centering
%\bgroup
%\setlength{\tabcolsep}{0.5em} % horizontal padding
\def\arraystretch{\tabvspace} %  vertical padding (default is 1)
\begin{tabular}{ l c c c c c c c }
\hline
\hline
\textbf{Full name} & \textbf{Short name} & \textbf{Latitude} & \textbf{Longitude} & \textbf{x-arm azimuth} & \textbf{y-arm azimuth} & \textbf{$\psd$} & \textbf{$f_{\rm start}$} \\
\hline
\multirow{3}*{Einstein Telescope (ET)} & E1 & $0.7615$ & $0.1833$ & $0.3392$ & $5.5752$ & ET-D & \SI{1}{\hertz} \\
                                       & E2 & $0.7629$ & $0.1841$ & $4.5280$ & $3.4808$ & ET-D & \SI{1}{\hertz}\\
                                       & E3 & $0.7627$ & $0.1819$ & $2.4336$ & $1.3864$ & ET-D & \SI{1}{\hertz}\\
\hline
\multirow{2}*{Cosmic Explorers (CEs)} & CE1 & $0.7613$ & $-2.0281$ & $1.5708$ & $0$ & baseline \SI{40}{\kilo\meter} & \SI{5}{\hertz}\\
                                      & CE2 & $-0.5811$ & $2.6021$ & $2.3562$ & $0.7854$ & baseline \SI{20}{\kilo\meter} & \SI{5}{\hertz}\\

\hline
\hline
\end{tabular}
\caption{Summary of the 3G GW detectors we consider in this study. Angles are rounded and expressed in \si{\radian}, while the last column refers to the lowest frequency of the associated power spectral density $\psd$.
}
\label{tab:detectors}
\end{table*}

We simulate each GW signal directly in the frequency domain with a frequency resolution of $df = \SI[exponent-base = 256, retain-unity-mantissa = false]{1e-1}{\hertz}$, starting from \SI{1}{\hertz} and up to a sharp cut at \SI{4096}{\hertz}. 
In the context of \num{3}G detectors, GW signals may last from some minutes to several hours. Earth's rotation will play a crucial role for the localization of a source in the sky, since each detector will observe a strain modulated by its unique antenna pattern function.
However, in this work we consider BBHs, which emit signals in the 3G band with typical duration that spans from a few seconds and up to an hour. We therefore trade Earth's rotation effects with computational efficiency, leading to conservative results in the rare limit of long lasting signals. We validate our choice by comparing the sky location uncertainty of a GW150914-like signal including and not including Earth's rotation, finding no significant improvement.

We further assume that each source may be resolved and its parameters estimated independently. While we may expect to observe some signals overlapping in time (see, e.g.,~\cite{samajdar2021biases,himemoto2021impacts,pizzati2022toward}), in this work we assume that by the time \num{3}G detectors will become operative we will be able to make robust parameter estimation for all detected compact binary signals, i.e.~we can benefit from all the BBH sources as dark sirens.

We model each signal, or injection, with the \texttt{IMRPhenomXHM} \cite{PhysRevD.102.064002} waveform in order to capture higher harmonics, which are expected to be important for asymmetric-mass binaries and can potentially break fundamental degeneracies between source parameters, most notably luminosity distance and inclination~\cite{Littenberg2012, Varma2014, Shaik2019}. This waveform model assumes nonprecessing binaries, so that each object's spin is characterized only by one spatial component parallel to the total orbital angular momentum of the system, as modeled in \cref{sec:spins}.

We make use of the publicly available \textsc{Python} library \textsc{PyCBC} \cite{alex_nitz_2021_5256134} to generate our injections and compute the quantities described in \cref{eqn:snr,eqn:fisher_matrix}.

\begin{figure*}
    \centering
    \includegraphics[width=0.4973\textwidth]{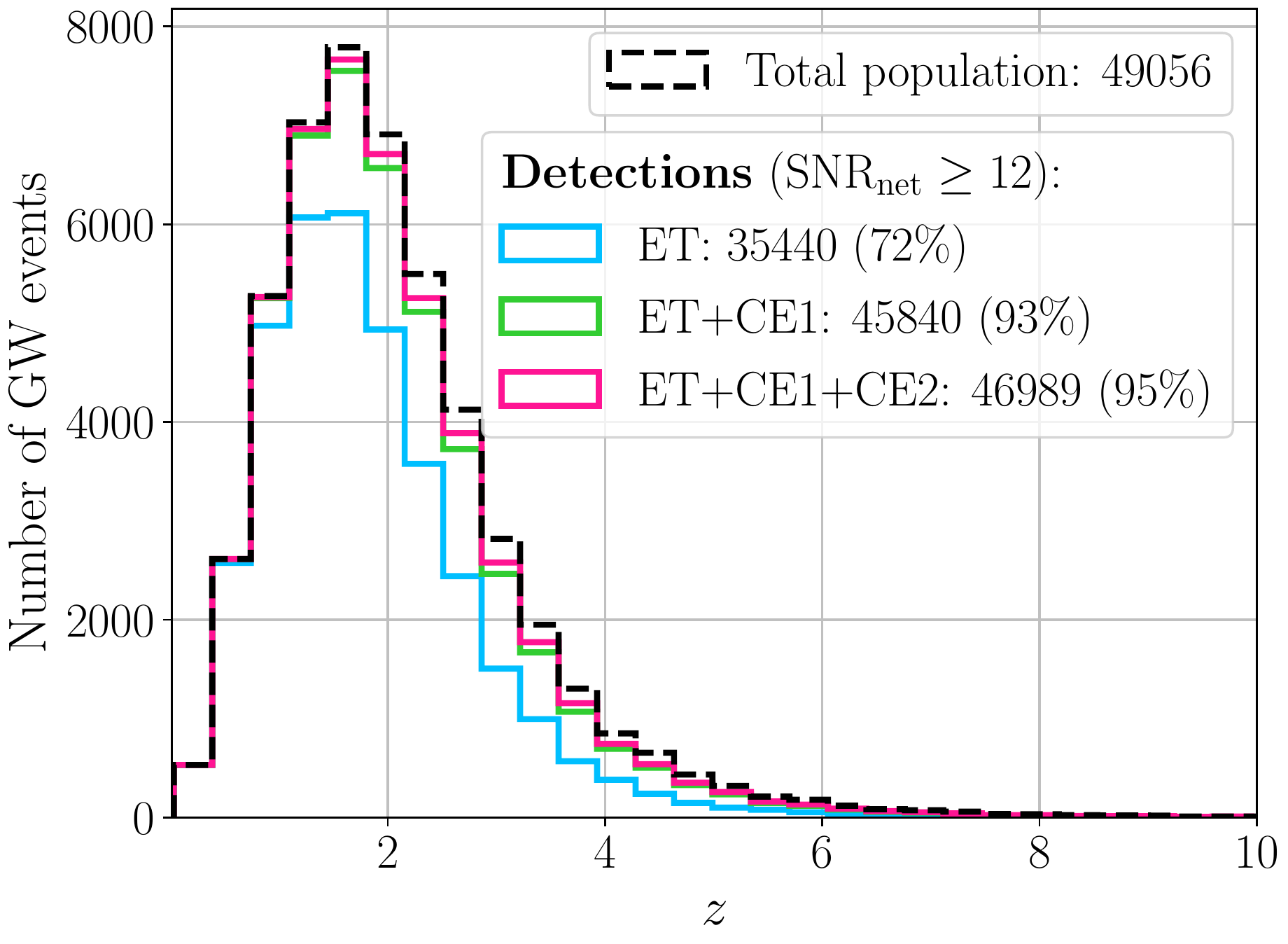}
    \includegraphics[width=0.4973\textwidth]{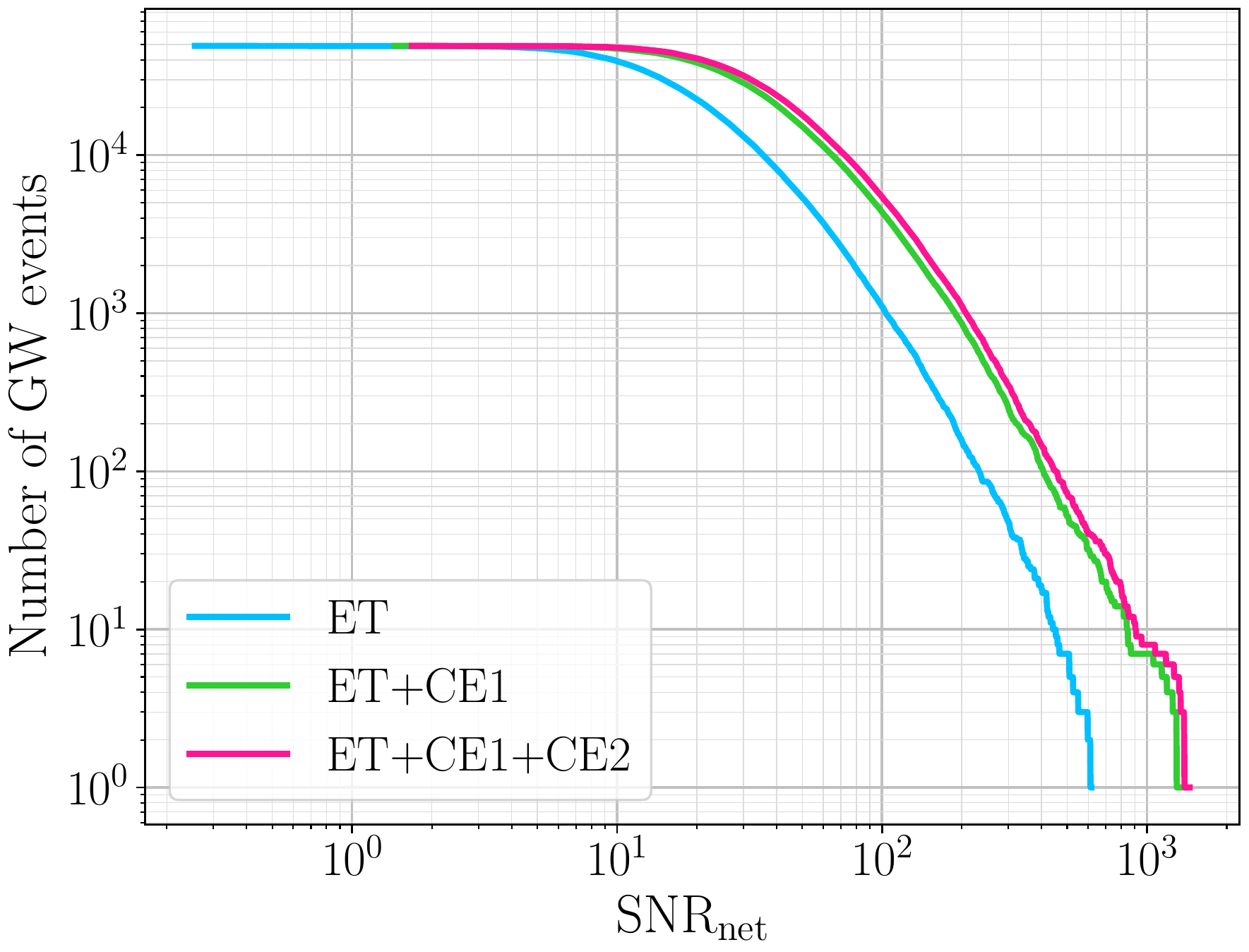}
    \caption{Left: Redshift distributions of the total and detected population for different networks in one year of observation (full duty cycle), colors as in legend. Right: Number of detected GW events left above a given $\SNR_{\rm net}$, colors as in legend.
    }
    \label{fig:total_and_detected_pops}
\end{figure*}

\subsection{The Fisher matrix approach}

The parameter estimation of a GW source often requires large computational resources due to the vast parameter space that needs to be explored with sampling methods.
The Fisher information matrix (FIM) framework offers a more accessible way to assess the measurement capabilities of detector networks~\cite{cutler1994gravitational} in the strong-signal limit by approximating the parameter posteriors to be Gaussian (under the assumption of Gaussian noise), thanks to the analytic computation of estimators that allow to estimate the expected uncertainties affecting the measured parameters.
We stress that this approach is robust under optimal conditions, notably high SNR, which is the typical situation that we consider in this work, and it represents the approximation of a much more complex statistical analysis.

The general expression for the output $s(t)$ of a detector can be written as the sum of a noise term $n(t)$ and a possible GW signal term $h(t)$:
\begin{equation}
    s(t) = n(t) + h(t) \, .
\end{equation}
Under the assumption that $n(t)$ is a stochastic, stationary, and Gaussian function of time, the loudness of the signal can be quantified through the signal-to-noise ratio (\snr), defined as
\begin{equation}
    \SNR = \inner{h}{h}^{1/2} \, .
\label{eqn:snr}
\end{equation}
Here the parentheses denote the inner product, which reads
\begin{equation}
\inner{A}{B} = 4 \, \text{Re} \int_{0}^{+\infty} df \, \frac{\tilde A^{*}(f) \tilde B(f)}{\psd (f)} \, ,
\label{eqn:inner}
\end{equation}
where the tilde symbol labels Fourier transformed quantities and a star denotes the complex conjugate. The inner product is weighted over the one-sided power spectral density $\psd$ of a detector, which quantifies the sensitivity of an interferometer per frequency bin and is measured in \si{\hertz^{-1}}. For a network of $M$ detectors, the total $\SNR_{\rm net}$ is the square root of the quadratic sum of the individual $\SNR$s:
\begin{equation}
    \SNR_{\rm net} = \sqrt{\sum_{k=1}^{M} \SNR_k^2} \, .
\end{equation}

In the presence of a high $\SNR$ signal, the posterior distribution for the $n$ source parameters $\{ \Theta_1, \, ... \, , \Theta_i, \, ... \, , \Theta_j, \, ... \, , \Theta_n\}$ can be well approximated to an $n$-dimensional Gaussian with covariance matrix
\begin{equation}
    \CM = \FM^{-1} \, ,
\label{eqn:covariance_matrix}
\end{equation}
where 
\begin{equation}
    \FM_{ij} = \inner{\de{i}h}{\de{j}h} \, ,
\label{eqn:fisher_matrix}
\end{equation}
is the Fisher matrix and $\de{i} \equiv \partial/\partial\Theta_i$ is the partial derivative with respect to the $i$th source parameter $\Theta_{i}$. When multiple detectors are involved, the Fisher matrix of the network is  given by the sum of the individual ones:
\begin{equation}
    \FM_{\rm net} = \sum_{k=1}^M (\FM)_k \, .
\end{equation}

This formalism allows to access the uncertainty of the $i$th parameter by trivially taking the square root of the diagonal element $\CM_{ii}$ of the covariance matrix, i.e.,
\begin{equation}
    \sigma_{i} = \sqrt{\CM_{ii}} \, .
\end{equation}

The computation of these quantities requires careful numerical implementation. Specifically, in order to obtain the Fisher matrix, we first need to evaluate the partial derivative of the waveform with respect to each source parameter. An effective way to do that is by performing a symmetric derivative:
\begin{equation}
    \de{i} h (\Theta_i) = \lim_{\delta_i \to 0^+} \frac{h(\Theta_i + \delta_i \Theta_i) - h(\Theta_i -\delta_i \Theta_i)}{2\delta_i \Theta_i} \, .
\label{eqn:symmetric_derivative}
\end{equation}
When computed numerically, the symmetric derivative is no longer a limit. Thus, to compute \cref{eqn:symmetric_derivative} we must choose the magnitude of the infinitesimal increment $\delta_i$ for each parameter. 
We compute $\FM_{ij}$ for different values of $\delta_i$ and $\delta_j$, and we validate our choice as soon as the Fisher matrix element becomes a stable function of the infinitesimal increments. We report the values\footnote{We stress that $\delta_i$ is a dimensionless value, and that the product $\delta_i \Theta_i$ enters in the computation of the derivative.} we adopted for each parameter in \cref{tab:selected_parameters}.
Once we obtain the Fisher matrix for each source, we compute the correlation matrix $\CM$ through \cref{eqn:covariance_matrix} and by adopting the lower-upper decomposition method \cite{PresTeukVettFlan92}.
We validate the inversion process and check the resulting matrix by evaluating
\begin{equation}
    \epsinv = \max_{i, \, j} |(\FM \cdot \CM)_{ij} - \mathbb{1}_{ij} | \, ,
\label{eqn:epsilon_inv}
\end{equation}
where $\mathbb{1}$ is the identity matrix. We consider the inversion successful if \cref{eqn:epsilon_inv} returns $\epsinv \le \num[retain-unity-mantissa = false]{1e-3}$. Moreover, we validated our implementation by comparing results with other public pipelines (e.g.~\textsc{GWFAST}\cite{Iacovelli_2022}, \textsc{GWFish}\cite{dupletsa2023gwfish}). Specifically, we made common injections for each library and compared $\SNR$ values, as well as the uncertainties on the source parameters, with a particular focus on luminosity distance and the sky location, finding broad agreement overall. These tests were also used as benchmark to check our choice of the infinitesimal increments $\delta_i$.

The quasicircular, nonprecessing waveform model \texttt{IMRPhenomXHM}, once projected in a detector, is characterized by \num{11} source parameters: the two individual masses $M_1$ and $M_2$, the luminosity distance $\dl$ of the source, the $z$-component of the two spins $\chi_{z_1}$ and $\chi_{z_2}$, the sky position parameters $\theta$ and $\varphi$, the inclination $\iota$ and polarization $\psi$ angles, the coalescence phase $\Phi_c$ and time $t_c$.
We characterize the Fisher matrix with a different set of parameters: in particular, the two individual masses are replaced by the redshifted chirp mass $\mchirp = (1 + z)(M_1 M_2)^{3/5} / (M_1+M_2)^{1/5}$ and the symmetric mass ration $\eta = (M_1 M_2)/(M_1+M_2)^2$. We further take the natural logarithm of $\mchirp$ and $\dl$ to obtain directly their relative errors in the correlation matrix. We define $\mu = \cos\theta$ as a parametrization of the declination, and the two individual $z$-oriented spins are replaced by two orthogonal symmetrical and asymmetrical combinations $\chi_S = (\chi_{z_1} + \chi_{z_2})/2$ and $\chi_A = (\chi_{z_1} - \chi_{z_2})/2$.
The full list of parameters is reported in \cref{tab:selected_parameters}. 
We consider a GW event as detected if $\SNR_{\rm net} \ge 12$ and then compute the Fisher matrix for the set of detected sources.

\begin{table}[ht]
\centering
%\bgroup
%\setlength{\tabcolsep}{0.5em} % horizontal padding
\def\arraystretch{\tabvspace} %  vertical padding (default is 1)
\begin{tabular}{ l c c }
\hline
\hline
\textbf{Quantity} & \textbf{Parametrization} & \textbf{$\delta_i$}\\
\hline
Redshifted chirp mass & $\ln \mchirp$ & \num[retain-unity-mantissa = false]{1e-6}\\

Symmetric mass ratio & $\eta$ & \num[retain-unity-mantissa = false]{1e-6}\\

Luminosity distance & $\ln \dl$ & \num[retain-unity-mantissa = false]{1e-5}\\

Symmetric spin & $\chi_S$ & \num[retain-unity-mantissa = false]{1e-5}\\

Asymmetric spin & $\chi_A$ & \num[retain-unity-mantissa = false]{1e-5}\\

RA  & $\varphi$ & \num[retain-unity-mantissa = false]{1e-3}\\

DEC & $\mu$ & \num[retain-unity-mantissa = false]{1e-2}\\

Inclination & $\iota$ & \num[retain-unity-mantissa = false]{1e-5}\\

Polarization & $\psi$ & \num[retain-unity-mantissa = false]{1e-6}\\

Coalescence phase & $\Phi_c$ & \texttt{None}\\

Coalescence time & $t_c$ & \texttt{None}\\
\hline
\hline
\end{tabular}
\caption{
Quantities computed in
the Fisher matrix (parametrizations defined in the main text), together with the associated infinitesimal increment $\delta_i$ used to compute \cref{eqn:symmetric_derivative}. The \texttt{None} label marks the parameters whose derivatives are performed analytically.
}
\label{tab:selected_parameters}
\end{table}

\begin{figure*}
    \includegraphics[width=1.\textwidth]{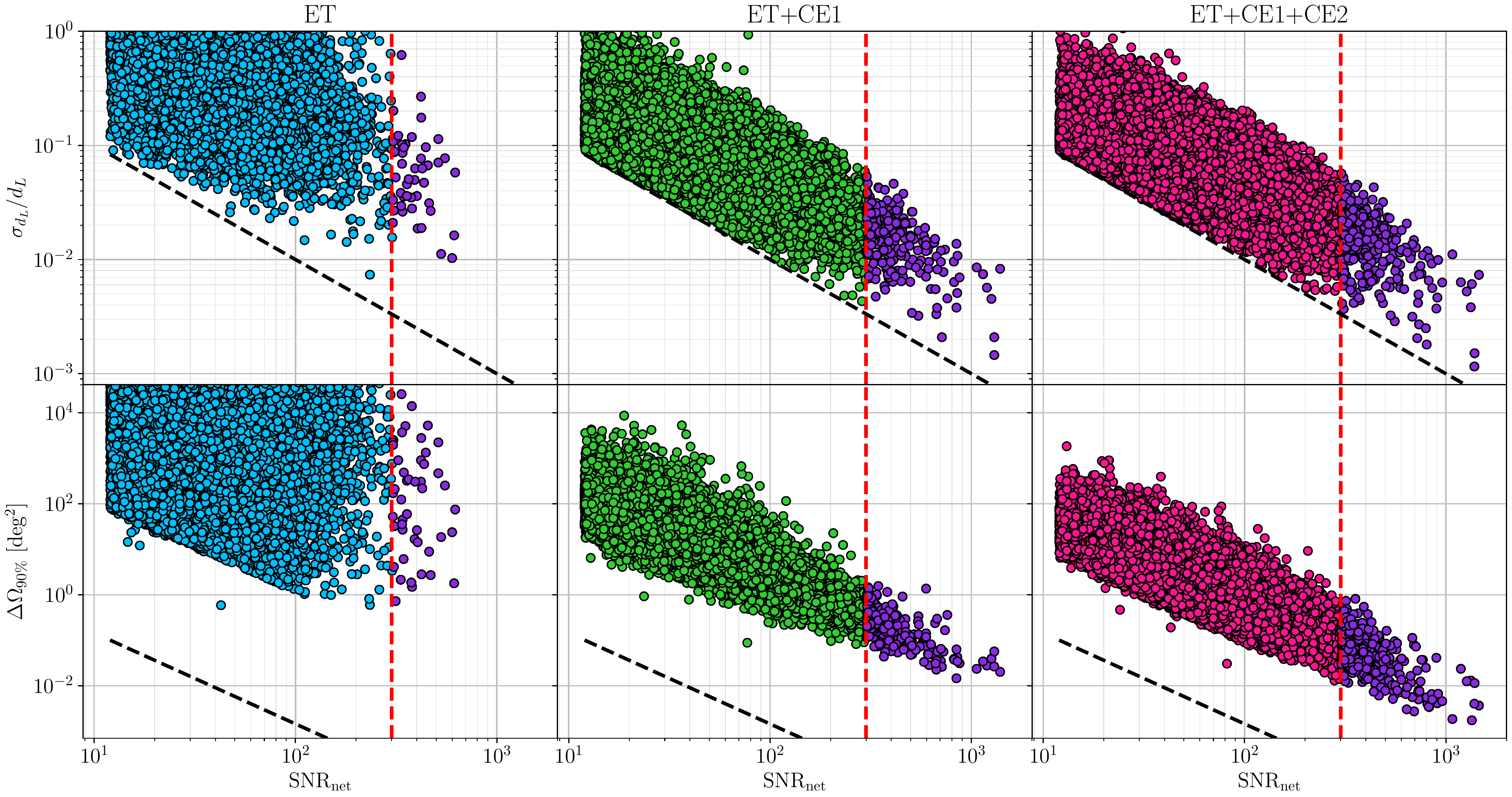}
    \caption{ 
    Correlation between GW parameter uncertainties and $\SNR_{\rm net}$ for different network configurations. Each circle represents a detected GW event in a specific network, labeled at the top. In the top row, the black dotted lines represent $\sigma_{\dl}/\dl = \SNR_{\rm net}^{-1}$, while in the bottom row they display the $90\%$ credible region $\Delta\Omega_{90\%} = -2 \pi \ln(1 - 90/100) \SNR_{\rm net}^{-2}$ converted in \si{\square\degree}. The red dotted vertical line marks the minimum $\SNR_{\rm net}$ threshold that we set to create localization error volumes. The purple-filled circles are the GW events above that threshold.
    }
    \label{fig:summary_PE}
\end{figure*}

We now discuss the FIM results. 
In \cref{sec:redshift_distribution} we estimated $\mathcal{O}(\num[retain-unity-mantissa = false]{1e5})$ GW sources per year, distributed over cosmological distances. 
Interestingly, we find that a combination of at least two 3G detectors will be able to detect the vast majority of the simulated population, while ET alone would still be able to detect a large fraction of the total BBHs, as reported in the left panel of \cref{fig:total_and_detected_pops}. These GW events will be detected potentially up to high redshift: such result would be the key for population studies to grow, leading to a better understanding of binary formation and evolution (see, e.g., \cite{taylor2018mining,mould2022deep}) across cosmic history with just \SI{1}{\year} of observations.
In the right panel of \cref{fig:total_and_detected_pops} we report the number of detections as a function of $\SNR_{\rm net}$. We see that a large number of the GW signals will be detected above $\SNR_{\rm net} \sim 100$, with the best cases given again by combinations of ET with at least one CE.
Under the FIM formalism, the parameters associated to these sources are characterized by Gaussian posterior distributions, with progressively smaller width as the $\SNR$ increases. 
Focusing on the localization of the source, which is crucial for dark siren studies, we expect $\sigma_{\dl}/\dl$ and the sky location area $\Delta\Omega$ to scale with $\SNR_{\rm net}^{-1}$ and $\SNR_{\rm net}^{-2}$ respectively, as one can see from \cref{eqn:sigma_dL_and_delta_omega}.
Our expectations are confirmed as illustrated in \cref{fig:summary_PE}, where we show that these trends are well recovered. The progressively narrower dispersion of the points remarks also the importance of having multiple detectors in the network of interferometers, as this allows to break degeneracies and triangulate the GW signal. Furthermore, higher harmonics in the waveform model help disentangling the luminosity distance and inclination contributions to the signal's amplitude, leading to better constraints on these parameters. We find reasonable agreement with similar studies which also make use of higher modes, see e.g.~\cite{borhanian2022listening,iacovelli2022forecasting,pieroni2022detectability}.

%%%%%%%%%%%%%%%%%%%%%%%%%%%%%%%%%%%%%%%%%%%%%%%%%%%%%%%%%%%%%%%%%%%%%%%%%%%%%%%%%%%%%%%%%%%%%%%%%%%%%%%%%%%%%%%%%%%%%%%%%%%%%%%%%%%%%%%%%%%%%%%%%%%%%%%%%%%%%%%%%%%%%%%%%%%%%%%%%%%%%%%%%%%%%%%%%%%%%%%%%%%%%%%%%%%%%%%%%%%%%%%%%%%%%%%%%%%%%%%%%%%%%%%%%%%%%%%%%%%%%%%%%%%%%%%%%%%%%%%%%%%%%%%%%%%%%%%%%%%%

\section{Cross-matching with galaxy catalogs}
\label{sec:errorboxes}

In this section we outline our simulation procedure for galaxy catalogs and explain how we cross-match localization error volumes with potential host galaxies of the GW source.

\subsection{Galaxy catalog}
\label{subsec:gal_cats}

The galaxy catalog employed in this work is generated by using \texttt{L-Galaxies} \citep{Henriques2015}, a state-of-the-art semi-analytical model (SAM) applied on top of the merger trees of dark matter simulations. Specifically, we run the SAM on the Millennium simulation \citep{Springel2005} whose halo mass resolution (\SI{2e10}{\msun}) and box size (\SI{685}{\mega\parsec}) offer a good compromise to trace the cosmological assembly of galaxies with stellar mass $M_{*} > \SI[retain-unity-mantissa=false]{1e9}{\msun}$. By using the procedure presented in \cite{IzquierdoVillalba2019} we transform the outputs of \texttt{L-Galaxies} into a customized lightcone which embraces one octant of the sky and contains the physical properties (such as mass, magnitudes, observed and cosmological redshift) of all the galaxies with $M_{*} > \SI[retain-unity-mantissa=false]{1e10}{\msun}$ up to $z = 3$. Regarding the cosmological parameters, the version of \texttt{L-Galaxies} used in this work re-scales the original values used in the Millennium (WMAP1 and 2dFGRS concordance cosmology) to match the ones of Planck first-year data \citep{PlanckCollaboration2014} (which is the fiducial cosmology defined in \cref{sec:redshift_distribution}).

In this work, we explore two different scenarios. 
In our main scenario, we study low-$z$ GW sources ($z<1$) whose galaxy fields could be traced by current galaxy catalogues provided by SDSS Legacy Survey~\citep{SDSS2000} or upcoming missions like EUCLID~\citep{EUCLID2012} and WFIRST~\citep{WFIRST2015}. On the other hand, we consider an optimistic scenario where we take advantage of the full redshift depth of the lightcone and analyze high-$z$ GW events ($z < 3$). In this case, future deep surveys like LSST will be able to provide complete photometric galaxy catalogs \citep{LSST2019}.

Galaxies with $M_{*} < \SI[retain-unity-mantissa=false]{1e10}{\msun}$ are more numerous than more massive galaxies, but harder to detect (see, e.g., \cite{Baldry2008,Yasuda2001,Rovilos2009, DominguezSanchez2011}).
Future surveys are likely to reflect this challenging issue (especially at high-$z$), leading to incompleteness effects in the low-mass regime similar to the ones reproduced in our mock catalog.
On the other hand, we expect that dwarf galaxies have a secondary role in cosmological inference given that their low stellar masses reduce their probability of hosting GW events.
Taking into account all this, we assume that BBHs cannot be hosted by galaxies with $M_{*} < \SI[retain-unity-mantissa=false]{1e10}{\msun}$, which could be implemented in our cosmological inference by adding a mass-dependent weight to each galaxy (see \cref{eqn:prior_zgw} below).
We refer to our catalog as ``complete'' under these assumptions.

\subsection{Gravitational-wave localization error volumes}\label{subsec:error_box}

We can now use the parameter uncertainties estimated with the FIM analysis to estimate the localization error volume, or ``error-box,'' of each GW event. Each error-box will be populated by the galaxies contained in our light cone. To this end, we are mainly interested in the measurements of the GW luminosity distance and the sky position.
Given the parametrization listed in \cref{tab:selected_parameters}, we recover their errors through
\begin{equation}
\begin{split}
    \frac{\sigma_{\dl}}{\dl} & = 
     \sqrt{\biggl(  \frac{\sigma_{\dl}}{\dl} \biggr)^2_{\rm GW} + \biggl( \frac{\sigma_{\dl}}{\dl} \biggr)^2_{\rm WL}} \, , \\
    \Delta\Omega_{\rm X\%} & = - 2 \pi \sqrt{\CM_{\mu\mu}\CM_{\varphi\varphi} - (\CM_{\mu\varphi})^2} \ln\biggl(1 - \frac{\rm X}{100}\biggr) \, ,
\end{split}
\label{eqn:sigma_dL_and_delta_omega}
\end{equation}
where $(\sigma_{\dl}/\dl)_{\rm GW} = \sqrt{\CM_{\ln\dl \, \ln\dl}}$ is the uncertainty coming from the Fisher matrix, while $(\sigma_{\dl}/\dl)_{\rm WL}$ keeps into account for the contribution of weak lensing (WL) to the measure. As described in \cite{Tamanini:2016zlh}, we model this term through\footnote{As noted in~\cite{Cusin:2020ezb}, we correct for a missing factor 1/2 in this expression with respect to the one reported in~\cite{Tamanini:2016zlh}. No demagnification is considered here, contrary, e.g., to~\cite{Speri:2020hwc}.}

\begin{equation}
    \biggl( \frac{\sigma_{\dl}}{\dl} \biggr)_{\rm WL} = 0.033 \biggl(\frac{1 - (1+z)^{-0.25}}{0.25}\biggr)^{1.8} \, .
    \label{eqn:weak_lensing}
\end{equation}

In \cref{eqn:sigma_dL_and_delta_omega}, $\Delta\Omega_{\rm X\%}$ is expressed in \si{\steradian} and $\rm X\%$ represents the percent confidence interval (CI) of the measure \cite{PhysRevD.81.082001,Iacovelli_2022}. 
We compute $\Delta\Omega_{90\%}$ so to be able to compare our results with 
most of what can be found in the literature.

Since our ensemble of GW events is generated independently
of the galaxy catalog, we need to cross-match the eligible binaries within the Millennium Universe. 
The following procedure is applied to a subset of the simulated GW events. The eligible BBHs must satisfy $\SNR_{\rm net} > 300$: this selection ensures great precision both in luminosity distance and sky location, two essential requirements that limit the potentially prohibitive number of galaxies $N_{\rm hosts}$ per GW localization error volume. 
However, ET alone does not provide very accurate sky localization and luminosity distance measurements. Therefore, when we consider ET alone, we further require binaries to feature $3\sigma_{\mu} = 3 \sqrt{\CM_{\mu\mu}}$ and $3\sigma_{\varphi} = 3\sqrt{\CM_{\varphi\varphi}}$ small enough to be able to fit within the angular aperture of the light cone. For a similar reason, we also require $3\sigma_{\dl}/\dl \le 1$. 

For each eligible dark siren, we follow the procedure detailed below: 

\begin{enumerate}

    \item Assuming our fiducial cosmology,
    we compute the redshift interval $z \pm \num{3}\sigma_z$ from the $\num{3}\sigma$ measurement of $\dl$, and we list all the galaxies whose cosmological redshift lies within this range. Among them, we extract a galaxy with probability given by $\mathcal{N}(\dl,\sigma^2_{\dl})$, and we label it  ``true host'' of the GW event. We denote the true host sky coordinates by ${\bf{\Theta}}_{\rm th}=\{ \mu_{\rm th}, \varphi_{\rm th}\}$.
    \item Next, we extract the center of the localization error volume ${\bf{\Theta}}_{c}=\{ \mu_{c}, \varphi_{c}\}$, namely we redefine the maximum of the 2D Gaussian in $\{ \mu, \varphi\}$, ensuring that the true host falls within a $3\sigma$ sky location region from it and is consistent with the GW sky location uncertainty computed in~\cref{sec:GWPE}. To do so, we draw ${\bf{\Theta}}_{c}$ from $\mathcal{N}({\bf{\Theta}}_{\rm th}$, $\CM_{\Delta\Omega}$), where $\CM_{\Delta\Omega}$ is the 2D sky location subcovariance matrix of the GW signal. This new point redefines the best direction (i.e.~the peak of the Gaussian) measured by the GW detector.
    \item We then compute the redshift boundaries of the localization error volume that we will use for the cosmological inference. This needs to take into account the full prior ranges of the cosmological parameters, otherwise we would implicitly assume a prior given by our fiducial cosmology.
    We refer to this new, much broader interval as $[z^{-}, \, z^{+}]$.
    In practice this is computed as $z^{-} = \min \left[ z - \num{3}\sigma_z \right]$ ($z^{+} = \max \left[ z + \num{3}\sigma_z \right]$) where the $\min$ ($\max$) is taken with respect to all possible values of the cosmological parameters within the allowed priors that we assume to be $h \in [0.6, 0.86]$, $\Om \in [0.04, 0.5]$.
    \item Since peculiar velocities affect galaxy redshift measurements, we need to model the related uncertainty. Following \cite{Laghi_2021, Muttoni:2021veo}, we characterize this effect as~\cite{hogg1999}
    \begin{equation}
        \sigma_{v_p}(z) = (1 + z)\frac{v_p}{c} \, ,
        \label{eqn:sigma_pv}
    \end{equation}
    where $v_p = \SI{700}{\kilo\meter\,\second^{-1}}$ is representative of the standard deviation of the radial peculiar velocity distribution of the galaxies in the catalog.
    The redshift interval is therefore updated to $[z^{-} - \sigma_{v_p}(z^{-}), \, z^{+} + \sigma_{v_p}(z^{+})]$.
    \item We populate the localization error volume with all the galaxies that fall within a $\num{3}\sigma$ sky location region from the center ${\bf{\Theta}}_{c}$ and inside $[z^{-} - \sigma_{v_p}(z^{-}), \, z^{+} + \sigma_{v_p}(z^{+})]$. These galaxies represent the potential host candidates of the GW event.
    \item For each potential host in the localization error volume, labeled by $j$ ($j=1,...,N_{\rm hosts}$), we compute a normalized galaxy ``weight'' $w_j$ according to its position in the sky relative to the center of the localization volume:
    \begin{equation}\label{eqn:weights}
    w_j \propto \mathcal{N}({\bf \Theta}_{c}, \CM_{\Delta \Omega})\big|_{(\mu_{j}, \varphi_{j})} \, .
    \end{equation}
\end{enumerate}

\begin{figure*}[t!]
    \includegraphics[width=1.\textwidth]{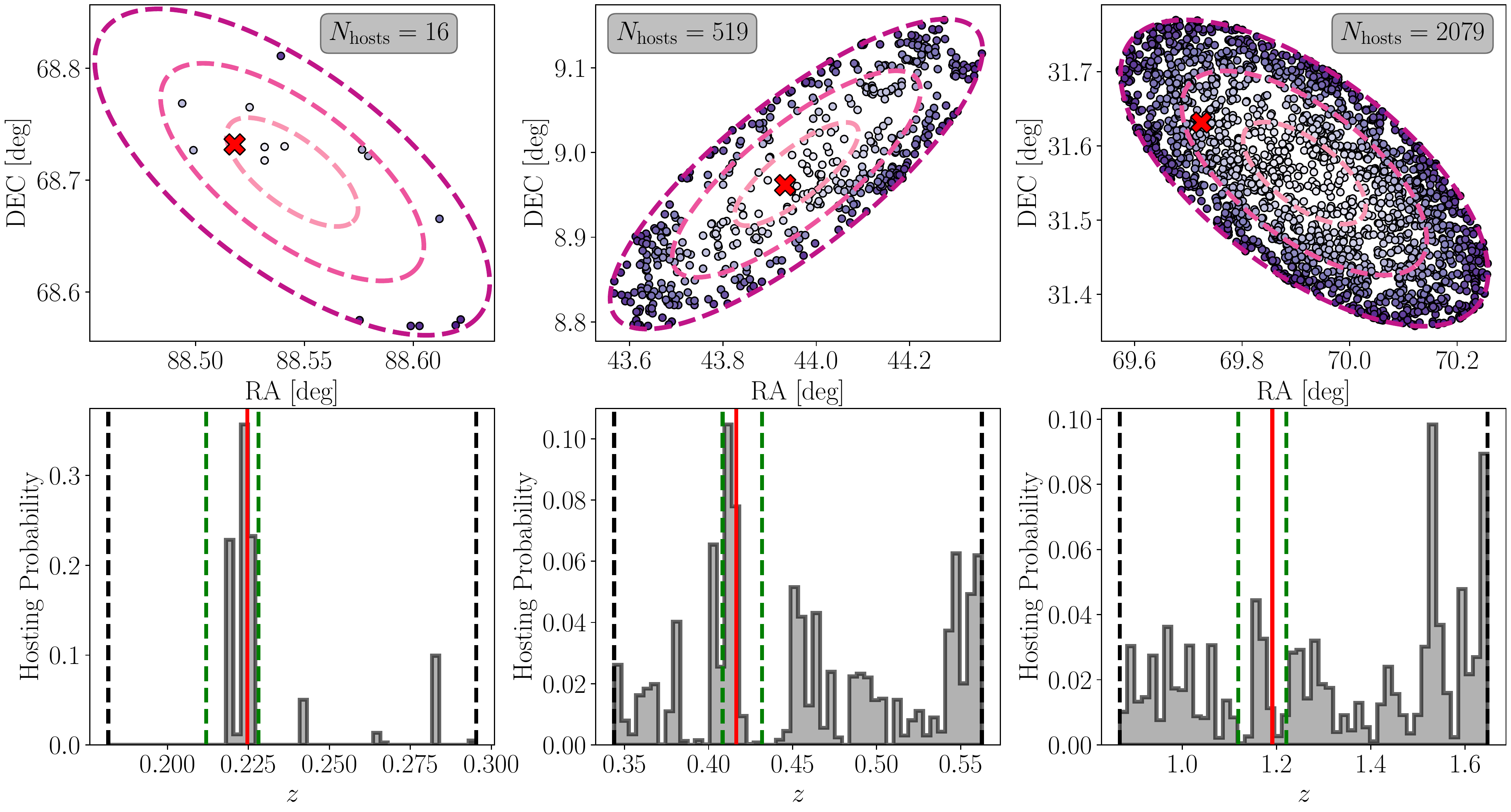}
    \caption{Graphic representation of three representative localization error volumes, displayed per column. The top panel shows the distribution of the $N_{\rm hosts}$ galaxies (circles) falling inside it on the RA-DEC plane, where the color-scale represents the magnitude of the hosting probability, as computed from \cref{eqn:weights}, from white (high) to purple (low), while the dashed ellipses denote the $1\sigma$, $2\sigma$, and $3\sigma$ probability contours. A red cross marks the selected true host. The bottom panel displays how these galaxies are distributed in redshift. Here the red solid line marks the selected true host, the green dashed lines denote the redshift interval obtained by inverting \cref{eqn:luminosity_distance_flat} assuming $\dl \pm 3\sigma_{\dl}$ and our fiducial cosmology, while the black dotted lines are the boundaries of the localization error volume that take into account the full prior range on the cosmological parameters and galaxy peculiar velocities, computed as described in \cref{subsec:error_box}.
    }
    \label{fig:error-box}
\end{figure*}

Once built, we only consider the localization volumes whose $z^{+} + \sigma_{v_p}(z^{+})$ value does not exceed a maximum redshift threshold, which is $z_{\rm max}=1$ for our fiducial case, and $z_{\rm max}=3$ for the optimistic one. We illustrate the outcome of this procedure in \cref{fig:error-box}.

\begin{figure*}
    \includegraphics[width=1.\textwidth]{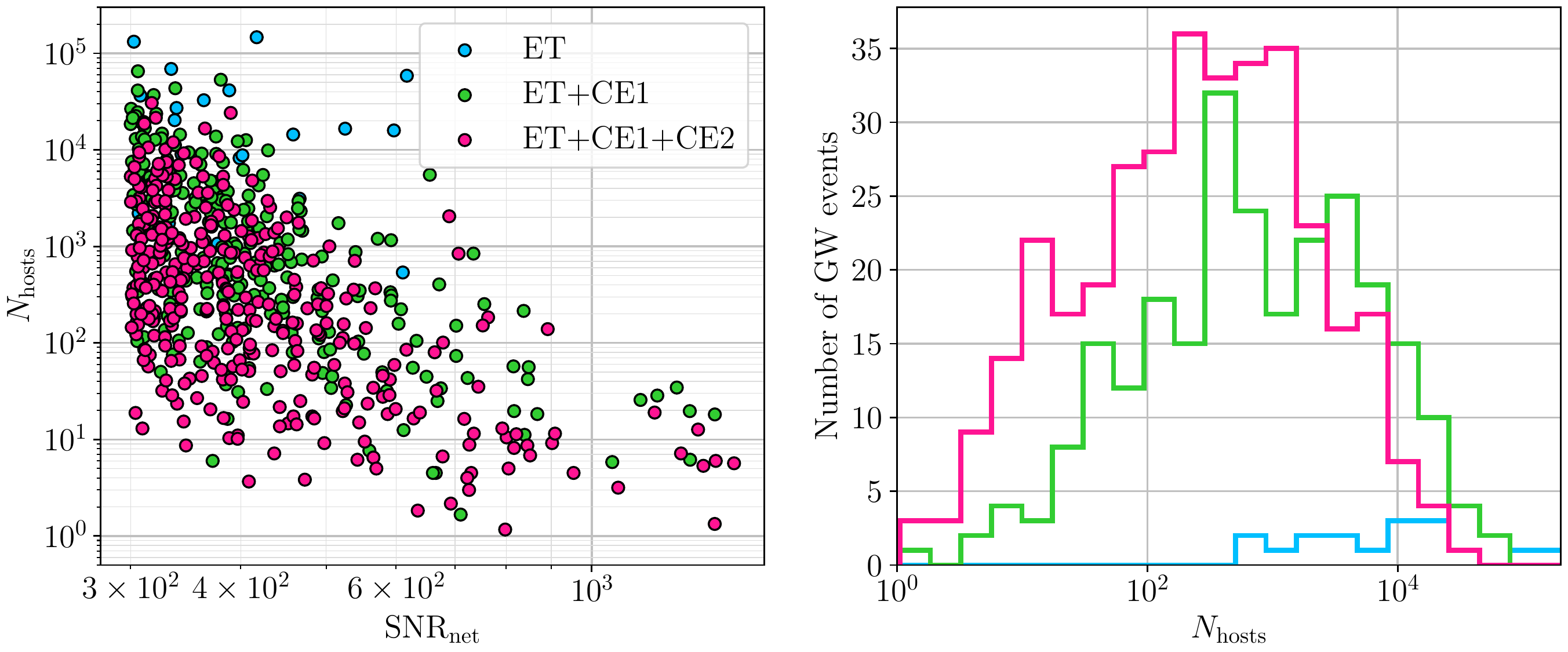}
    \caption{
    Left: correlation between the number of hosts $N_{\rm hosts}$ per error-box and the $\SNR_{\rm net}$ of the associated dark siren. Each circle represents the localization error volume of a GW event that satisfies the conditions detailed in \cref{subsec:error_box} in a specific network (labels in the legend). For this reason, the \num{3} networks do not share the same number of GW events. Right: distribution of $N_{\rm hosts}$ in the generated localization error volumes, colors as in the left plot. The values of $N_{\rm hosts}$ in both panels are averaged over the \num{6} realizations considered in this study.
    }
    \label{fig:N_hosts_statistics}
\end{figure*}

Each network observes the BBHs population with different $\SNR_{\rm net}$ and recovers different values of the uncertainties. We therefore expect different number of localization error volumes for each configuration of detectors, as shown in \cref{fig:N_hosts_statistics}, where we also display how $N_{\rm hosts}$ scales with the $\SNR_{\rm net}$. Nonetheless, it is worth noting that the same GW event can satisfy the conditions detailed above in more than one network.

%%%%%%%%%%%%%%%%%%%%%%%%%%%%%%%%%%%%%%%%%%%%%%%%%%%%%%%%%%%%%%%%%%%%%%%%%%%%%%%%%%%%%%%%%%%%%%%%%%%%%%%%%%%%%%%%%%%%%%%%%%%%%%%%%%%%%%%%%%%%%%%%%%%%%%%%%%%%%%%%%%%%%%%%%%%%%%%%%%%%%%%%%%%%%%%%%%%%%%%%%%%%%%%%%%%%%%%%%%%%%%%%%%%%%%%%%%%%%%%%%%%%%%%%%%%%%%%%%%%%%%%%%%%%%%%%%%%%%%%%%%%%%%%%%%%%%%%%%%%%

\section{Cosmological inference}
\label{sec:cosmoinference}

We adopt the \emph{dark siren} approach to infer the cosmological parameters.
We thus cross-match GW luminosity distance information with EM galaxy catalog redshifts through the luminosity distance-redshift relation in a \emph{flat} $\lcdm$ model, as explained in~\cref{sec:errorboxes}. In what follows we discuss the Bayesian framework apt to make the inference of the cosmological parameters.

\subsection{Bayesian formulation}

We are interested in using a set of GW observations $\mathcal{D} = \{ \mathcal{D}_i \}_{i=1}^N$ to jointly constrain the cosmological parameters $\Omega=\{ h,\, \Om\}$ in a flat $\lcdm$ model, namely the two parameters that determine the dynamics of the Universe at the background level.
We represent the $\lcdm$ cosmological model by $\mathcal{H}$ and any background information that is useful for the inference problem with $I$, including the redshift information about the galaxy catalog.
Our mathematical framework is mainly based on~\cite{Laghi_2021}, of which we summarize the main concepts here (see also~\cite{DelPozzo:2011yh,DelPozzo:2015bna,DelPozzo:2017kme}).

From Bayes' theorem, the posterior probability distribution can be computed as 
\begin{equation}\label{eqn:bayes}
    p(\Omega | \mathcal{D} , \mathcal{H}, I) \propto p(\Omega | \mathcal{H}, I) \, p(\mathcal{D} | \Omega , \mathcal{H}, I) \text{,}
\end{equation}
where $p(\Omega | \mathcal{H}, I)$ is the prior probability distribution for the cosmological parameters $h$ and $\Om$, while $p(\mathcal{D} | \Omega , \mathcal{H}, I)$ is the likelihood function for the GW dataset $\mathcal{D}$, that, for statistically independent events, can be written as:
\begin{equation}\label{eqn:lk}
    p(\mathcal{D} | \Omega , \mathcal{H}) = \prod_{i=1}^{N} p(\mathcal{D}_i | \Omega , \mathcal{H}) . 
\end{equation}

We choose our dataset according to the $\SNR_{\rm net}$ of the event.
The assumption of an $\SNR_{\rm net}$ detection threshold as a proxy for the detectability of an event has the effect to constrain the range of GW luminosity distances, which, in turn, can bias the estimate of the cosmological parameters. Accounting for the GW selection effect requires to normalise the single-event likelihood by means of a selection function, which is estimated as the single-event likelihood integrated over all datasets that would be classified as detected according to our detection statistic~\cite{mandel2019extracting,vitale2022inferring}.
However, in this study we focus on high-SNR events, which allows us to partially simplify the inference problem, not including selection bias corrections and reducing the already significant computational cost of the analysis. 
We find indeed that, for all the $\SNR_{\rm net}$ thresholds considered in this analysis, selection biases are subdominant compared to statistical uncertainties (see~\cref{sec:results}). The correction for GW selection effects typically requires to assume a detection model and a population model; since we ignore selection effects, our cosmological inference is thus independent of any assumption on population models. 
The reason why we do not observe any sensible systematic error can be explained by the fact that, due to the precision of the GW parameter estimation in the $\SNR_{\rm net}$ limit, the selection function is only weakly dependent on the cosmological parameters; thus, the selection function, which can be seen as a correction factor to \cref{eqn:bayes}, can be approximated as an overall constant which does not significantly change from event to event. 
As we include less informative events in our analysis, that is, lowering the $\SNR_{\rm net}$ threshold, this argument does not hold and selection biases can become comparable to statistical uncertainties, significantly affecting posterior estimates (see, e.g., ~\cite{2022arXiv221208694G}).
In our simulation we find that lowering the $\SNR_{\rm net}$ threshold to 200 already yields a non-negligible bias.
Moreover, the inclusion of GW events with $\SNR_{\rm net} < 300$ pose as well challenges from the point of view of the computational cost of the analysis, as discussed in~\cref{subsec:numerics}. For these reasons, in this study we limit ourselves to the analysis of the most informative events with $\SNR_{\rm net} > 300$.

Another potential source of bias may come from the incompleteness of the galaxy catalog \cite{chen17,Finke:2021aom,Gray2022,2022arXiv221208694G}. In this analysis we use a light cone that by construction contains all the galaxies with $M_{*} > \SI[retain-unity-mantissa=false]{1e10}{\msun}$ up to $z = 3$, irrespective of their magnitude: this allows us to simplify the formalism and assume the completeness of the galaxy catalog up to the redshift covered by the light cone (see~\cref{subsec:gal_cats} for a discussion on the validity of this assumption in the context of 3G detectors).

After marginalization over the GW nuisance parameters which are not relevant for this analysis, and assuming that we can neglect the correlation between the detector measurements of the angular coordinates and distance, so that the joint GW likelihood on sky position and distance parameters factorizes, the single-event ``quasilikelihood''~\cite{jaynes2003} can be written as~\cite{Laghi_2021}
\begin{equation}\label{eqn:single-event-lk1}
\begin{split}
p(\mathcal{D}_i\,|\,\Omega,\mathcal{H}, I) \propto
\int  &dd_L \,dz_{\rm GW}\,\,
p(d_L \,|\, z_{\rm GW},\Omega,\mathcal{H}, I) \times \\
& \!\!\!\!\times p(z_{\rm GW} \,|\, \Omega,\mathcal{H}, I)\,
p(\bar{d}_L \,|\, d_L,z_{\rm GW},\mathcal{H}, I)\text{.}
\end{split}
\end{equation}
The term $p(d_L \,|\, z_{\rm GW},\Omega,\mathcal{H}, I)$
is the probability of obtaining the luminosity distance assuming we know the source redshift and the cosmological parameters:
\begin{equation}
    p(d_L\,|\,z_{\rm GW},\Omega,\mathcal{H}, I) = \delta(d_L - d(z_{\rm GW}, \Omega)).
\end{equation}
The GW redshift prior term $p(z_{\rm GW} \,|\, \Omega,\mathcal{H}, I)$ reflects the properties of the localization volume and the potential host galaxies that are within it:
\begin{equation}\label{eqn:prior_zgw}
    p(z_{\rm GW}\,|\,\Omega,\mathcal{H}, I)  \propto
    \sum_{j=1}^{N_{\rm hosts}} w_j \,
    \mathcal{N}(z_{\rm GW},\sigma^2_{v_p})\big|_{z_j}\,,
\end{equation}
where we account for the peculiar velocity uncertainties of each galaxy assuming Gaussian functions in redshift with $\sigma_{v_p}(z)$ given by \cref{eqn:sigma_pv}. Here we are also including the weights computed in \cref{eqn:weights}, that are derived from the marginalization of the quasilikelihood over the GW angular coordinates $\varphi$ and $\mu$, that we assume to coincide with the angular coordinates of each of the potential hosts.
We note that while in principle one could assign galaxy weights according to, e.g., astrophysical properties of the galaxies~\cite{fishbach}, here we only use information coming from the GW marginal distribution over the sky position angles (see Sec.~\ref{subsec:gal_cats} however).
We remark that since we account for peculiar velocity uncertainties, we also infer the redshift of the GW events, which we then marginalise over to get the cosmological posterior samples in \cref{eqn:bayes}. 

Finally, the remaining term $p(\bar{d}_L \,|\, d_L, \Omega, z_{\rm GW},\mathcal{H}, I)$ represents the detector quasilikelihood in the luminosity distance for the $i$-th event, as measured by a network of $M$ detectors, 
\begin{equation}
    p(\bar{d}_L \,|\, d_L, z_{\rm GW},\mathcal{H}, I) = \prod_{k=1}^{M} p(\bar{d}_L^{\,(k)} \,|\, d_L, z_{\rm GW},\mathcal{H}, I)\,,
\end{equation}
which we approximate as a Gaussian distribution~\cite{DelPozzo:2017kme, Laghi_2021, Muttoni:2021veo} centered on the best estimate $\bar{d}_L$ with uncertainties coming from the Fisher analysis presented in \cref{sec:GWPE}.
After the integration over $d_L$, we have:
\begin{equation}\label{eqn:lk_dL}
    p(\bar{d}_L \,|\, d(z_{\rm GW}, \Omega), z_{\rm GW},\mathcal{H}, I) 
    \propto
    \mathcal{N}(\bar{d}_L, \sigma^2_{\dl})\big|_{d(z_{\rm GW}, \Omega)}\,,
\end{equation}
and \cref{eqn:single-event-lk1} becomes
\begin{equation}\label{eqn:single-event-lk2}
    \begin{split}
    p(\mathcal{D}_i\,|\,\Omega,\mathcal{H}, I) \propto
    \int &dz_{\rm GW}\, \mathcal{N}(\bar{d}_L, \sigma^2_{\dl})\big|_{d(z_{\rm GW}, \Omega)}\times\\
    &\times \!\!\sum_{j=1}^{N_{\rm hosts}} w_j \,
    \mathcal{N}(z_{\rm GW},\sigma^2_{v_p})\big|_{z_j}\,,
    \end{split}
\end{equation}
which we compute through a nested sampling algorithm.

\subsection{Numerical implementation}\label{subsec:numerics}

We estimate the posterior distribution in \cref{eqn:bayes} with \textsc{cosmoLISA}~\cite{cosmoLISA}, a public pipeline for the Bayesian inference of the cosmological parameters with simulated GW observations. 
\textsc{cosmoLISA} computes the single-event likelihood \cref{eqn:single-event-lk2} and the posterior samples in \cref{eqn:bayes} by making use of a nested sampling algorithm as implemented in \textsc{CPNest}~\cite{CPNest}, a public package optimized to parallelise nested sampling computation.
In a typical \textsc{cosmoLISA} run, we employ 5 nested samplings in parallel of 1000 live points each, making a total ensemble of 5000 live points per run. 
At each step of each nested sampling, we independently evolve 6 live points via MCMC with maximum number of steps equal to 5000.
Although rather expensive from a computational point of view, we found that this is the ideal setup to face the complexity of the likelihood and to get numerically stable and reliable results.
In this work we used the commit \texttt{3760800} of the branch \texttt{massively\_parallel} of \textsc{CPNest}.
For the cosmological prior we assume uniform distributions in the same range of values used for the production of the localization error volumes, while each GW event redshift is marginalized in the redshift interval defined by its localization error volume (see \cref{sec:errorboxes}).
Since the marginalization over redshift is done through nested sampling, the dimension of the parameter space to be explored (or equivalently, the dimension of the integral to be computed to estimate the evidence) increases with the number of events, which makes the analysis of large number of dark sirens impracticable from a computational point of view.
Moreover, the computational cost raises quickly as we lower the $\SNR_{\rm net}$ threshold, since we add GW sources having large localization error volumes with potentially several thousands of galaxies inside.
To overcome this problem, which makes the analysis of $N \gtrsim \mathcal{O}(40)$ GW events prohibitive, here we adopt a different procedure compared to~\cite{Laghi_2021}: when considering a large number of events that would be too computationally costly to analyse with a single \texttt{cosmoLISA} run, we split our dataset in random subsets of less than $\mathcal{O}(40)$ events that we analyse separately.
Then we model $p(\Omega | \mathcal{D} , \mathcal{H}, I)$ by combining the posterior samples for $\Omega$ obtained from these data subsets via a Dirichlet process Gaussian mixture model, a fully Bayesian non-parametric method to reconstruct probability density functions out of a finite number of samples (see, e.g.,~\cite{DelPozzo2018}), using the approximate variational algorithm~\cite{Blei2006} as implemented in~\cite{haines_dpgmm}. 
We tested the robustness of the posterior reconstruction by checking, for some representative cases,  that the results obtained with this procedure are equivalent to those obtained  without splitting the dataset.

%%%%%%%%%%%%%%%%%%%%%%%%%%%%%%%%%%%%%%%%%%%%%%%%%%%%%%%%%%%%%%%%%%%%%%%%%%%%%%%%%%%%%%%%%%%%%%%%%%%%%%%%%%%%%%%%%%%%%%%%%%%%%%%%%%%%%%%%%%%%%%%%%%%%%%%%%%%%%%%%%%%%%%%%%%%%%%%%%%%%%%%%%%%%%%%%%%%%%%%%%%%%%%%%%%%%%%%%%%%%%%%%%%%%%%%%%%%%%%%%%%%%%%%%%%%%%%%%%%%%%%%%%%%%%%%%%%%%%%%%%%%%%%%%%%%%%%%%%%%%

\section{Results}
\label{sec:results}

In what follows we present our estimates of the cosmological parameters for 1 year of network observation (assuming full duty cycle).
For the GW population produced and analyzed in \cref{sec:GWPE}, we repeat the production of the localization error volumes described in \cref{subsec:error_box} six times to consider a reasonable sample of realizations in the galaxy distribution, and then analyse each of them to get cosmological posterior samples, obtaining precisions on the measure for the cosmological parameters that we average over the six realizations to get a fiducial mean precision at 68\% and 90\% CI. First, we consider a fiducial scenario defined by a selection threshold equal to $\SNR_{\rm net}=300$ and a galaxy catalog complete up to $z=1$ (\cref{subsec:results_fiducial}).
We then investigate how results change if we consider: different increasing values of $\SNR_{\rm net}$ (\cref{subsec:results_different_SNR_cuts}); the inference of $h$ only, assuming $\Om$ known (\cref{subsec:results_h_only}); single-host dark sirens only (\cref{subsec:results_single_host}). Finally, we question how forecasts change in an optimistic scenario where we have a complete galaxy catalog up to $z=3$ (\cref{subsec:results_z<3}).
In the following report of results, we are excluding those of ET alone. In fact, from \cref{fig:summary_PE,fig:N_hosts_statistics} it emerges that a single detector in a triangular configuration is not able to localize sufficiently well the sources even at large $\SNR$, leading to a few error-boxes with (tens of) thousands galaxies. The inference on this dataset does not contain any more information with respect to the prior and it is therefore considered uninformative.
Unless otherwise specified, in the following we will draw our main conclusions based on the 90\% CI estimates provided by each network of detectors for each parameter, whose precision is estimated from the mean half-width and median of the posterior distribution.

\subsection{Fiducial case: \texorpdfstring{$\SNR_{\rm net} > 300$}{SNR} at \texorpdfstring{$z < 1$}{z}}\label{subsec:results_fiducial}

We define our fiducial scenario by imposing an $\SNR_{\rm net}$ threshold of 300 and a galaxy catalog complete up to $z=1$.
As discussed in \cref{sec:cosmoinference}, the choice $\SNR_{\rm net} = 300$ is a trade-off between our availability of computational resources and the validity of our formulation of the likelihood.
The redshift catalog limitation is assumed based on realistic expectations of galaxy surveys in the era of 3G detectors, according to which completeness of all-sky readily available galaxy catalogs rapidly declines after $z=1$ (in \cref{subsec:results_z<3} we will investigate an optimistic scenario in which we assume that a dedicated deep survey capable of yielding a complete catalog up to $z=3$ along the GW sky localization cone is available for each dark siren).
In this scenario we are effectively selecting the best-localized events (see \cref{fig:summary_PE}).

\begin{figure}
    \centering
    \includegraphics[width=0.485\textwidth]{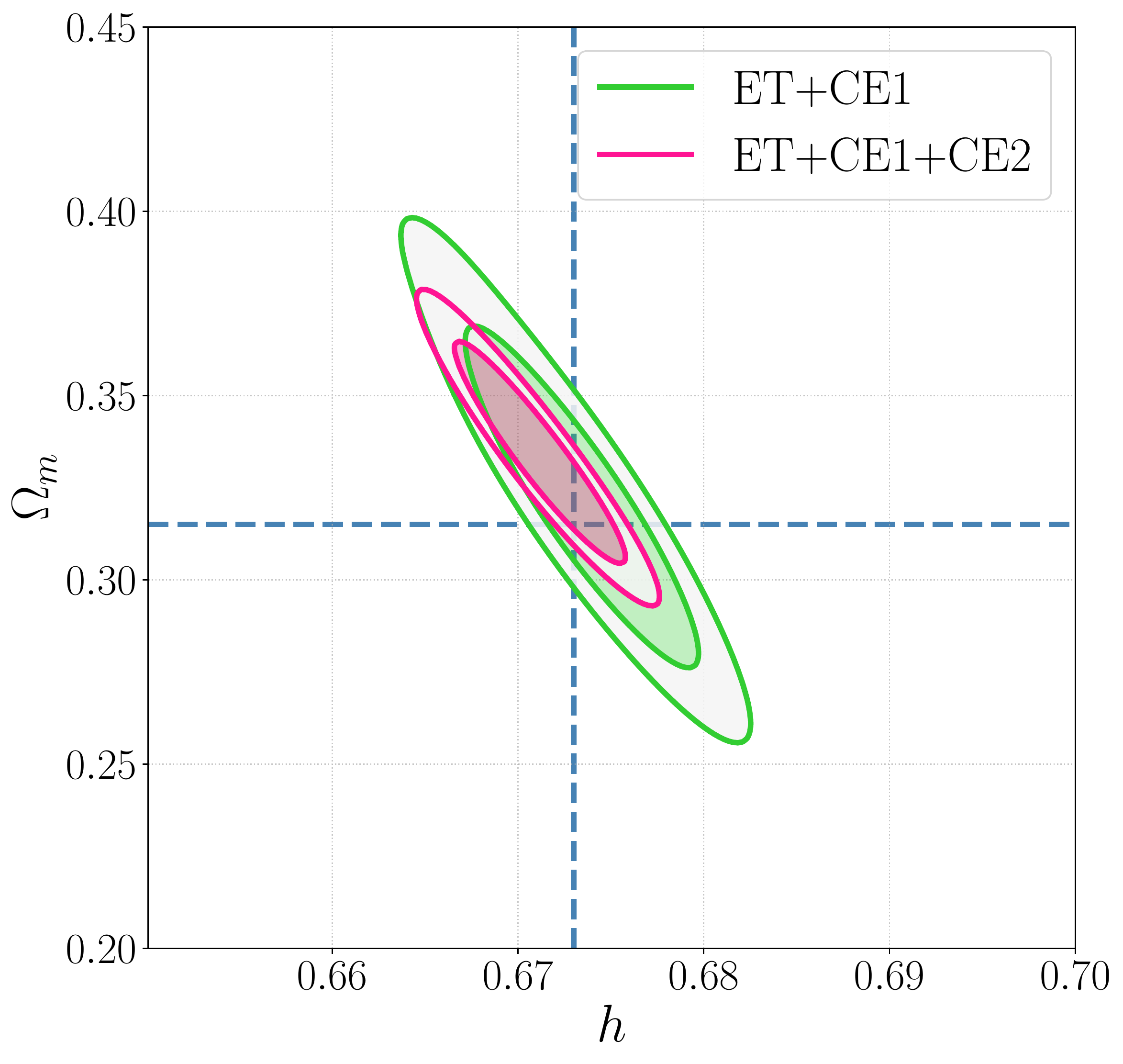}
    \caption{
    Posterior distributions (68\% and 90\% credible regions) in the $h-\Om$ plane for 1 year of observation, from the analysis of a representative realization of the localization error volumes, as described in \cref{subsec:results_fiducial} (fiducial scenario, full duty cycle). In each panel, the cyan dotted lines represent the fiducial cosmology.
    }
    \label{fig:corner_plots_fiducial}
\end{figure}

Results for our fiducial scenario are presented in~\cref{tab:cosmo_results}, where we report a mean precision averaged over the analysis of the six different realizations of the localization error volumes, while in \cref{fig:corner_plots_fiducial,fig:h_Om_BBH_latest} we show joint posteriors obtained from a representative realization of the localization error volumes, chosen as the one, among all of the six realizations analyzed, that has the median precision for $h$.
From the joint inference of $h$ and  $\Om$, we find that the network ET+CE1 is able to constrain $h$ at the $1.1\%$ with $N = 207$ BBHs. A slight improvement is observed with the network ET+CE1+CE2, where $N=278$ dark sirens can produce a measure of $h$ at the $0.8\%$ level.
These numbers suggest that if we want to reach a $\sim$1\% precision on the measure of $h$, we need to consider at least a network made of two 3G detectors. Adding a third detector to the network may allow us to reach a subpercent precision.
This level of precision on $\hubble$ would allow us to solve the Hubble tension within one full year of 3G observations, assuming the tension persists until the 3G era.

The situation for $\Om$ is different. A network of at least two detectors allows us to reach a $14.4\%$ precision, which, in the optimistic scenario of the network ET+CE1+CE2, could go down to a $10\%$-level measure. 

Our fiducial results are slightly better than the ones recently reported in~\cite{Zhu:2023jti}.
In there a dark sirens analysis similar to ours has been considered for the ET+CE1 scenario only, with average forecast constraints reaching $\sim$1\% for $\hubble$ and $\sim$20\% for $\Omega_m$ (at 68\% CI) with 300 BBHs.
This discrepancy can be attributed to the different settings of the simulations.
For example, contrary to our setup, in~\cite{Zhu:2023jti} higher GW modes in the GW signal are not included.
Given their importance in obtaining accurate sky localization volumes, thanks to their role in breaking degeneracies between GW waveform parameters, this may explain the slightly more optimistic results obtained in our analysis.

\subsection{Higher \texorpdfstring{$\SNR_{\rm net}$}{SNR} thresholds at \texorpdfstring{$z < 1$}{z}}
\label{subsec:results_different_SNR_cuts}

\begin{figure*}[t!]
    \centering
    \includegraphics[width=.8\textwidth]{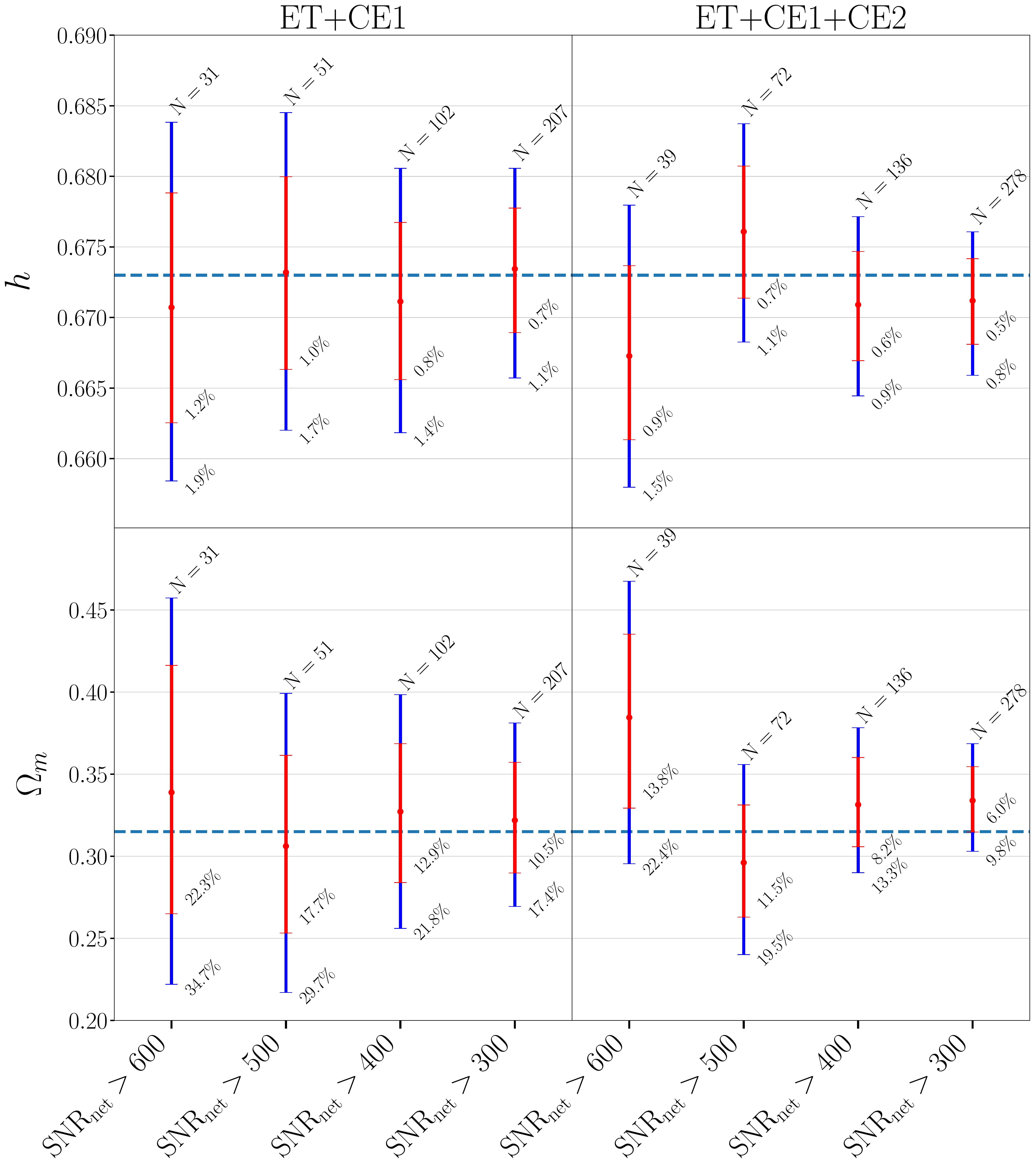}
    \caption{Comparison of the precision for the joint inference of $h$ and $\Om$ using $N$ dark sirens with a galaxy catalog complete up to $z<1$ from the events of a representative realization of the localization error volumes, as defined in \cref{subsec:results_fiducial,subsec:results_different_SNR_cuts}. Red (blue) intervals show 68\% (90\%) CI (precision shown next to them), considering different detector networks and $\SNR_{\rm net}$ thresholds for 1 year of observation.
    }
    \label{fig:h_Om_BBH_latest}
\end{figure*}

In \cref{fig:h_Om_BBH_latest} we show how our forecasts change as a function of increasing $\SNR_{\rm net}$ threshold values, in order to characterize the importance of the loudest observed dark sirens for cosmological inference.
We choose some representative threshold values of $\SNR_{\rm net} = 600, 500, 400$ and repeat the cosmological analysis for each respective dataset of the same realization.
Starting from the highest $\SNR_{\rm net}$ threshold, in the case $\SNR_{\rm net} > 600$, the network ET+CE1 ($N=31$) and ET+CE1+CE2 ($N=39$) lead to $1.9\%$ and $1.5\%$ constraints on $h$, respectively.
These results are only slightly worse than those obtained in the case $\SNR_{\rm net} > 500$, while $\Om$ is constrained at the 34.7\% (ET+CE1) and 22.4\% (ET+CE1+CE2). 
Including events at $\SNR_{\rm net} > 500$ and $\SNR_{\rm net} > 400$, we increase the number of events (see \cref{fig:h_Om_BBH_latest}) and, as expected, we get better constraints for both $\lcdm$ parameters.
The shrinkage evolution of the marginalised posteriors is evident from \cref{fig:h_Om_BBH_latest}, where we report the number $N$ of events passing the $\SNR_{\rm net}$ threshold and the precision for each case. As for the fiducial scenario, here we also report results from the realization that gives the median precision on $h$.
Overall, we can see how results for $h$ are less dependent on the number of events with respect to $\Om$. This can be explained by the fact that the measurement of $h$ mainly depends on the observation of nearby events, which are mostly characterized by high $\SNR_{\rm net}$. Constraints on $\Om$ are instead more dependent on mid-high redshift dark sirens, therefore the inclusion of lower-$\SNR_{\rm net}$ events has more impact.
Our analysis suggests that most of the cosmological predicting power of 3G BBH dark sirens is contained in high SNR events; yet, to find the most accurate forecasts one should include also lower SNR events, eventually considering all events above detection threshold.
Such a complete analysis however is prohibitive with the computational resources at our disposal.
Our methods of inference need to be further developed and optimized before such a study will be possible, but nevertheless the approach considered here is a good compromise that provides sufficiently accurate forecast estimations at a relatively affordable computational cost.

\begin{table*}[t]
\centering
%\bgroup
%\setlength{\tabcolsep}{0.5em} % horizontal padding
\def\arraystretch{\tabvspace} %  vertical padding (default is 1)
\begin{tabular}{ l || c | c || c | c | c | c || c | c | c }
\hline
\hline
\multirow{3}{*}{\textbf{Network}} & \multicolumn{2}{c||}{$N$} & \multicolumn{4}{c||}{$\Delta h / h$ ($\%$)} & \multicolumn{3}{c}{$\Delta \Om / \Om$ ($\%$)} \\\cline{2-10}
& \multirow{2}{*}{$z<1$} & \multirow{2}{*}{$z<3$} &  \multirow{2}{*}{$z<1$} & $z<1$ & $z<1$ & \multirow{2}{*}{$z<3$} & \multirow{2}{*}{$z<1$} & $z<1$ & \multirow{2}{*}{$z<3$} \\
& & & & fixed $\Om$ & single-host & & & single-host & \\
 \hline
 ET+CE1 & \num{207} & \num{248}  &  0.6 (1.1) & 0.2 (0.4) & \num{3.3}-\num{7.1} (\num{5.6}-\num{11.2}) & 0.7 (1.1) &  8.8 (14.4) & - & 8.8 (14.6)\\
 ET+CE1+CE2 & \num{278} & \num{348} & 0.5 (0.8) & 0.2 (0.3) & \num{1.7}-\num{2.1} (\num{2.7}-\num{3.3}) & 0.4 (0.7) & 6.1 (10.0) & - & 5.3 (8.7)\\
\hline
\hline
\end{tabular}
\caption{
For each network of detectors (column 1), we report the number $N$ of dark sirens with $\SNR_{\rm net} > 300$ used in~\cref{sec:results} for 1 year of full observation, assuming a galaxy catalog complete up to $z<1$ and $z<3$ (column 2-3). We report $68\%$ ($90\%$) CI for $h$ and $\Om$ (column 4-7 and 8-10) assuming $\SNR_{\rm net} > 300$ and: using a complete galaxy catalog up to $z<1$, inferring both parameters or assuming $\Om$ known, analysing single-host dark sirens only, and using a complete galaxy catalog up to $z<3$. The quantities $\Delta h$ ($\Delta \Om$) and $h$ ($\Om$) are the mean half-width of the posterior distribution and the median, respectively. We report a mean precision averaged over the six different realizations analyzed. For the single-host analysis (columns 6 and 9) the average number of events were \num{1} (ET+CE1) and \num{5} (ET+CE1+CE2) (see \cref{subsec:results_single_host}).
}
\label{tab:cosmo_results}
\end{table*}

\subsection{\texorpdfstring{$\SNR_{\rm net} > 300$}{SNR} at \texorpdfstring{$z<1$}{z}: Assuming \texorpdfstring{$\Om$}{O} known}\label{subsec:results_h_only}

Here we repeat the analysis with the same dataset used in our fiducial scenario assuming that we know $\Om$ exactly.
This reduces our cosmological model from two to only one parameter to infer.
We perform this analysis mainly to compare with other results in the literature, but as a further motivation a scenario in which the Hubble tension persists to the 3G era while $\Om$ is measured with high precision by EM observations is not excluded.
For ET+CE1 and ET+CE1+CE2, we find similar subpercent precision, around 0.3\%, even if the latter network observes more events. In fact, these events are mostly at high redshift, thus they do not contribute significantly to the measure of $h$.

We compare again our results with the ones we can find in the literature.
Reference~\cite{Song:2022siz} claims that the ET+CE1+CE2 network can deliver a surprising $\mathcal{O}(0.001\%)$ constraint on $\hubble$ within 5 year of observations of BBHs at $z<0.3$.
This differs by two orders of magnitude from the numbers we reported above for 1 year of observations.
Such a discrepancy is clearly due to differences in the two simulations.
For example, by comparing Fig.~6 in~\cite{Song:2022siz} with our \cref{fig:N_hosts_statistics}, it is clear that on average the number of galaxies contained within a BBH sky localization volume is much smaller in~\cite{Song:2022siz}, where basically it never exceeds 10 with the majority of GW events [$\mathcal{O}(100)$] having one single potential host galaxy, than in our setup, where we count on average hundreds of galaxies per GW event with only a handful of BBHs having 10 galaxies or less.
From this comparison we clearly understand that the forecasts provided in~\cite{Song:2022siz} are extremely optimistic if compared to our study.

\subsection{\texorpdfstring{$\SNR_{\rm net} > 300$}{SNR} at \texorpdfstring{$z<1$}{z}: Single-host dark sirens only}\label{subsec:results_single_host}

In case a dark siren has only one potential galaxy host falling within the localization error volume, we may consider them as ``effective bright sirens.'' 
These ``golden sirens'' are expected to be powerful probes of the cosmological parameters, since the redshift information comes from a single galaxy. 
The only caveat is that such golden sirens are not expected to be very numerous, since in general they are characterized by having localization error volumes small enough to contain just one galaxy. 
Moreover they are observed preferentially at low redshift since on average the higher the distance to the source, the larger its sky localization volume, and consequently the less likely there is only one galaxy within.
Nevertheless, given their similarity with bright sirens, it is of interest to understand how useful these golden events can be in the inference of the cosmological parameters.
In general, all the realizations analyzed here have at least one golden dark sirens (see the first bin on the x-axis of the right plot in \cref{fig:N_hosts_statistics}).
In all the two network configurations, we find that 
single-host dark sirens cannot constrain $\Om$. This is not surprising given the low-redshift of these golden events: $z<0.08$ for ET+CE1, and $z<0.22$ for ET+CE1+CE2.
Yet, these GW events can constrain the Hubble constant $h$ in all the two network configurations. 
For the network ET+CE1, we have only $N=1$ single-host GW event, which therefore allows for constraints on $h$ at the level of $5.6\%$-$11.2\%$, while ET+CE1+CE2 gives better results than ET+CE1, with on average $N=5$ observations and $h$ constrained with a $2.7\%$-$3.3\%$ precision. 
We can now compare our results with the ones reported by similar studies in the literature exploiting golden sirens observed by a 3G network~\cite{Borhanian:2020vyr,Gupta:2022fwd}.
Reference~\cite{Borhanian:2020vyr} in particular consider several populations of BBH golden sirens and reports constraints on $\hubble$ that can reach $\mathcal{O}(0.1\%)$ at 68\% CI or better within 2 years of observations with ET+CE1+CE2 and only considering events at $z<0.1$.
If compared with the numbers we report above, our results are more than one order of magnitude worse than the one reported in~\cite{Borhanian:2020vyr}.
This is not surprising and may be due to several reasons, in particular to the different assumptions that have been employed in the two different studies which overall are more optimistic in~\cite{Borhanian:2020vyr} than in our study. Among them, we can cite the lack of redshift uncertainty, the linear Hubble law is used with $\hubble$ as the only parameter to be inferred and different BBH populations. Most importantly, however, the main motivation behind our differences lies in the number of single-host events employed in the inference: our study suggests that the average rate of single-host events in the ET+CE1 (ET+CE1+CE2) network is \SI{1}{\per\year} (\SI{5}{\per\year}), while~\cite{Borhanian:2020vyr} reports \SI{22}{\per\year} in the most sensitive network. We find this rate consistent with the one we obtain once we repeat the error-box generation process without extending the redshift boundaries in the last step (i.e., without priors on the cosmological parameters), which corresponds on average to \SI{26}{\per\year} in ET+CE1+CE2.
The more realistic simulations performed here suggest that golden sirens from 3G detectors will not be able to constrain $\hubble$ at the subpercent level, but nonetheless reach an interesting $\mathcal{O}(1\%)$ precision.

\subsection{\texorpdfstring{$\SNR_{\rm net} > 300$}{SNR} at \texorpdfstring{$z<3$}{z}}
\label{subsec:results_z<3}

As presented above, our fiducial scenario considers a galaxy catalog complete at $z<1$.
This is a somehow conservative scenario in which we can perform 3G cosmological analyses only with readily available all-sky galaxy catalogs, which we assume will be complete up to $z=1$ in the 3G era.
Nevertheless in a more optimistic scenario one could foresee that dedicated deep-field surveys will be performed along the sky-localization cone of each BBH detected with $\SNR_{\rm net} >300$.
As we have shown above this $\SNR_{\rm net}$ threshold yields at most a few hundreds BBH detections per year, specifically with the ET+CE1+CE2 network.
Providing a deep-field galaxy survey for each of these events may seem unfeasible, but one must remember that the follow-up survey can be taken even years after 3G detectors have obtained the GW data.
This means that, provided adequate EM telescope resources will be available during or after the 3G detector era, such a scenario can be considered realistic.

Importantly, we make the simplifying assumption that instrumental errors of high-$z$ galaxies are negligible: this will not probably be the case, since typically at high-$z$ we will have photometric redshifts. Nonetheless, we can think of this idealized case as a very optimistic scenario where we assume we can perfectly correct for instrumental uncertainties and we can test the full performance of 3G detectors at high-$z$.

\begin{figure}
    \centering
    \includegraphics[width=0.5\textwidth]{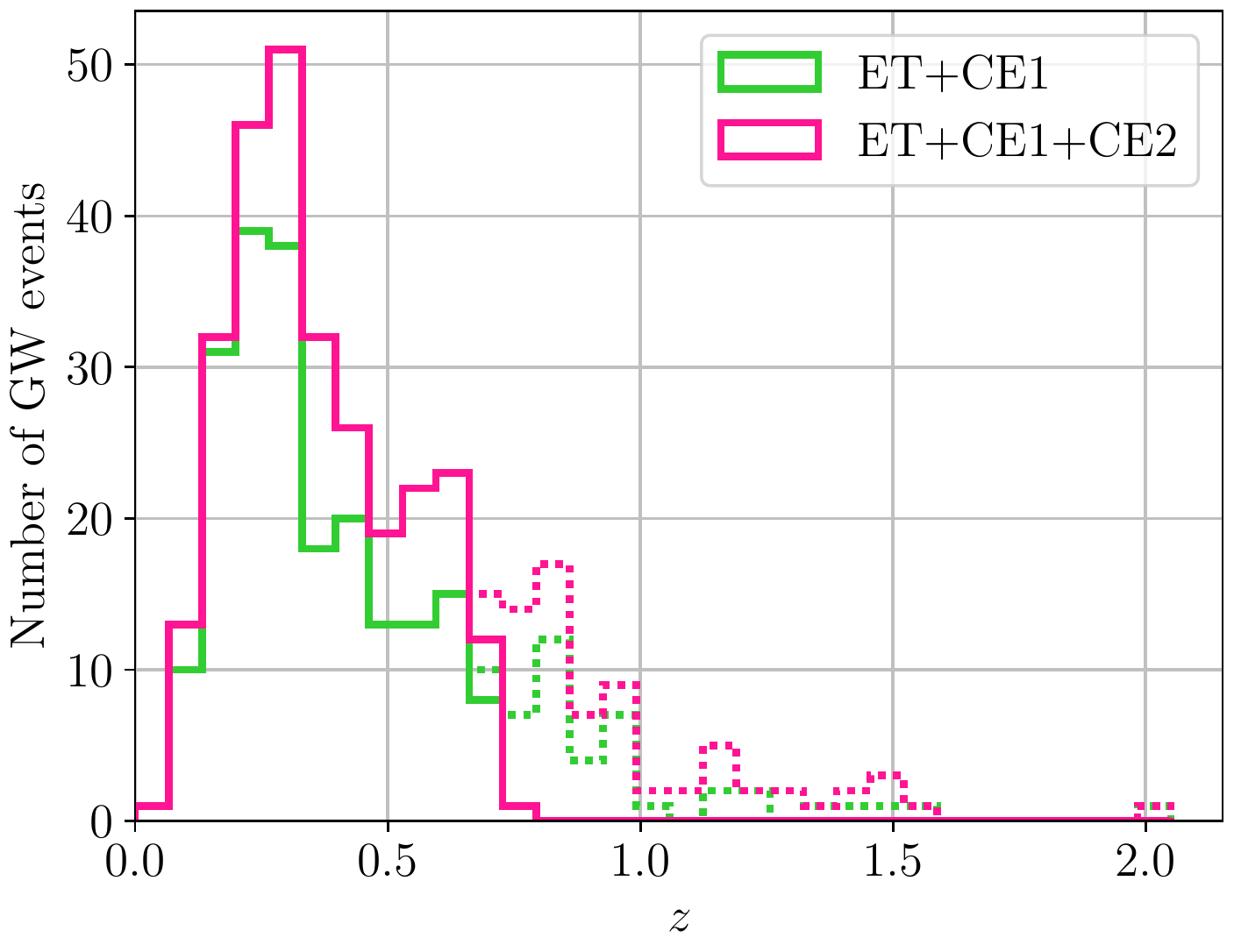}
    \caption{Redshift distribution of the $\SNR_{\rm net} > 300$ dark sirens employed in the inference for the different network configurations, colors as in legend. The solid histograms refer to the fiducial $z < 1$ scenario, while the dotted histograms extend the distributions up to the optimistic $z < 3$ case.}
    \label{fig:z_dark_sirens}
\end{figure}

In general, we expect some improvement on the measure of $\Om$, which is more sensitive to the higher-$z$ observations reported in \cref{fig:z_dark_sirens}. With a relatively larger number of high-redshift dark sirens (see \cref{tab:cosmo_results}), we find that constraints on $\hubble$ substantially coincides with those obtained in our fiducial scenario ($z<1$), while we obtain only slightly stringier constraints for $\Om$, with the network ET+CE1+CE1, which may reach a precision of 8.7\% at 90\% CI, starting from a $z<1$ result of 10.0\%.

These results suggest that BBHs at $z>1$ will not substantially contribute to measurements of $\hubble$ and $\Om$.
This is certainly due to the lower accuracy with which BBHs at high redshift, which on average have a lower SNR, can be localized in the Universe.
The wider sky localization volume will in fact contain a large number of potential host galaxies, conveying basically no information on the posteriors of the cosmological parameters.

\section{Discussion}
\label{sec:discussion}

The ensemble of results we reported above constitutes the most up-to-date realistic cosmological forecasts with 3G BBH dark sirens.
In this section we discuss few important issues and limitations of our investigations.

\subsection{Expectations from longer observations}

In \cref{sec:redshift_distribution} we reported an expected BBH cosmic merger rate of $\sim$50000 per year.
Although the results presented in this study come from $1$ year of data recording, 3G detectors are going to be the cornerstone of GW observations for the next decades and operate for several years.
An analysis similar to ours over a multiyear dataset of 3G dark siren observations would be prohibitive with the computational resources at our disposal.
Nevertheless we can obtain a simple estimation starting from the results we obtained for one year of data and extrapolating them assuming they scale as the square root of observational time, which is directly related to the number of observed dark sirens.
We stress that this is clearly an oversimplification useful only to provide us with approximate estimates.

Starting from the numbers in \cref{tab:cosmo_results}, we compute the constraints on $\hubble$ and $\Omega_m$ expected in 3, 5, and 10 years of observations for our fiducial scenario.
The resulting estimates are reported in \cref{tab:multi-year-results}.
We can clearly see that all 3G detector networks will be able to reach subpercent constraints on $\hubble$ within few years of observations, delivering $\mathcal{O}(0.1\%)$ precisions.
On the other hand, constraints on $\Omega_m$ hover around few $\%$ irrespectively of the observational time.

\begin{table*}[t]
\centering
%\bgroup
%\setlength{\tabcolsep}{0.5em} % horizontal padding
\def\arraystretch{\tabvspace} %  vertical padding (default is 1)
\begin{tabular}{ l || c | c | c || c | c | c }
\hline
\hline
 \multirow{2}{*}{\textbf{Network}} & \multicolumn{3}{c||}{$\Delta h / h$ ($\%$)} & \multicolumn{3}{c}{$\Delta \Om / \Om$ ($\%$)} \\\cline{2-7}
 & \SI{3}{\year} & \SI{5}{\year} & \SI{10}{\year} & \SI{3}{\year} & \SI{5}{\year} & \SI{10}{\year} \\
 \hline
ET+CE1 & 0.4 (0.6) & 0.3 (0.5) & 0.2 (0.4) & 5.1 (8.3) & 3.9 (6.4) & 2.8 (4.6) \\
ET+CE1+CE2 & 0.3 (0.5) & 0.2 (0.4) & 0.1 (0.3) & 3.5 (5.8) & 2.7 (4.5) & 1.9 (3.2) \\
\hline
\hline
\end{tabular}
\caption{Expected cosmological constraints at the 68\% (90\%) CI for multiyear 3G observations estimated from the 1 year fiducial results (see \cref{tab:cosmo_results}) with a simple scaling proportional to the square root of the observational time.}
\label{tab:multi-year-results}
\end{table*}

\subsection{Comparison with the literature}\label{subsec:compare_literature}

In \cref{sec:results} we confronted our estimates with the ones reported by comparable investigations that we found in the literature. Here we briefly summarize the results of these comparisons.
Our analysis with ET+CE1+CE2 and single-host events only, leads to a few $\%$ precision on $\hubble$, which is worse by an order of magnitude than what reported in~\cite{Borhanian:2020vyr}, as discussed in~\cref{subsec:results_single_host}.
Furthermore assuming $\Om$ known, our constraints on $\hubble$ with the detector network ET+CE1+CE2 are on the order of the subpercent, which is two orders of magnitude lower than what found in~\cite{Song:2022siz} (see discussion in~\cref{subsec:results_h_only}).
Our findings are in better agreement with, though slightly better than, the ones reported in~\cite{Zhu:2023jti}, which provides cosmological forecasts for the network ET+CE1 only.
These results are not surprising.
Our simulation overall represents a more realistic setup than the ones considered in~\cite{Borhanian:2020vyr,Song:2022siz}, implying that less optimistic cosmological constraints were expected.
On the other hand our simulation is better comparable with the approach taken by~\cite{Zhu:2023jti}, with order one discrepancies in the reported cosmological constraints probably due to the different assumptions considered by the two investigations.
By building on these previous forecasts, our results certainly help to better define the dark siren cosmological science case of 3G detectors and to understand their potential and limitation.

\subsection{Gravitational-wave systematic effects}

Our forecasts assume that we will be able to account for systematic errors coming for example from approximate waveforms or the uncertainty in the calibration of the detectors~\cite{Huang:2022rdg}.
In particular in order to achieve the measurement of the $\lcdm$ parameters reported in this study, specifically subpercent measurements of $\hubble$, we need to keep these systematics under control with a precision greater than the accuracy with which cosmological parameters are measured.
This requires for example that waveform models may need to be calibrated with an improved accuracy by as much as three orders of magnitude with respect to current waveforms, especially for high SNR BBHs as the ones considered in this work~\cite{Purrer:2019jcp,Hu:2022rjq}.
This clearly poses a challenge for future waveform models, which will need to be significantly more accurate and physically more complete in order to meet the scientific objectives of 3G detectors. 
Similarly 3G interferometers must be precisely calibrated in order to avoid propagation of systematics to the inference of cosmological parameters~\cite{Essick:2022vzl}.
This may pose a technical challenge to obtain highly precise measurements of $\hubble$, although other sources of systematics are expected to be more problematic~\cite{Payne:2020myg}.
Finally, an additional source of systematics may come from environmental effects perturbing the dynamics of the GW source: for example the vicinity of a perturbing third body, the presence of gas surrounding the source or gravitational lensing, the impact of peculiar velocities on very close-by events.
Such effects can lead to an erroneous estimation of the distance or sky localization of the source, which in turn will yield a wrong measurement of the cosmological parameters~\cite{Zhu:2023jti}.
From these remarks it is clear that much theoretical and experimental work is still needed in order to achieve a subpercent measurement of $\hubble$, calling for an intense development effort over the next decade.

\subsection{Limitations and future perspectives}

Here we discuss the limitations of our study together with suggestions for future improvements.
The impact of these limitations on the cosmological inference with dark sirens will be the object of future studies.

We characterized BBH GW signals with the \texttt{IMRPhenomXHM} waveform model, which describes nonprecessing binary systems with BH parallel spin vectors. This assumption allowed us to reduce the total number of source parameters - and therefore the dimension of the Fisher matrix - from \num{15} to \num{11}, limiting computational cost and potential issues in the FIM inversion process. However, precessing systems are expected to form in nature, and precessing waveform models (e.g.~\texttt{IMRPhenomXPHM} \cite{pratten2021computationally}) should be used.
This in turn highlights the need to update the sampling distribution of the spins as well, making the BBH population more accurate and consistent with the latest available results, if observations will suggest evidence for precessing systems (see e.g.~the models described in \cite{ligo2021population}). 
We neglected Earth's rotation effects on the observed signal. While BBHs are not particularly influenced, we underline the need to take time-varying antenna pattern functions into account, especially to properly model very light BBH systems.
From the instrument point of view, we considered GW detectors to be fully operative during the whole period of observation. In reality, detectors may undergo provisional maintenance and upgrade works that inevitably force the discontinuation of data recording.
For this reason, each detector should be characterized by its own duty cycle. Furthermore, the GW emission of some compact binary coalescences are expected to overlap in the time-domain data strain $s(t)$ of a detector, posing a challenging task for the extraction of the individual GW signals. In this work, we assumed no overlap of BBH signals: while 
recent studies showed that it seems possible to run a reliable parameter estimation on overlapping events~\cite{samajdar2021biases,himemoto2021impacts,pizzati2022toward}), one should quantify how many of these events 3G detectors might observe and how robustly these signals could be resolved, with the potential effect of reducing the number of dark sirens available for cosmological inference.

Concerning the galaxy catalog, we report the main upgrades that can enhance the current simulations. To mimic incompleteness effects in our catalog, we have performed a stellar mass cut. However, galaxy mass is not an easy quantity to determine from EM surveys given the degeneracies involved in the galaxy template fitting method. To avoid this limitation and account catalogue incompleteness in a more precise way, it would be convenient to use a direct observable like the luminosity (or magnitude). To guide the reader, current galaxy catalogues of SDSS Legacy Survey are complete up to optical magnitudes of $25.1$ which will increase up to $27.5$ for the future LSST survey. 
Moreover, future simulations should account in a more accurate way for incompleteness and selection effects ruling the catalogs provided by current and future surveys. In this way, it will be possible to account for missing dwarf (or, in terms of luminosity, faint) galaxies in the GW localization error volumes.
On the other hand, on top of the sky location of each galaxy, its luminosity could be also used as an extra condition to model the probability of housing a GW event. Bright galaxies are expected to have higher chances of hosting compact binary coalescences, and this information could be used to improve the cosmological inference methodology.

The inference on the cosmological parameters is carried out on high $\SNR_{\rm net}$ events. This cut meets our requirements in terms of available computational resources and formulation of the inference problem, notably the exclusion of GW selection effects.
Indeed we observed that by lowering our $\SNR_{\rm net}$ threshold, the estimates start to systematically deviate from the true values. Furthermore, recent studies where systematic effects are thoroughly discussed report the need to perform joint source population and cosmological inference, since a separate treatment can impact the final measurement accuracy~\cite{2021ApJ...909..218A,mastrogiovanni_2021}. One could therefore make a similar analysis on a much larger sample of GW events (i.e.~by lowering the $\SNR_{\rm net}$ threshold), provided that selection and systematic effects are properly accounted for.

\section{Conclusion}
\label{sec:conclusion}

In this work we studied the scientific potential of dark sirens in the context of next generation, 3G ground-based detectors such as ET and CE. We performed parameter estimation on GW signals emitted by a mock population of BBHs through the FIM formalism in different configurations of detectors, namely ET, ET+CE1, ET+CE1+CE2, assuming $\SI{1}{\year}$ of continuous observations. 
We then selected and employed high $\SNR_{\rm net}$ GW events as dark siren candidates in a Bayesian framework to recover joint posterior distributions on the set of $\lcdm$ cosmological parameters, namely $\hubble$ and $\Om$.
Our main results are based on a fiducial scenario in which we assumed galaxy surveys to be complete up to $z = 1$ by the 3G detector era. Under these premises, we found that the best constraints are obtained by the ET+CE1+CE2 network, where $\hubble$ ($\Om$) is recovered at a promising $0.8\%$ ($10.0\%$) at $90\%$ CI (cf.~\cref{subsec:results_fiducial}). On the other hand, we find that ET alone is not able to provide informative results.

Assuming $\Om$ is known perfectly a priori, a network made of ET and at least one CE (with a 40 km baseline) can lead to a $0.4\%$ precision in the measure of $\hubble$ (cf.~\cref{subsec:results_h_only}). 

Furthermore, we characterized the constraining power
of well-localized BBHs by comparing the precision on $\hubble$ obtained from single-host dark sirens only. 
We find that the precision in ET+CE1 (ET+CE1+CE2) ranges between $5.6\%$ ($2.7\%$) and $11.2\%$ ($3.3\%$)(cf.~\cref{subsec:results_single_host}). 
Finally, we considered an optimistic scenario where deep sky surveys may be employed to reach and scan galaxies up to a $z=3$ horizon, finding no significant improvement on $\hubble$ and modest improvement on $\Om$ in all the network configurations (cf.~\cref{subsec:results_z<3}).

Our results suggest that the synergy between multiple 3G detectors is crucial to reach sub-$\%$ precision on $\hubble$. 
Even if we expect EM estimates to improve in the next decade, we highlight that GW-based observations can offer an independent way to measure $\hubble$.

Furthermore, we underline that the dark siren statistical method could offer significant information on $\Om$, even if the number of high-redshift GW events is usually low as they are usually not well localized and thus hardly satisfy the necessary conditions for them to be used as dark sirens.
Nevertheless we find that a network of 3G detectors can still constrain $\Om$ at $\sim 10\%$ of precision at $90\%$ CI.
High-redshift dark sirens could nonetheless provide interesting information on alternative cosmological models, especially if they predict deviations at high redshift, and thus help in testing dark energy or modified gravity.
Further investigations are needed in order to understand the full potential of dark sirens at high redhsift.

To conclude, 3G detectors have the potential to greatly improve our understanding of the Universe, especially by providing stringent constraints on cosmological parameters, thus ushering us in the era of precision GW cosmology.
Further work is needed in order to define the full cosmological science case for 3G detectors, notably on the integration of different standard siren methods, but our forecasts show already the promising results that we will obtain from dark sirens only.

\acknowledgements
The authors would like to thank the anonymous referee for providing useful comments to the first version of this study.
The authors would like to thank Archisman Ghosh for feedback on the manuscript.
D.L. thanks Walter Del Pozzo for stimulating discussions.
N.M. and D.L. thank Chang Liu for useful comments.
Support for D.L. was partially provided by CNES through a CNES Postdoctoral Fellowship grant.
D.L., N.T. and S.M.~acknowledge support form the French space agency CNES in the framework of LISA. 
N.M., D.L. and N.T.~acknowledge support from an ANR Tremplin ERC Grant (No. ANR-20-ERC9-0006-01). N.M. acknowledges support from the Swiss National Science Foundation, grant No. 200020$\_$191957, and from the SwissMap National Center for Competence in Research.
D.I.V. acknowledges the financial support provided under the European Union’s H2020 ERC Consolidator Grant ``Binary Massive Black Hole Astrophysics'' (B Massive, Grant Agreement: 818691) and from INFN H45J18000450006.
Most of the numerical analyses have been performed at the IN2P3 computing centre (CC-IN2P3) in Lyon (Villeurbanne), which we thank for assistance and computational resources.

%-----------------------------
% BACK MATTER

\newpage
\bibliography{bibliography.bib}% Produces the bibliography via BibTeX.

\begin{thebibliography}{140}%
\makeatletter
\providecommand \@ifxundefined [1]{%
 \@ifx{#1\undefined}
}%
\providecommand \@ifnum [1]{%
 \ifnum #1\expandafter \@firstoftwo
 \else \expandafter \@secondoftwo
 \fi
}%
\providecommand \@ifx [1]{%
 \ifx #1\expandafter \@firstoftwo
 \else \expandafter \@secondoftwo
 \fi
}%
\providecommand \natexlab [1]{#1}%
\providecommand \enquote  [1]{``#1''}%
\providecommand \bibnamefont  [1]{#1}%
\providecommand \bibfnamefont [1]{#1}%
\providecommand \citenamefont [1]{#1}%
\providecommand \href@noop [0]{\@secondoftwo}%
\providecommand \href [0]{\begingroup \@sanitize@url \@href}%
\providecommand \@href[1]{\@@startlink{#1}\@@href}%
\providecommand \@@href[1]{\endgroup#1\@@endlink}%
\providecommand \@sanitize@url [0]{\catcode `\\12\catcode `\$12\catcode
  `\&12\catcode `\#12\catcode `\^12\catcode `\_12\catcode `\%12\relax}%
\providecommand \@@startlink[1]{}%
\providecommand \@@endlink[0]{}%
\providecommand \url  [0]{\begingroup\@sanitize@url \@url }%
\providecommand \@url [1]{\endgroup\@href {#1}{\urlprefix }}%
\providecommand \urlprefix  [0]{URL }%
\providecommand \Eprint [0]{\href }%
\providecommand \doibase [0]{http://dx.doi.org/}%
\providecommand \selectlanguage [0]{\@gobble}%
\providecommand \bibinfo  [0]{\@secondoftwo}%
\providecommand \bibfield  [0]{\@secondoftwo}%
\providecommand \translation [1]{[#1]}%
\providecommand \BibitemOpen [0]{}%
\providecommand \bibitemStop [0]{}%
\providecommand \bibitemNoStop [0]{.\EOS\space}%
\providecommand \EOS [0]{\spacefactor3000\relax}%
\providecommand \BibitemShut  [1]{\csname bibitem#1\endcsname}%
\let\auto@bib@innerbib\@empty
%</preamble>
\bibitem [{\citenamefont {{Turner}}(2022)}]{2022ARNPS..72....1T}%
  \BibitemOpen
  \bibfield  {author} {\bibinfo {author} {\bibfnamefont {M.~S.}\ \bibnamefont
  {{Turner}}},\ }\href {\doibase 10.1146/annurev-nucl-111119-041046} {\bibfield
   {journal} {\bibinfo  {journal} {Annual Review of Nuclear and Particle
  Science}\ }\textbf {\bibinfo {volume} {72}},\ \bibinfo {pages} {1} (\bibinfo
  {year} {2022})},\ \Eprint {http://arxiv.org/abs/2201.04741} {arXiv:2201.04741
  [astro-ph.CO]} \BibitemShut {NoStop}%
\bibitem [{\citenamefont {Aghanim}\ \emph
  {et~al.}(2020{\natexlab{a}})\citenamefont {Aghanim} \emph
  {et~al.}}]{Planck:2018nkj}%
  \BibitemOpen
  \bibfield  {author} {\bibinfo {author} {\bibfnamefont {N.}~\bibnamefont
  {Aghanim}} \emph {et~al.} (\bibinfo {collaboration} {Planck}),\ }\href
  {\doibase 10.1051/0004-6361/201833880} {\bibfield  {journal} {\bibinfo
  {journal} {Astron. Astrophys.}\ }\textbf {\bibinfo {volume} {641}},\ \bibinfo
  {pages} {A1} (\bibinfo {year} {2020}{\natexlab{a}})},\ \Eprint
  {http://arxiv.org/abs/1807.06205} {arXiv:1807.06205 [astro-ph.CO]}
  \BibitemShut {NoStop}%
\bibitem [{\citenamefont {Aghanim}\ \emph
  {et~al.}(2020{\natexlab{b}})\citenamefont {Aghanim} \emph
  {et~al.}}]{Planck:2018vyg}%
  \BibitemOpen
  \bibfield  {author} {\bibinfo {author} {\bibfnamefont {N.}~\bibnamefont
  {Aghanim}} \emph {et~al.} (\bibinfo {collaboration} {Planck}),\ }\href
  {\doibase 10.1051/0004-6361/201833910} {\bibfield  {journal} {\bibinfo
  {journal} {Astron. Astrophys.}\ }\textbf {\bibinfo {volume} {641}},\ \bibinfo
  {pages} {A6} (\bibinfo {year} {2020}{\natexlab{b}})},\ \bibinfo {note}
  {[Erratum: Astron.Astrophys. 652, C4 (2021)]},\ \Eprint
  {http://arxiv.org/abs/1807.06209} {arXiv:1807.06209 [astro-ph.CO]}
  \BibitemShut {NoStop}%
\bibitem [{\citenamefont {Riess}\ \emph {et~al.}(2022)\citenamefont {Riess}
  \emph {et~al.}}]{Riess:2021jrx}%
  \BibitemOpen
  \bibfield  {author} {\bibinfo {author} {\bibfnamefont {A.~G.}\ \bibnamefont
  {Riess}} \emph {et~al.},\ }\href {\doibase 10.3847/2041-8213/ac5c5b}
  {\bibfield  {journal} {\bibinfo  {journal} {Astrophys. J. Lett.}\ }\textbf
  {\bibinfo {volume} {934}},\ \bibinfo {pages} {L7} (\bibinfo {year} {2022})},\
  \Eprint {http://arxiv.org/abs/2112.04510} {arXiv:2112.04510 [astro-ph.CO]}
  \BibitemShut {NoStop}%
\bibitem [{\citenamefont {Scolnic}\ \emph {et~al.}(2018)\citenamefont {Scolnic}
  \emph {et~al.}}]{Pan-STARRS1:2017jku}%
  \BibitemOpen
  \bibfield  {author} {\bibinfo {author} {\bibfnamefont {D.~M.}\ \bibnamefont
  {Scolnic}} \emph {et~al.} (\bibinfo {collaboration} {Pan-STARRS1}),\ }\href
  {\doibase 10.3847/1538-4357/aab9bb} {\bibfield  {journal} {\bibinfo
  {journal} {Astrophys. J.}\ }\textbf {\bibinfo {volume} {859}},\ \bibinfo
  {pages} {101} (\bibinfo {year} {2018})},\ \Eprint
  {http://arxiv.org/abs/1710.00845} {arXiv:1710.00845 [astro-ph.CO]}
  \BibitemShut {NoStop}%
\bibitem [{\citenamefont {Abbott}\ \emph {et~al.}(2022)\citenamefont {Abbott}
  \emph {et~al.}}]{DES:2021wwk}%
  \BibitemOpen
  \bibfield  {author} {\bibinfo {author} {\bibfnamefont {T.~M.~C.}\
  \bibnamefont {Abbott}} \emph {et~al.} (\bibinfo {collaboration} {DES}),\
  }\href {\doibase 10.1103/PhysRevD.105.023520} {\bibfield  {journal} {\bibinfo
   {journal} {Phys. Rev. D}\ }\textbf {\bibinfo {volume} {105}},\ \bibinfo
  {pages} {023520} (\bibinfo {year} {2022})},\ \Eprint
  {http://arxiv.org/abs/2105.13549} {arXiv:2105.13549 [astro-ph.CO]}
  \BibitemShut {NoStop}%
\bibitem [{\citenamefont {Amon}\ \emph {et~al.}(2022)\citenamefont {Amon} \emph
  {et~al.}}]{DES:2021bvc}%
  \BibitemOpen
  \bibfield  {author} {\bibinfo {author} {\bibfnamefont {A.}~\bibnamefont
  {Amon}} \emph {et~al.} (\bibinfo {collaboration} {DES}),\ }\href {\doibase
  10.1103/PhysRevD.105.023514} {\bibfield  {journal} {\bibinfo  {journal}
  {Phys. Rev. D}\ }\textbf {\bibinfo {volume} {105}},\ \bibinfo {pages}
  {023514} (\bibinfo {year} {2022})},\ \Eprint
  {http://arxiv.org/abs/2105.13543} {arXiv:2105.13543 [astro-ph.CO]}
  \BibitemShut {NoStop}%
\bibitem [{\citenamefont {Abbott}\ \emph {et~al.}(2016)\citenamefont {Abbott}
  \emph {et~al.}}]{LIGOScientific:2016aoc}%
  \BibitemOpen
  \bibfield  {author} {\bibinfo {author} {\bibfnamefont {B.~P.}\ \bibnamefont
  {Abbott}} \emph {et~al.} (\bibinfo {collaboration} {LIGO Scientific,
  Virgo}),\ }\href {\doibase 10.1103/PhysRevLett.116.061102} {\bibfield
  {journal} {\bibinfo  {journal} {Phys. Rev. Lett.}\ }\textbf {\bibinfo
  {volume} {116}},\ \bibinfo {pages} {061102} (\bibinfo {year} {2016})},\
  \Eprint {http://arxiv.org/abs/1602.03837} {arXiv:1602.03837 [gr-qc]}
  \BibitemShut {NoStop}%
\bibitem [{\citenamefont {{Schutz}}(1986)}]{schutz86}%
  \BibitemOpen
  \bibfield  {author} {\bibinfo {author} {\bibfnamefont {B.~F.}\ \bibnamefont
  {{Schutz}}},\ }\href {\doibase 10.1038/323310a0} {\bibfield  {journal}
  {\bibinfo  {journal} {\nat}\ }\textbf {\bibinfo {volume} {323}},\ \bibinfo
  {pages} {310} (\bibinfo {year} {1986})}\BibitemShut {NoStop}%
\bibitem [{\citenamefont {{Holz}}\ and\ \citenamefont
  {{Hughes}}(2005)}]{2005ApJ...629...15H}%
  \BibitemOpen
  \bibfield  {author} {\bibinfo {author} {\bibfnamefont {D.~E.}\ \bibnamefont
  {{Holz}}}\ and\ \bibinfo {author} {\bibfnamefont {S.~A.}\ \bibnamefont
  {{Hughes}}},\ }\href {\doibase 10.1086/431341} {\bibfield  {journal}
  {\bibinfo  {journal} {\apj}\ }\textbf {\bibinfo {volume} {629}},\ \bibinfo
  {pages} {15} (\bibinfo {year} {2005})},\ \Eprint
  {http://arxiv.org/abs/astro-ph/0504616} {astro-ph/0504616} \BibitemShut
  {NoStop}%
\bibitem [{\citenamefont {Dalal}\ \emph {et~al.}(2006)\citenamefont {Dalal},
  \citenamefont {Holz}, \citenamefont {Hughes},\ and\ \citenamefont
  {Jain}}]{Dalal:2006qt}%
  \BibitemOpen
  \bibfield  {author} {\bibinfo {author} {\bibfnamefont {N.}~\bibnamefont
  {Dalal}}, \bibinfo {author} {\bibfnamefont {D.~E.}\ \bibnamefont {Holz}},
  \bibinfo {author} {\bibfnamefont {S.~A.}\ \bibnamefont {Hughes}}, \ and\
  \bibinfo {author} {\bibfnamefont {B.}~\bibnamefont {Jain}},\ }\href {\doibase
  10.1103/PhysRevD.74.063006} {\bibfield  {journal} {\bibinfo  {journal} {Phys.
  Rev. D}\ }\textbf {\bibinfo {volume} {74}},\ \bibinfo {pages} {063006}
  (\bibinfo {year} {2006})},\ \Eprint {http://arxiv.org/abs/astro-ph/0601275}
  {arXiv:astro-ph/0601275} \BibitemShut {NoStop}%
\bibitem [{\citenamefont {Abbott}\ \emph
  {et~al.}(2017{\natexlab{a}})\citenamefont {Abbott}, \citenamefont {Abbott},
  \citenamefont {Abbott}, \citenamefont {Acernese}, \citenamefont {Ackley}
  \emph {et~al.}}]{ligobns}%
  \BibitemOpen
  \bibfield  {author} {\bibinfo {author} {\bibfnamefont {B.~P.}\ \bibnamefont
  {Abbott}}, \bibinfo {author} {\bibfnamefont {R.}~\bibnamefont {Abbott}},
  \bibinfo {author} {\bibfnamefont {T.~D.}\ \bibnamefont {Abbott}}, \bibinfo
  {author} {\bibfnamefont {F.}~\bibnamefont {Acernese}}, \bibinfo {author}
  {\bibfnamefont {K.}~\bibnamefont {Ackley}},  \emph {et~al.} (\bibinfo
  {collaboration} {LIGO Scientific Collaboration and Virgo Collaboration}),\
  }\href {\doibase 10.1103/PhysRevLett.119.161101} {\bibfield  {journal}
  {\bibinfo  {journal} {Phys. Rev. Lett.}\ }\textbf {\bibinfo {volume} {119}},\
  \bibinfo {pages} {161101} (\bibinfo {year} {2017}{\natexlab{a}})}\BibitemShut
  {NoStop}%
\bibitem [{\citenamefont {Abbott}\ \emph
  {et~al.}(2017{\natexlab{b}})\citenamefont {Abbott}, \citenamefont {Abbott},
  \citenamefont {Abbott}, \citenamefont {Acernese}, \citenamefont {Ackley}
  \emph {et~al.}}]{MMApaper}%
  \BibitemOpen
  \bibfield  {author} {\bibinfo {author} {\bibfnamefont {B.~P.}\ \bibnamefont
  {Abbott}}, \bibinfo {author} {\bibfnamefont {R.}~\bibnamefont {Abbott}},
  \bibinfo {author} {\bibfnamefont {T.~D.}\ \bibnamefont {Abbott}}, \bibinfo
  {author} {\bibfnamefont {F.}~\bibnamefont {Acernese}}, \bibinfo {author}
  {\bibfnamefont {K.}~\bibnamefont {Ackley}},  \emph {et~al.},\ }\href
  {\doibase 10.3847/2041-8213/aa91c9} {\bibfield  {journal} {\bibinfo
  {journal} {\apjl}\ }\textbf {\bibinfo {volume} {848}},\ \bibinfo {eid} {L12}
  (\bibinfo {year} {2017}{\natexlab{b}})},\ \Eprint
  {http://arxiv.org/abs/1710.05833} {arXiv:1710.05833 [astro-ph.HE]}
  \BibitemShut {NoStop}%
\bibitem [{\citenamefont {{Abbott}}\ \emph {et~al.}(2017)\citenamefont
  {{Abbott}}, \citenamefont {{Abbott}}, \citenamefont {{Abbott}}, \citenamefont
  {{Acernese}}, \citenamefont {{Ackley}}, \citenamefont {{Adams}},
  \citenamefont {{Adams}}, \citenamefont {{Addesso}}, \citenamefont
  {{Adhikari}}, \citenamefont {{Adya}},\ and\ \citenamefont
  {et~al.}}]{2017Natur.551...85A}%
  \BibitemOpen
  \bibfield  {author} {\bibinfo {author} {\bibfnamefont {B.~P.}\ \bibnamefont
  {{Abbott}}}, \bibinfo {author} {\bibfnamefont {R.}~\bibnamefont {{Abbott}}},
  \bibinfo {author} {\bibfnamefont {T.~D.}\ \bibnamefont {{Abbott}}}, \bibinfo
  {author} {\bibfnamefont {F.}~\bibnamefont {{Acernese}}}, \bibinfo {author}
  {\bibfnamefont {K.}~\bibnamefont {{Ackley}}}, \bibinfo {author}
  {\bibfnamefont {C.}~\bibnamefont {{Adams}}}, \bibinfo {author} {\bibfnamefont
  {T.}~\bibnamefont {{Adams}}}, \bibinfo {author} {\bibfnamefont
  {P.}~\bibnamefont {{Addesso}}}, \bibinfo {author} {\bibfnamefont {R.~X.}\
  \bibnamefont {{Adhikari}}}, \bibinfo {author} {\bibfnamefont {V.~B.}\
  \bibnamefont {{Adya}}}, \ and\ \bibinfo {author} {\bibnamefont {et~al.}},\
  }\href {\doibase 10.1038/nature24471} {\bibfield  {journal} {\bibinfo
  {journal} {\nat}\ }\textbf {\bibinfo {volume} {551}},\ \bibinfo {pages} {85}
  (\bibinfo {year} {2017})},\ \Eprint {http://arxiv.org/abs/1710.05835}
  {arXiv:1710.05835} \BibitemShut {NoStop}%
\bibitem [{\citenamefont {{Chen}}\ \emph {et~al.}(2018)\citenamefont {{Chen}},
  \citenamefont {{Fishbach}},\ and\ \citenamefont {{Holz}}}]{chen17}%
  \BibitemOpen
  \bibfield  {author} {\bibinfo {author} {\bibfnamefont {H.-Y.}\ \bibnamefont
  {{Chen}}}, \bibinfo {author} {\bibfnamefont {M.}~\bibnamefont {{Fishbach}}},
  \ and\ \bibinfo {author} {\bibfnamefont {D.~E.}\ \bibnamefont {{Holz}}},\
  }\href {\doibase 10.1038/s41586-018-0606-0} {\bibfield  {journal} {\bibinfo
  {journal} {\nat}\ }\textbf {\bibinfo {volume} {562}},\ \bibinfo {pages} {545}
  (\bibinfo {year} {2018})},\ \Eprint {http://arxiv.org/abs/1712.06531}
  {arXiv:1712.06531 [astro-ph.CO]} \BibitemShut {NoStop}%
\bibitem [{\citenamefont {Feeney}\ \emph {et~al.}(2019)\citenamefont {Feeney},
  \citenamefont {Peiris}, \citenamefont {Williamson}, \citenamefont {Nissanke},
  \citenamefont {Mortlock}, \citenamefont {Alsing},\ and\ \citenamefont
  {Scolnic}}]{Feeney:2018mkj}%
  \BibitemOpen
  \bibfield  {author} {\bibinfo {author} {\bibfnamefont {S.~M.}\ \bibnamefont
  {Feeney}}, \bibinfo {author} {\bibfnamefont {H.~V.}\ \bibnamefont {Peiris}},
  \bibinfo {author} {\bibfnamefont {A.~R.}\ \bibnamefont {Williamson}},
  \bibinfo {author} {\bibfnamefont {S.~M.}\ \bibnamefont {Nissanke}}, \bibinfo
  {author} {\bibfnamefont {D.~J.}\ \bibnamefont {Mortlock}}, \bibinfo {author}
  {\bibfnamefont {J.}~\bibnamefont {Alsing}}, \ and\ \bibinfo {author}
  {\bibfnamefont {D.}~\bibnamefont {Scolnic}},\ }\href {\doibase
  10.1103/PhysRevLett.122.061105} {\bibfield  {journal} {\bibinfo  {journal}
  {Phys. Rev. Lett.}\ }\textbf {\bibinfo {volume} {122}},\ \bibinfo {pages}
  {061105} (\bibinfo {year} {2019})},\ \Eprint
  {http://arxiv.org/abs/1802.03404} {arXiv:1802.03404 [astro-ph.CO]}
  \BibitemShut {NoStop}%
\bibitem [{\citenamefont {Feeney}\ \emph {et~al.}(2021)\citenamefont {Feeney},
  \citenamefont {Peiris}, \citenamefont {Nissanke},\ and\ \citenamefont
  {Mortlock}}]{Feeney:2020kxk}%
  \BibitemOpen
  \bibfield  {author} {\bibinfo {author} {\bibfnamefont {S.~M.}\ \bibnamefont
  {Feeney}}, \bibinfo {author} {\bibfnamefont {H.~V.}\ \bibnamefont {Peiris}},
  \bibinfo {author} {\bibfnamefont {S.~M.}\ \bibnamefont {Nissanke}}, \ and\
  \bibinfo {author} {\bibfnamefont {D.~J.}\ \bibnamefont {Mortlock}},\ }\href
  {\doibase 10.1103/PhysRevLett.126.171102} {\bibfield  {journal} {\bibinfo
  {journal} {Phys. Rev. Lett.}\ }\textbf {\bibinfo {volume} {126}},\ \bibinfo
  {pages} {171102} (\bibinfo {year} {2021})},\ \Eprint
  {http://arxiv.org/abs/2012.06593} {arXiv:2012.06593 [astro-ph.CO]}
  \BibitemShut {NoStop}%
\bibitem [{\citenamefont {Vitale}\ and\ \citenamefont
  {Chen}(2018)}]{Vitale:2018wlg}%
  \BibitemOpen
  \bibfield  {author} {\bibinfo {author} {\bibfnamefont {S.}~\bibnamefont
  {Vitale}}\ and\ \bibinfo {author} {\bibfnamefont {H.-Y.}\ \bibnamefont
  {Chen}},\ }\href {\doibase 10.1103/PhysRevLett.121.021303} {\bibfield
  {journal} {\bibinfo  {journal} {Phys. Rev. Lett.}\ }\textbf {\bibinfo
  {volume} {121}},\ \bibinfo {pages} {021303} (\bibinfo {year} {2018})},\
  \Eprint {http://arxiv.org/abs/1804.07337} {arXiv:1804.07337 [astro-ph.CO]}
  \BibitemShut {NoStop}%
\bibitem [{\citenamefont {Chen}\ \emph {et~al.}(2021)\citenamefont {Chen},
  \citenamefont {Cowperthwaite}, \citenamefont {Metzger},\ and\ \citenamefont
  {Berger}}]{Chen:2020zoq}%
  \BibitemOpen
  \bibfield  {author} {\bibinfo {author} {\bibfnamefont {H.-Y.}\ \bibnamefont
  {Chen}}, \bibinfo {author} {\bibfnamefont {P.~S.}\ \bibnamefont
  {Cowperthwaite}}, \bibinfo {author} {\bibfnamefont {B.~D.}\ \bibnamefont
  {Metzger}}, \ and\ \bibinfo {author} {\bibfnamefont {E.}~\bibnamefont
  {Berger}},\ }\href {\doibase 10.3847/2041-8213/abdab0} {\bibfield  {journal}
  {\bibinfo  {journal} {Astrophys. J. Lett.}\ }\textbf {\bibinfo {volume}
  {908}},\ \bibinfo {pages} {L4} (\bibinfo {year} {2021})},\ \Eprint
  {http://arxiv.org/abs/2011.01211} {arXiv:2011.01211 [astro-ph.CO]}
  \BibitemShut {NoStop}%
\bibitem [{\citenamefont {Del~Pozzo}(2012)}]{PhysRevD.86.043011}%
  \BibitemOpen
  \bibfield  {author} {\bibinfo {author} {\bibfnamefont {W.}~\bibnamefont
  {Del~Pozzo}},\ }\href {\doibase 10.1103/PhysRevD.86.043011} {\bibfield
  {journal} {\bibinfo  {journal} {Phys. Rev. D}\ }\textbf {\bibinfo {volume}
  {86}},\ \bibinfo {pages} {043011} (\bibinfo {year} {2012})}\BibitemShut
  {NoStop}%
\bibitem [{\citenamefont {Fishbach}\ \emph {et~al.}(2019)\citenamefont
  {Fishbach}, \citenamefont {Gray}, \citenamefont {Hernandez}, \citenamefont
  {Qi}, \citenamefont {Sur}, \citenamefont {Acernese}, \citenamefont {Aiello},
  \citenamefont {Allocca}, \citenamefont {Aloy}, \citenamefont {Amato} \emph
  {et~al.}}]{fishbach}%
  \BibitemOpen
  \bibfield  {author} {\bibinfo {author} {\bibfnamefont {M.}~\bibnamefont
  {Fishbach}}, \bibinfo {author} {\bibfnamefont {R.}~\bibnamefont {Gray}},
  \bibinfo {author} {\bibfnamefont {I.~M.}\ \bibnamefont {Hernandez}}, \bibinfo
  {author} {\bibfnamefont {H.}~\bibnamefont {Qi}}, \bibinfo {author}
  {\bibfnamefont {A.}~\bibnamefont {Sur}}, \bibinfo {author} {\bibfnamefont
  {F.}~\bibnamefont {Acernese}}, \bibinfo {author} {\bibfnamefont
  {L.}~\bibnamefont {Aiello}}, \bibinfo {author} {\bibfnamefont
  {A.}~\bibnamefont {Allocca}}, \bibinfo {author} {\bibfnamefont
  {M.}~\bibnamefont {Aloy}}, \bibinfo {author} {\bibfnamefont {A.}~\bibnamefont
  {Amato}},  \emph {et~al.},\ }\href {\doibase
  https://doi.org/10.3847/2041-8213/aaf96e} {\bibfield  {journal} {\bibinfo
  {journal} {The Astrophysical Journal Letters}\ }\textbf {\bibinfo {volume}
  {871}},\ \bibinfo {pages} {L13} (\bibinfo {year} {2019})}\BibitemShut
  {NoStop}%
\bibitem [{\citenamefont {{Gray}}\ \emph {et~al.}(2020)\citenamefont {{Gray}},
  \citenamefont {{Hernandez}}, \citenamefont {{Qi}}, \citenamefont {{Sur}},
  \citenamefont {{Brady}}, \citenamefont {{Chen}}, \citenamefont {{Farr}},
  \citenamefont {{Fishbach}}, \citenamefont {{Gair}}, \citenamefont {{Ghosh}},
  \citenamefont {{Holz}}, \citenamefont {{Mastrogiovanni}}, \citenamefont
  {{Messenger}}, \citenamefont {{Steer}},\ and\ \citenamefont
  {{Veitch}}}]{2020PhRvD.101l2001G}%
  \BibitemOpen
  \bibfield  {author} {\bibinfo {author} {\bibfnamefont {R.}~\bibnamefont
  {{Gray}}}, \bibinfo {author} {\bibfnamefont {I.~M.}\ \bibnamefont
  {{Hernandez}}}, \bibinfo {author} {\bibfnamefont {H.}~\bibnamefont {{Qi}}},
  \bibinfo {author} {\bibfnamefont {A.}~\bibnamefont {{Sur}}}, \bibinfo
  {author} {\bibfnamefont {P.~R.}\ \bibnamefont {{Brady}}}, \bibinfo {author}
  {\bibfnamefont {H.-Y.}\ \bibnamefont {{Chen}}}, \bibinfo {author}
  {\bibfnamefont {W.~M.}\ \bibnamefont {{Farr}}}, \bibinfo {author}
  {\bibfnamefont {M.}~\bibnamefont {{Fishbach}}}, \bibinfo {author}
  {\bibfnamefont {J.~R.}\ \bibnamefont {{Gair}}}, \bibinfo {author}
  {\bibfnamefont {A.}~\bibnamefont {{Ghosh}}}, \bibinfo {author} {\bibfnamefont
  {D.~E.}\ \bibnamefont {{Holz}}}, \bibinfo {author} {\bibfnamefont
  {S.}~\bibnamefont {{Mastrogiovanni}}}, \bibinfo {author} {\bibfnamefont
  {C.}~\bibnamefont {{Messenger}}}, \bibinfo {author} {\bibfnamefont {D.~A.}\
  \bibnamefont {{Steer}}}, \ and\ \bibinfo {author} {\bibfnamefont
  {J.}~\bibnamefont {{Veitch}}},\ }\href {\doibase 10.1103/PhysRevD.101.122001}
  {\bibfield  {journal} {\bibinfo  {journal} {\prd}\ }\textbf {\bibinfo
  {volume} {101}},\ \bibinfo {eid} {122001} (\bibinfo {year} {2020})},\ \Eprint
  {http://arxiv.org/abs/1908.06050} {arXiv:1908.06050 [gr-qc]} \BibitemShut
  {NoStop}%
\bibitem [{\citenamefont {Finke}\ \emph {et~al.}(2021)\citenamefont {Finke},
  \citenamefont {Foffa}, \citenamefont {Iacovelli}, \citenamefont {Maggiore},\
  and\ \citenamefont {Mancarella}}]{Finke:2021aom}%
  \BibitemOpen
  \bibfield  {author} {\bibinfo {author} {\bibfnamefont {A.}~\bibnamefont
  {Finke}}, \bibinfo {author} {\bibfnamefont {S.}~\bibnamefont {Foffa}},
  \bibinfo {author} {\bibfnamefont {F.}~\bibnamefont {Iacovelli}}, \bibinfo
  {author} {\bibfnamefont {M.}~\bibnamefont {Maggiore}}, \ and\ \bibinfo
  {author} {\bibfnamefont {M.}~\bibnamefont {Mancarella}},\ }\href {\doibase
  10.1088/1475-7516/2021/08/026} {\bibfield  {journal} {\bibinfo  {journal}
  {JCAP}\ }\textbf {\bibinfo {volume} {08}},\ \bibinfo {pages} {026} (\bibinfo
  {year} {2021})},\ \Eprint {http://arxiv.org/abs/2101.12660} {arXiv:2101.12660
  [astro-ph.CO]} \BibitemShut {NoStop}%
\bibitem [{\citenamefont {Gray}\ \emph {et~al.}(2022)\citenamefont {Gray},
  \citenamefont {Messenger},\ and\ \citenamefont {Veitch}}]{Gray2022}%
  \BibitemOpen
  \bibfield  {author} {\bibinfo {author} {\bibfnamefont {R.}~\bibnamefont
  {Gray}}, \bibinfo {author} {\bibfnamefont {C.}~\bibnamefont {Messenger}}, \
  and\ \bibinfo {author} {\bibfnamefont {J.}~\bibnamefont {Veitch}},\ }\href
  {\doibase 10.1093/mnras/stac366} {\bibfield  {journal} {\bibinfo  {journal}
  {Monthly Notices of the Royal Astronomical Society}\ }\textbf {\bibinfo
  {volume} {512}},\ \bibinfo {pages} {1127} (\bibinfo {year} {2022})},\ \Eprint
  {http://arxiv.org/abs/https://academic.oup.com/mnras/article-pdf/512/1/1127/45303118/stac366.pdf}
  {https://academic.oup.com/mnras/article-pdf/512/1/1127/45303118/stac366.pdf}
  \BibitemShut {NoStop}%
\bibitem [{\citenamefont {Leandro}\ \emph
  {et~al.}(2022{\natexlab{a}})\citenamefont {Leandro}, \citenamefont {Marra},\
  and\ \citenamefont {Sturani}}]{PhysRevD.105.023523}%
  \BibitemOpen
  \bibfield  {author} {\bibinfo {author} {\bibfnamefont {H.}~\bibnamefont
  {Leandro}}, \bibinfo {author} {\bibfnamefont {V.}~\bibnamefont {Marra}}, \
  and\ \bibinfo {author} {\bibfnamefont {R.}~\bibnamefont {Sturani}},\ }\href
  {\doibase 10.1103/PhysRevD.105.023523} {\bibfield  {journal} {\bibinfo
  {journal} {Phys. Rev. D}\ }\textbf {\bibinfo {volume} {105}},\ \bibinfo
  {pages} {023523} (\bibinfo {year} {2022}{\natexlab{a}})}\BibitemShut
  {NoStop}%
\bibitem [{\citenamefont {Gair}\ \emph {et~al.}(2022)\citenamefont {Gair},
  \citenamefont {Ghosh}, \citenamefont {Gray}, \citenamefont {Holz},
  \citenamefont {Mastrogiovanni}, \citenamefont {Mukherjee}, \citenamefont
  {Palmese}, \citenamefont {Tamanini}, \citenamefont {Baker}, \citenamefont
  {Beirnaert} \emph {et~al.}}]{2022arXiv221208694G}%
  \BibitemOpen
  \bibfield  {author} {\bibinfo {author} {\bibfnamefont {J.~R.}\ \bibnamefont
  {Gair}}, \bibinfo {author} {\bibfnamefont {A.}~\bibnamefont {Ghosh}},
  \bibinfo {author} {\bibfnamefont {R.}~\bibnamefont {Gray}}, \bibinfo {author}
  {\bibfnamefont {D.~E.}\ \bibnamefont {Holz}}, \bibinfo {author}
  {\bibfnamefont {S.}~\bibnamefont {Mastrogiovanni}}, \bibinfo {author}
  {\bibfnamefont {S.}~\bibnamefont {Mukherjee}}, \bibinfo {author}
  {\bibfnamefont {A.}~\bibnamefont {Palmese}}, \bibinfo {author} {\bibfnamefont
  {N.}~\bibnamefont {Tamanini}}, \bibinfo {author} {\bibfnamefont
  {T.}~\bibnamefont {Baker}}, \bibinfo {author} {\bibfnamefont
  {F.}~\bibnamefont {Beirnaert}},  \emph {et~al.},\ }\href@noop {} {\
  (\bibinfo {year} {2022})},\ \Eprint {http://arxiv.org/abs/2212.08694}
  {arXiv:2212.08694 [astro-ph.CO]} \BibitemShut {NoStop}%
\bibitem [{\citenamefont {Soares-Santos}\ \emph {et~al.}(2019)\citenamefont
  {Soares-Santos} \emph {et~al.}}]{2019ApJ...876L...7S}%
  \BibitemOpen
  \bibfield  {author} {\bibinfo {author} {\bibfnamefont {M.}~\bibnamefont
  {Soares-Santos}} \emph {et~al.} (\bibinfo {collaboration} {DES, LIGO
  Scientific, Virgo}),\ }\href {\doibase 10.3847/2041-8213/ab14f1} {\bibfield
  {journal} {\bibinfo  {journal} {Astrophys. J. Lett.}\ }\textbf {\bibinfo
  {volume} {876}},\ \bibinfo {pages} {L7} (\bibinfo {year} {2019})},\ \Eprint
  {http://arxiv.org/abs/1901.01540} {arXiv:1901.01540 [astro-ph.CO]}
  \BibitemShut {NoStop}%
\bibitem [{\citenamefont {Palmese}\ \emph {et~al.}(2020)\citenamefont {Palmese}
  \emph {et~al.}}]{2020ApJ...900L..33P}%
  \BibitemOpen
  \bibfield  {author} {\bibinfo {author} {\bibfnamefont {A.}~\bibnamefont
  {Palmese}} \emph {et~al.} (\bibinfo {collaboration} {DES}),\ }\href {\doibase
  10.3847/2041-8213/abaeff} {\bibfield  {journal} {\bibinfo  {journal}
  {Astrophys. J. Lett.}\ }\textbf {\bibinfo {volume} {900}},\ \bibinfo {pages}
  {L33} (\bibinfo {year} {2020})},\ \Eprint {http://arxiv.org/abs/2006.14961}
  {arXiv:2006.14961 [astro-ph.CO]} \BibitemShut {NoStop}%
\bibitem [{\citenamefont {Abbott}\ \emph
  {et~al.}(2021{\natexlab{a}})\citenamefont {Abbott} \emph
  {et~al.}}]{2021ApJ...909..218A}%
  \BibitemOpen
  \bibfield  {author} {\bibinfo {author} {\bibfnamefont {B.~P.}\ \bibnamefont
  {Abbott}} \emph {et~al.} (\bibinfo {collaboration} {LIGO Scientific, Virgo,
  VIRGO}),\ }\href {\doibase 10.3847/1538-4357/abdcb7} {\bibfield  {journal}
  {\bibinfo  {journal} {Astrophys. J.}\ }\textbf {\bibinfo {volume} {909}},\
  \bibinfo {pages} {218} (\bibinfo {year} {2021}{\natexlab{a}})},\ \Eprint
  {http://arxiv.org/abs/1908.06060} {arXiv:1908.06060 [astro-ph.CO]}
  \BibitemShut {NoStop}%
\bibitem [{\citenamefont {Abbott}\ \emph
  {et~al.}(2021{\natexlab{b}})\citenamefont {Abbott} \emph
  {et~al.}}]{2021arXiv211103604T}%
  \BibitemOpen
  \bibfield  {author} {\bibinfo {author} {\bibfnamefont {R.}~\bibnamefont
  {Abbott}} \emph {et~al.} (\bibinfo {collaboration} {LIGO Scientific, VIRGO,
  KAGRA}),\ }\href@noop {} {\  (\bibinfo {year} {2021}{\natexlab{b}})},\
  \Eprint {http://arxiv.org/abs/2111.03604} {arXiv:2111.03604 [astro-ph.CO]}
  \BibitemShut {NoStop}%
\bibitem [{\citenamefont {{Palmese}}\ \emph {et~al.}(2021)\citenamefont
  {{Palmese}}, \citenamefont {{Bom}}, \citenamefont {{Mucesh}},\ and\
  \citenamefont {{Hartley}}}]{2021arXiv211106445P}%
  \BibitemOpen
  \bibfield  {author} {\bibinfo {author} {\bibfnamefont {A.}~\bibnamefont
  {{Palmese}}}, \bibinfo {author} {\bibfnamefont {C.~R.}\ \bibnamefont
  {{Bom}}}, \bibinfo {author} {\bibfnamefont {S.}~\bibnamefont {{Mucesh}}}, \
  and\ \bibinfo {author} {\bibfnamefont {W.~G.}\ \bibnamefont {{Hartley}}},\
  }\href@noop {} {\bibfield  {journal} {\bibinfo  {journal} {arXiv e-prints}\
  ,\ \bibinfo {eid} {arXiv:2111.06445}} (\bibinfo {year} {2021})},\ \Eprint
  {http://arxiv.org/abs/2111.06445} {arXiv:2111.06445 [astro-ph.CO]}
  \BibitemShut {NoStop}%
\bibitem [{\citenamefont {Oguri}(2016)}]{PhysRevD.93.083511}%
  \BibitemOpen
  \bibfield  {author} {\bibinfo {author} {\bibfnamefont {M.}~\bibnamefont
  {Oguri}},\ }\href {\doibase 10.1103/PhysRevD.93.083511} {\bibfield  {journal}
  {\bibinfo  {journal} {Phys. Rev. D}\ }\textbf {\bibinfo {volume} {93}},\
  \bibinfo {pages} {083511} (\bibinfo {year} {2016})}\BibitemShut {NoStop}%
\bibitem [{\citenamefont {Mukherjee}\ \emph {et~al.}(2020)\citenamefont
  {Mukherjee}, \citenamefont {Wandelt},\ and\ \citenamefont
  {Silk}}]{Mukherjee:2019wcg}%
  \BibitemOpen
  \bibfield  {author} {\bibinfo {author} {\bibfnamefont {S.}~\bibnamefont
  {Mukherjee}}, \bibinfo {author} {\bibfnamefont {B.~D.}\ \bibnamefont
  {Wandelt}}, \ and\ \bibinfo {author} {\bibfnamefont {J.}~\bibnamefont
  {Silk}},\ }\href {\doibase 10.1093/mnras/staa827} {\bibfield  {journal}
  {\bibinfo  {journal} {Mon. Not. Roy. Astron. Soc.}\ }\textbf {\bibinfo
  {volume} {494}},\ \bibinfo {pages} {1956} (\bibinfo {year} {2020})},\ \Eprint
  {http://arxiv.org/abs/1908.08951} {arXiv:1908.08951 [astro-ph.CO]}
  \BibitemShut {NoStop}%
\bibitem [{\citenamefont {Mukherjee}\ \emph {et~al.}(2021)\citenamefont
  {Mukherjee}, \citenamefont {Wandelt}, \citenamefont {Nissanke},\ and\
  \citenamefont {Silvestri}}]{Mukherjee:2020hyn}%
  \BibitemOpen
  \bibfield  {author} {\bibinfo {author} {\bibfnamefont {S.}~\bibnamefont
  {Mukherjee}}, \bibinfo {author} {\bibfnamefont {B.~D.}\ \bibnamefont
  {Wandelt}}, \bibinfo {author} {\bibfnamefont {S.~M.}\ \bibnamefont
  {Nissanke}}, \ and\ \bibinfo {author} {\bibfnamefont {A.}~\bibnamefont
  {Silvestri}},\ }\href {\doibase 10.1103/PhysRevD.103.043520} {\bibfield
  {journal} {\bibinfo  {journal} {Phys. Rev. D}\ }\textbf {\bibinfo {volume}
  {103}},\ \bibinfo {pages} {043520} (\bibinfo {year} {2021})},\ \Eprint
  {http://arxiv.org/abs/2007.02943} {arXiv:2007.02943 [astro-ph.CO]}
  \BibitemShut {NoStop}%
\bibitem [{\citenamefont {Bera}\ \emph {et~al.}(2020)\citenamefont {Bera},
  \citenamefont {Rana}, \citenamefont {More},\ and\ \citenamefont
  {Bose}}]{Bera:2020jhx}%
  \BibitemOpen
  \bibfield  {author} {\bibinfo {author} {\bibfnamefont {S.}~\bibnamefont
  {Bera}}, \bibinfo {author} {\bibfnamefont {D.}~\bibnamefont {Rana}}, \bibinfo
  {author} {\bibfnamefont {S.}~\bibnamefont {More}}, \ and\ \bibinfo {author}
  {\bibfnamefont {S.}~\bibnamefont {Bose}},\ }\href {\doibase
  10.3847/1538-4357/abb4e0} {\bibfield  {journal} {\bibinfo  {journal}
  {Astrophys. J.}\ }\textbf {\bibinfo {volume} {902}},\ \bibinfo {pages} {79}
  (\bibinfo {year} {2020})},\ \Eprint {http://arxiv.org/abs/2007.04271}
  {arXiv:2007.04271 [astro-ph.CO]} \BibitemShut {NoStop}%
\bibitem [{\citenamefont {Diaz}\ and\ \citenamefont
  {Mukherjee}(2022)}]{Diaz:2021pem}%
  \BibitemOpen
  \bibfield  {author} {\bibinfo {author} {\bibfnamefont {C.~C.}\ \bibnamefont
  {Diaz}}\ and\ \bibinfo {author} {\bibfnamefont {S.}~\bibnamefont
  {Mukherjee}},\ }\href {\doibase 10.1093/mnras/stac208} {\bibfield  {journal}
  {\bibinfo  {journal} {Mon. Not. Roy. Astron. Soc.}\ }\textbf {\bibinfo
  {volume} {511}},\ \bibinfo {pages} {2782} (\bibinfo {year} {2022})},\ \Eprint
  {http://arxiv.org/abs/2107.12787} {arXiv:2107.12787 [astro-ph.CO]}
  \BibitemShut {NoStop}%
\bibitem [{\citenamefont {Balaudo}\ \emph {et~al.}(2022)\citenamefont
  {Balaudo}, \citenamefont {Garoffolo}, \citenamefont {Martinelli},
  \citenamefont {Mukherjee},\ and\ \citenamefont
  {Silvestri}}]{Balaudo:2022znx}%
  \BibitemOpen
  \bibfield  {author} {\bibinfo {author} {\bibfnamefont {A.}~\bibnamefont
  {Balaudo}}, \bibinfo {author} {\bibfnamefont {A.}~\bibnamefont {Garoffolo}},
  \bibinfo {author} {\bibfnamefont {M.}~\bibnamefont {Martinelli}}, \bibinfo
  {author} {\bibfnamefont {S.}~\bibnamefont {Mukherjee}}, \ and\ \bibinfo
  {author} {\bibfnamefont {A.}~\bibnamefont {Silvestri}},\ }\href@noop {} {\
  (\bibinfo {year} {2022})},\ \Eprint {http://arxiv.org/abs/2210.06398}
  {arXiv:2210.06398 [astro-ph.CO]} \BibitemShut {NoStop}%
\bibitem [{\citenamefont {{Chernoff}}\ and\ \citenamefont
  {{Finn}}(1993)}]{1993ApJ...411L...5C}%
  \BibitemOpen
  \bibfield  {author} {\bibinfo {author} {\bibfnamefont {D.~F.}\ \bibnamefont
  {{Chernoff}}}\ and\ \bibinfo {author} {\bibfnamefont {L.~S.}\ \bibnamefont
  {{Finn}}},\ }\href {\doibase 10.1086/186898} {\bibfield  {journal} {\bibinfo
  {journal} {\apjl}\ }\textbf {\bibinfo {volume} {411}},\ \bibinfo {pages} {L5}
  (\bibinfo {year} {1993})},\ \Eprint {http://arxiv.org/abs/gr-qc/9304020}
  {arXiv:gr-qc/9304020 [gr-qc]} \BibitemShut {NoStop}%
\bibitem [{\citenamefont {Taylor}\ \emph
  {et~al.}(2012{\natexlab{a}})\citenamefont {Taylor}, \citenamefont {Gair},\
  and\ \citenamefont {Mandel}}]{Taylor_2012}%
  \BibitemOpen
  \bibfield  {author} {\bibinfo {author} {\bibfnamefont {S.~R.}\ \bibnamefont
  {Taylor}}, \bibinfo {author} {\bibfnamefont {J.~R.}\ \bibnamefont {Gair}}, \
  and\ \bibinfo {author} {\bibfnamefont {I.}~\bibnamefont {Mandel}},\ }\href
  {\doibase 10.1103/physrevd.85.023535} {\bibfield  {journal} {\bibinfo
  {journal} {Physical Review D}\ }\textbf {\bibinfo {volume} {85}} (\bibinfo
  {year} {2012}{\natexlab{a}}),\ 10.1103/physrevd.85.023535}\BibitemShut
  {NoStop}%
\bibitem [{\citenamefont {Farr}\ \emph {et~al.}(2019)\citenamefont {Farr},
  \citenamefont {Fishbach}, \citenamefont {Ye},\ and\ \citenamefont
  {Holz}}]{Farr_2019}%
  \BibitemOpen
  \bibfield  {author} {\bibinfo {author} {\bibfnamefont {W.~M.}\ \bibnamefont
  {Farr}}, \bibinfo {author} {\bibfnamefont {M.}~\bibnamefont {Fishbach}},
  \bibinfo {author} {\bibfnamefont {J.}~\bibnamefont {Ye}}, \ and\ \bibinfo
  {author} {\bibfnamefont {D.~E.}\ \bibnamefont {Holz}},\ }\href {\doibase
  10.3847/2041-8213/ab4284} {\bibfield  {journal} {\bibinfo  {journal} {The
  Astrophysical Journal}\ }\textbf {\bibinfo {volume} {883}},\ \bibinfo {pages}
  {L42} (\bibinfo {year} {2019})}\BibitemShut {NoStop}%
\bibitem [{\citenamefont {{Mar{\'\i}a Ezquiaga}}\ and\ \citenamefont
  {{Holz}}(2020)}]{2020arXiv200602211M}%
  \BibitemOpen
  \bibfield  {author} {\bibinfo {author} {\bibfnamefont {J.}~\bibnamefont
  {{Mar{\'\i}a Ezquiaga}}}\ and\ \bibinfo {author} {\bibfnamefont {D.~E.}\
  \bibnamefont {{Holz}}},\ }\href@noop {} {\bibfield  {journal} {\bibinfo
  {journal} {arXiv}\ ,\ \bibinfo {eid} {arXiv:2006.02211}} (\bibinfo {year}
  {2020})},\ \Eprint {http://arxiv.org/abs/2006.02211} {arXiv:2006.02211
  [astro-ph.HE]} \BibitemShut {NoStop}%
\bibitem [{\citenamefont {{Mastrogiovanni}}\ \emph {et~al.}(2021)\citenamefont
  {{Mastrogiovanni}}, \citenamefont {{Leyde}}, \citenamefont {{Karathanasis}},
  \citenamefont {{Chassande-Mottin}}, \citenamefont {{Steer}}, \citenamefont
  {{Gair}}, \citenamefont {{Ghosh}}, \citenamefont {{Gray}}, \citenamefont
  {{Mukherjee}},\ and\ \citenamefont {{Rinaldi}}}]{mastrogiovanni_2021}%
  \BibitemOpen
  \bibfield  {author} {\bibinfo {author} {\bibfnamefont {S.}~\bibnamefont
  {{Mastrogiovanni}}}, \bibinfo {author} {\bibfnamefont {K.}~\bibnamefont
  {{Leyde}}}, \bibinfo {author} {\bibfnamefont {C.}~\bibnamefont
  {{Karathanasis}}}, \bibinfo {author} {\bibfnamefont {E.}~\bibnamefont
  {{Chassande-Mottin}}}, \bibinfo {author} {\bibfnamefont {D.~A.}\ \bibnamefont
  {{Steer}}}, \bibinfo {author} {\bibfnamefont {J.}~\bibnamefont {{Gair}}},
  \bibinfo {author} {\bibfnamefont {A.}~\bibnamefont {{Ghosh}}}, \bibinfo
  {author} {\bibfnamefont {R.}~\bibnamefont {{Gray}}}, \bibinfo {author}
  {\bibfnamefont {S.}~\bibnamefont {{Mukherjee}}}, \ and\ \bibinfo {author}
  {\bibfnamefont {S.}~\bibnamefont {{Rinaldi}}},\ }\href {\doibase
  10.1103/PhysRevD.104.062009} {\bibfield  {journal} {\bibinfo  {journal}
  {\prd}\ }\textbf {\bibinfo {volume} {104}},\ \bibinfo {eid} {062009}
  (\bibinfo {year} {2021})},\ \Eprint {http://arxiv.org/abs/2103.14663}
  {arXiv:2103.14663 [gr-qc]} \BibitemShut {NoStop}%
\bibitem [{\citenamefont {Mukherjee}(2022)}]{Mukherjee:2021rtw}%
  \BibitemOpen
  \bibfield  {author} {\bibinfo {author} {\bibfnamefont {S.}~\bibnamefont
  {Mukherjee}},\ }\href {\doibase 10.1093/mnras/stac2152} {\bibfield  {journal}
  {\bibinfo  {journal} {Mon. Not. Roy. Astron. Soc.}\ }\textbf {\bibinfo
  {volume} {515}},\ \bibinfo {pages} {5495} (\bibinfo {year} {2022})},\ \Eprint
  {http://arxiv.org/abs/2112.10256} {arXiv:2112.10256 [astro-ph.CO]}
  \BibitemShut {NoStop}%
\bibitem [{\citenamefont {{Leyde}}\ \emph {et~al.}(2022)\citenamefont
  {{Leyde}}, \citenamefont {{Mastrogiovanni}}, \citenamefont {{Steer}},
  \citenamefont {{Chassande-Mottin}},\ and\ \citenamefont
  {{Karathanasis}}}]{2022JCAP...09..012L}%
  \BibitemOpen
  \bibfield  {author} {\bibinfo {author} {\bibfnamefont {K.}~\bibnamefont
  {{Leyde}}}, \bibinfo {author} {\bibfnamefont {S.}~\bibnamefont
  {{Mastrogiovanni}}}, \bibinfo {author} {\bibfnamefont {D.~A.}\ \bibnamefont
  {{Steer}}}, \bibinfo {author} {\bibfnamefont {E.}~\bibnamefont
  {{Chassande-Mottin}}}, \ and\ \bibinfo {author} {\bibfnamefont
  {C.}~\bibnamefont {{Karathanasis}}},\ }\href {\doibase
  10.1088/1475-7516/2022/09/012} {\bibfield  {journal} {\bibinfo  {journal}
  {\jcap}\ }\textbf {\bibinfo {volume} {2022}},\ \bibinfo {eid} {012} (\bibinfo
  {year} {2022})},\ \Eprint {http://arxiv.org/abs/2203.11680} {arXiv:2203.11680
  [gr-qc]} \BibitemShut {NoStop}%
\bibitem [{\citenamefont {Ezquiaga}\ and\ \citenamefont
  {Holz}(2022)}]{Ezquiaga_2022}%
  \BibitemOpen
  \bibfield  {author} {\bibinfo {author} {\bibfnamefont {J.~M.}\ \bibnamefont
  {Ezquiaga}}\ and\ \bibinfo {author} {\bibfnamefont {D.~E.}\ \bibnamefont
  {Holz}},\ }\href {\doibase 10.1103/physrevlett.129.061102} {\bibfield
  {journal} {\bibinfo  {journal} {Physical Review Letters}\ }\textbf {\bibinfo
  {volume} {129}} (\bibinfo {year} {2022}),\
  10.1103/physrevlett.129.061102}\BibitemShut {NoStop}%
\bibitem [{\citenamefont {Karathanasis}\ \emph {et~al.}(2022)\citenamefont
  {Karathanasis}, \citenamefont {Mukherjee},\ and\ \citenamefont
  {Mastrogiovanni}}]{Karathanasis:2022rtr}%
  \BibitemOpen
  \bibfield  {author} {\bibinfo {author} {\bibfnamefont {C.}~\bibnamefont
  {Karathanasis}}, \bibinfo {author} {\bibfnamefont {S.}~\bibnamefont
  {Mukherjee}}, \ and\ \bibinfo {author} {\bibfnamefont {S.}~\bibnamefont
  {Mastrogiovanni}},\ }\href@noop {} {\  (\bibinfo {year} {2022})},\ \Eprint
  {http://arxiv.org/abs/2204.13495} {arXiv:2204.13495 [astro-ph.CO]}
  \BibitemShut {NoStop}%
\bibitem [{\citenamefont {Mancarella}\ \emph {et~al.}(2022)\citenamefont
  {Mancarella}, \citenamefont {Borghi}, \citenamefont {Foffa}, \citenamefont
  {Genoud-Prachex}, \citenamefont {Iacovelli}, \citenamefont {Maggiore},
  \citenamefont {Moresco},\ and\ \citenamefont {Schulz}}]{Mancarella:2022cnu}%
  \BibitemOpen
  \bibfield  {author} {\bibinfo {author} {\bibfnamefont {M.}~\bibnamefont
  {Mancarella}}, \bibinfo {author} {\bibfnamefont {N.}~\bibnamefont {Borghi}},
  \bibinfo {author} {\bibfnamefont {S.}~\bibnamefont {Foffa}}, \bibinfo
  {author} {\bibfnamefont {E.}~\bibnamefont {Genoud-Prachex}}, \bibinfo
  {author} {\bibfnamefont {F.}~\bibnamefont {Iacovelli}}, \bibinfo {author}
  {\bibfnamefont {M.}~\bibnamefont {Maggiore}}, \bibinfo {author}
  {\bibfnamefont {M.}~\bibnamefont {Moresco}}, \ and\ \bibinfo {author}
  {\bibfnamefont {M.}~\bibnamefont {Schulz}},\ }\href {\doibase
  10.22323/1.414.0127} {\bibfield  {journal} {\bibinfo  {journal} {PoS}\
  }\textbf {\bibinfo {volume} {ICHEP2022}},\ \bibinfo {pages} {127} (\bibinfo
  {year} {2022})},\ \Eprint {http://arxiv.org/abs/2211.15512} {arXiv:2211.15512
  [gr-qc]} \BibitemShut {NoStop}%
\bibitem [{\citenamefont {{Ezquiaga}}\ and\ \citenamefont
  {{Holz}}(2022)}]{2022PhRvL.129f1102E}%
  \BibitemOpen
  \bibfield  {author} {\bibinfo {author} {\bibfnamefont {J.~M.}\ \bibnamefont
  {{Ezquiaga}}}\ and\ \bibinfo {author} {\bibfnamefont {D.~E.}\ \bibnamefont
  {{Holz}}},\ }\href {\doibase 10.1103/PhysRevLett.129.061102} {\bibfield
  {journal} {\bibinfo  {journal} {\prl}\ }\textbf {\bibinfo {volume} {129}},\
  \bibinfo {eid} {061102} (\bibinfo {year} {2022})},\ \Eprint
  {http://arxiv.org/abs/2202.08240} {arXiv:2202.08240 [astro-ph.CO]}
  \BibitemShut {NoStop}%
\bibitem [{\citenamefont {Punturo}\ \emph {et~al.}(2010)\citenamefont
  {Punturo}, \citenamefont {Abernathy}, \citenamefont {Acernese}, \citenamefont
  {Allen}, \citenamefont {Andersson}, \citenamefont {Arun}, \citenamefont
  {Barone}, \citenamefont {Barr}, \citenamefont {Barsuglia}, \citenamefont
  {Beker} \emph {et~al.}}]{2010CQGra..27s4002P}%
  \BibitemOpen
  \bibfield  {author} {\bibinfo {author} {\bibfnamefont {M.}~\bibnamefont
  {Punturo}}, \bibinfo {author} {\bibfnamefont {M.}~\bibnamefont {Abernathy}},
  \bibinfo {author} {\bibfnamefont {F.}~\bibnamefont {Acernese}}, \bibinfo
  {author} {\bibfnamefont {B.}~\bibnamefont {Allen}}, \bibinfo {author}
  {\bibfnamefont {N.}~\bibnamefont {Andersson}}, \bibinfo {author}
  {\bibfnamefont {K.}~\bibnamefont {Arun}}, \bibinfo {author} {\bibfnamefont
  {F.}~\bibnamefont {Barone}}, \bibinfo {author} {\bibfnamefont
  {B.}~\bibnamefont {Barr}}, \bibinfo {author} {\bibfnamefont {M.}~\bibnamefont
  {Barsuglia}}, \bibinfo {author} {\bibfnamefont {M.}~\bibnamefont {Beker}},
  \emph {et~al.},\ }\href {\doibase 10.1088/0264-9381/27/19/194002} {\bibfield
  {journal} {\bibinfo  {journal} {Classical and Quantum Gravity}\ }\textbf
  {\bibinfo {volume} {27}},\ \bibinfo {pages} {194002} (\bibinfo {year}
  {2010})}\BibitemShut {NoStop}%
\bibitem [{\citenamefont {Hild}\ \emph {et~al.}(2011)\citenamefont {Hild},
  \citenamefont {Abernathy}, \citenamefont {Acernese}, \citenamefont
  {Amaro-Seoane}, \citenamefont {Andersson}, \citenamefont {Arun},
  \citenamefont {Barone}, \citenamefont {Barr}, \citenamefont {Barsuglia},
  \citenamefont {Beker} \emph {et~al.}}]{2011CQGra..28i4013H}%
  \BibitemOpen
  \bibfield  {author} {\bibinfo {author} {\bibfnamefont {S.}~\bibnamefont
  {Hild}}, \bibinfo {author} {\bibfnamefont {M.}~\bibnamefont {Abernathy}},
  \bibinfo {author} {\bibfnamefont {F.~e.}\ \bibnamefont {Acernese}}, \bibinfo
  {author} {\bibfnamefont {P.}~\bibnamefont {Amaro-Seoane}}, \bibinfo {author}
  {\bibfnamefont {N.}~\bibnamefont {Andersson}}, \bibinfo {author}
  {\bibfnamefont {K.}~\bibnamefont {Arun}}, \bibinfo {author} {\bibfnamefont
  {F.}~\bibnamefont {Barone}}, \bibinfo {author} {\bibfnamefont
  {B.}~\bibnamefont {Barr}}, \bibinfo {author} {\bibfnamefont {M.}~\bibnamefont
  {Barsuglia}}, \bibinfo {author} {\bibfnamefont {M.}~\bibnamefont {Beker}},
  \emph {et~al.},\ }\href {\doibase 10.1088/0264-9381/28/9/094013} {\bibfield
  {journal} {\bibinfo  {journal} {Classical and Quantum gravity}\ }\textbf
  {\bibinfo {volume} {28}},\ \bibinfo {pages} {094013} (\bibinfo {year}
  {2011})}\BibitemShut {NoStop}%
\bibitem [{\citenamefont {{Maggiore}}\ \emph {et~al.}(2020)\citenamefont
  {{Maggiore}}, \citenamefont {{Van Den Broeck}}, \citenamefont {{Bartolo}},
  \citenamefont {{Belgacem}}, \citenamefont {{Bertacca}}, \citenamefont
  {{Bizouard}}, \citenamefont {{Branchesi}}, \citenamefont {{Clesse}},
  \citenamefont {{Foffa}}, \citenamefont {{Garc{\'\i}a-Bellido}}, \citenamefont
  {{Grimm}}, \citenamefont {{Harms}}, \citenamefont {{Hinderer}}, \citenamefont
  {{Matarrese}}, \citenamefont {{Palomba}}, \citenamefont {{Peloso}},
  \citenamefont {{Ricciardone}},\ and\ \citenamefont
  {{Sakellariadou}}}]{2020JCAP...03..050M}%
  \BibitemOpen
  \bibfield  {author} {\bibinfo {author} {\bibfnamefont {M.}~\bibnamefont
  {{Maggiore}}}, \bibinfo {author} {\bibfnamefont {C.}~\bibnamefont {{Van Den
  Broeck}}}, \bibinfo {author} {\bibfnamefont {N.}~\bibnamefont {{Bartolo}}},
  \bibinfo {author} {\bibfnamefont {E.}~\bibnamefont {{Belgacem}}}, \bibinfo
  {author} {\bibfnamefont {D.}~\bibnamefont {{Bertacca}}}, \bibinfo {author}
  {\bibfnamefont {M.~A.}\ \bibnamefont {{Bizouard}}}, \bibinfo {author}
  {\bibfnamefont {M.}~\bibnamefont {{Branchesi}}}, \bibinfo {author}
  {\bibfnamefont {S.}~\bibnamefont {{Clesse}}}, \bibinfo {author}
  {\bibfnamefont {S.}~\bibnamefont {{Foffa}}}, \bibinfo {author} {\bibfnamefont
  {J.}~\bibnamefont {{Garc{\'\i}a-Bellido}}}, \bibinfo {author} {\bibfnamefont
  {S.}~\bibnamefont {{Grimm}}}, \bibinfo {author} {\bibfnamefont
  {J.}~\bibnamefont {{Harms}}}, \bibinfo {author} {\bibfnamefont
  {T.}~\bibnamefont {{Hinderer}}}, \bibinfo {author} {\bibfnamefont
  {S.}~\bibnamefont {{Matarrese}}}, \bibinfo {author} {\bibfnamefont
  {C.}~\bibnamefont {{Palomba}}}, \bibinfo {author} {\bibfnamefont
  {M.}~\bibnamefont {{Peloso}}}, \bibinfo {author} {\bibfnamefont
  {A.}~\bibnamefont {{Ricciardone}}}, \ and\ \bibinfo {author} {\bibfnamefont
  {M.}~\bibnamefont {{Sakellariadou}}},\ }\href {\doibase
  10.1088/1475-7516/2020/03/050} {\bibfield  {journal} {\bibinfo  {journal}
  {\jcap}\ }\textbf {\bibinfo {volume} {2020}},\ \bibinfo {eid} {050} (\bibinfo
  {year} {2020})},\ \Eprint {http://arxiv.org/abs/1912.02622} {arXiv:1912.02622
  [astro-ph.CO]} \BibitemShut {NoStop}%
\bibitem [{\citenamefont {Reitze}\ \emph {et~al.}(2019)\citenamefont {Reitze},
  \citenamefont {Adhikari}, \citenamefont {Ballmer}, \citenamefont {Barish},
  \citenamefont {Barsotti}, \citenamefont {Billingsley}, \citenamefont {Brown},
  \citenamefont {Chen}, \citenamefont {Coyne}, \citenamefont {Eisenstein},
  \citenamefont {Evans}, \citenamefont {Fritschel}, \citenamefont {Hall},
  \citenamefont {Lazzarini}, \citenamefont {Lovelace}, \citenamefont {Read},
  \citenamefont {Sathyaprakash}, \citenamefont {Shoemaker}, \citenamefont
  {Smith}, \citenamefont {Torrie}, \citenamefont {Vitale}, \citenamefont
  {Weiss}, \citenamefont {Wipf},\ and\ \citenamefont
  {Zucker}}]{Reitze2019Cosmic}%
  \BibitemOpen
  \bibfield  {author} {\bibinfo {author} {\bibfnamefont {D.}~\bibnamefont
  {Reitze}}, \bibinfo {author} {\bibfnamefont {R.~X.}\ \bibnamefont
  {Adhikari}}, \bibinfo {author} {\bibfnamefont {S.}~\bibnamefont {Ballmer}},
  \bibinfo {author} {\bibfnamefont {B.}~\bibnamefont {Barish}}, \bibinfo
  {author} {\bibfnamefont {L.}~\bibnamefont {Barsotti}}, \bibinfo {author}
  {\bibfnamefont {G.}~\bibnamefont {Billingsley}}, \bibinfo {author}
  {\bibfnamefont {D.~A.}\ \bibnamefont {Brown}}, \bibinfo {author}
  {\bibfnamefont {Y.}~\bibnamefont {Chen}}, \bibinfo {author} {\bibfnamefont
  {D.}~\bibnamefont {Coyne}}, \bibinfo {author} {\bibfnamefont
  {R.}~\bibnamefont {Eisenstein}}, \bibinfo {author} {\bibfnamefont
  {M.}~\bibnamefont {Evans}}, \bibinfo {author} {\bibfnamefont
  {P.}~\bibnamefont {Fritschel}}, \bibinfo {author} {\bibfnamefont {E.~D.}\
  \bibnamefont {Hall}}, \bibinfo {author} {\bibfnamefont {A.}~\bibnamefont
  {Lazzarini}}, \bibinfo {author} {\bibfnamefont {G.}~\bibnamefont {Lovelace}},
  \bibinfo {author} {\bibfnamefont {J.}~\bibnamefont {Read}}, \bibinfo {author}
  {\bibfnamefont {B.~S.}\ \bibnamefont {Sathyaprakash}}, \bibinfo {author}
  {\bibfnamefont {D.}~\bibnamefont {Shoemaker}}, \bibinfo {author}
  {\bibfnamefont {J.}~\bibnamefont {Smith}}, \bibinfo {author} {\bibfnamefont
  {C.}~\bibnamefont {Torrie}}, \bibinfo {author} {\bibfnamefont
  {S.}~\bibnamefont {Vitale}}, \bibinfo {author} {\bibfnamefont
  {R.}~\bibnamefont {Weiss}}, \bibinfo {author} {\bibfnamefont
  {C.}~\bibnamefont {Wipf}}, \ and\ \bibinfo {author} {\bibfnamefont
  {M.}~\bibnamefont {Zucker}},\ }\href {https://baas.aas.org/pub/2020n7i035}
  {\bibfield  {journal} {\bibinfo  {journal} {Bulletin of the AAS}\ }\textbf
  {\bibinfo {volume} {51}} (\bibinfo {year} {2019})}\BibitemShut {NoStop}%
\bibitem [{\citenamefont {Evans}\ \emph {et~al.}(2021)\citenamefont {Evans}
  \emph {et~al.}}]{Evans:2021gyd}%
  \BibitemOpen
  \bibfield  {author} {\bibinfo {author} {\bibfnamefont {M.}~\bibnamefont
  {Evans}} \emph {et~al.},\ }\href@noop {} {\  (\bibinfo {year} {2021})},\
  \Eprint {http://arxiv.org/abs/2109.09882} {arXiv:2109.09882 [astro-ph.IM]}
  \BibitemShut {NoStop}%
\bibitem [{\citenamefont {Kalogera}\ \emph {et~al.}(2021)\citenamefont
  {Kalogera} \emph {et~al.}}]{Kalogera:2021bya}%
  \BibitemOpen
  \bibfield  {author} {\bibinfo {author} {\bibfnamefont {V.}~\bibnamefont
  {Kalogera}} \emph {et~al.},\ }\href@noop {} {\  (\bibinfo {year} {2021})},\
  \Eprint {http://arxiv.org/abs/2111.06990} {arXiv:2111.06990 [gr-qc]}
  \BibitemShut {NoStop}%
\bibitem [{\citenamefont {Cai}\ and\ \citenamefont {Yang}(2017)}]{Cai:2016sby}%
  \BibitemOpen
  \bibfield  {author} {\bibinfo {author} {\bibfnamefont {R.-G.}\ \bibnamefont
  {Cai}}\ and\ \bibinfo {author} {\bibfnamefont {T.}~\bibnamefont {Yang}},\
  }\href {\doibase 10.1103/PhysRevD.95.044024} {\bibfield  {journal} {\bibinfo
  {journal} {Phys. Rev. D}\ }\textbf {\bibinfo {volume} {95}},\ \bibinfo
  {pages} {044024} (\bibinfo {year} {2017})},\ \Eprint
  {http://arxiv.org/abs/1608.08008} {arXiv:1608.08008 [astro-ph.CO]}
  \BibitemShut {NoStop}%
\bibitem [{\citenamefont {Zhao}\ and\ \citenamefont
  {Wen}(2018)}]{Zhao:2017cbb}%
  \BibitemOpen
  \bibfield  {author} {\bibinfo {author} {\bibfnamefont {W.}~\bibnamefont
  {Zhao}}\ and\ \bibinfo {author} {\bibfnamefont {L.}~\bibnamefont {Wen}},\
  }\href {\doibase 10.1103/PhysRevD.97.064031} {\bibfield  {journal} {\bibinfo
  {journal} {Phys. Rev. D}\ }\textbf {\bibinfo {volume} {97}},\ \bibinfo
  {pages} {064031} (\bibinfo {year} {2018})},\ \Eprint
  {http://arxiv.org/abs/1710.05325} {arXiv:1710.05325 [astro-ph.CO]}
  \BibitemShut {NoStop}%
\bibitem [{\citenamefont {Belgacem}\ \emph
  {et~al.}(2019{\natexlab{a}})\citenamefont {Belgacem}, \citenamefont {Dirian},
  \citenamefont {Foffa}, \citenamefont {Howell}, \citenamefont {Maggiore},\
  and\ \citenamefont {Regimbau}}]{Belgacem:2019tbw}%
  \BibitemOpen
  \bibfield  {author} {\bibinfo {author} {\bibfnamefont {E.}~\bibnamefont
  {Belgacem}}, \bibinfo {author} {\bibfnamefont {Y.}~\bibnamefont {Dirian}},
  \bibinfo {author} {\bibfnamefont {S.}~\bibnamefont {Foffa}}, \bibinfo
  {author} {\bibfnamefont {E.~J.}\ \bibnamefont {Howell}}, \bibinfo {author}
  {\bibfnamefont {M.}~\bibnamefont {Maggiore}}, \ and\ \bibinfo {author}
  {\bibfnamefont {T.}~\bibnamefont {Regimbau}},\ }\href {\doibase
  10.1088/1475-7516/2019/08/015} {\bibfield  {journal} {\bibinfo  {journal}
  {JCAP}\ }\textbf {\bibinfo {volume} {08}},\ \bibinfo {pages} {015} (\bibinfo
  {year} {2019}{\natexlab{a}})},\ \Eprint {http://arxiv.org/abs/1907.01487}
  {arXiv:1907.01487 [astro-ph.CO]} \BibitemShut {NoStop}%
\bibitem [{\citenamefont {de~Souza}\ \emph {et~al.}(2022)\citenamefont
  {de~Souza}, \citenamefont {Sturani},\ and\ \citenamefont
  {Alcaniz}}]{deSouza:2021xtg}%
  \BibitemOpen
  \bibfield  {author} {\bibinfo {author} {\bibfnamefont {J.~M.~S.}\
  \bibnamefont {de~Souza}}, \bibinfo {author} {\bibfnamefont {R.}~\bibnamefont
  {Sturani}}, \ and\ \bibinfo {author} {\bibfnamefont {J.}~\bibnamefont
  {Alcaniz}},\ }\href {\doibase 10.1088/1475-7516/2022/03/025} {\bibfield
  {journal} {\bibinfo  {journal} {JCAP}\ }\textbf {\bibinfo {volume} {03}},\
  \bibinfo {pages} {025} (\bibinfo {year} {2022})},\ \Eprint
  {http://arxiv.org/abs/2110.13316} {arXiv:2110.13316 [gr-qc]} \BibitemShut
  {NoStop}%
\bibitem [{\citenamefont {Califano}\ \emph {et~al.}(2022)\citenamefont
  {Califano}, \citenamefont {de~Martino}, \citenamefont {Vernieri},\ and\
  \citenamefont {Capozziello}}]{Califano:2022syd}%
  \BibitemOpen
  \bibfield  {author} {\bibinfo {author} {\bibfnamefont {M.}~\bibnamefont
  {Califano}}, \bibinfo {author} {\bibfnamefont {I.}~\bibnamefont
  {de~Martino}}, \bibinfo {author} {\bibfnamefont {D.}~\bibnamefont
  {Vernieri}}, \ and\ \bibinfo {author} {\bibfnamefont {S.}~\bibnamefont
  {Capozziello}},\ }\href@noop {} {\  (\bibinfo {year} {2022})},\ \Eprint
  {http://arxiv.org/abs/2208.13999} {arXiv:2208.13999 [astro-ph.CO]}
  \BibitemShut {NoStop}%
\bibitem [{\citenamefont {Dhani}\ \emph {et~al.}(2022)\citenamefont {Dhani},
  \citenamefont {Borhanian}, \citenamefont {Gupta},\ and\ \citenamefont
  {Sathyaprakash}}]{Dhani:2022ulg}%
  \BibitemOpen
  \bibfield  {author} {\bibinfo {author} {\bibfnamefont {A.}~\bibnamefont
  {Dhani}}, \bibinfo {author} {\bibfnamefont {S.}~\bibnamefont {Borhanian}},
  \bibinfo {author} {\bibfnamefont {A.}~\bibnamefont {Gupta}}, \ and\ \bibinfo
  {author} {\bibfnamefont {B.}~\bibnamefont {Sathyaprakash}},\ }\href@noop {}
  {\  (\bibinfo {year} {2022})},\ \Eprint {http://arxiv.org/abs/2212.13183}
  {arXiv:2212.13183 [gr-qc]} \BibitemShut {NoStop}%
\bibitem [{\citenamefont {Alfradique}\ \emph {et~al.}(2022)\citenamefont
  {Alfradique}, \citenamefont {Quartin}, \citenamefont {Amendola},
  \citenamefont {Castro},\ and\ \citenamefont {Toubiana}}]{Alfradique:2022tox}%
  \BibitemOpen
  \bibfield  {author} {\bibinfo {author} {\bibfnamefont {V.}~\bibnamefont
  {Alfradique}}, \bibinfo {author} {\bibfnamefont {M.}~\bibnamefont {Quartin}},
  \bibinfo {author} {\bibfnamefont {L.}~\bibnamefont {Amendola}}, \bibinfo
  {author} {\bibfnamefont {T.}~\bibnamefont {Castro}}, \ and\ \bibinfo {author}
  {\bibfnamefont {A.}~\bibnamefont {Toubiana}},\ }\href {\doibase
  10.1093/mnras/stac2920} {\bibfield  {journal} {\bibinfo  {journal} {Mon. Not.
  Roy. Astron. Soc.}\ }\textbf {\bibinfo {volume} {517}},\ \bibinfo {pages}
  {5449} (\bibinfo {year} {2022})},\ \Eprint {http://arxiv.org/abs/2205.14034}
  {arXiv:2205.14034 [astro-ph.CO]} \BibitemShut {NoStop}%
\bibitem [{\citenamefont {Gupta}(2022)}]{Gupta:2022fwd}%
  \BibitemOpen
  \bibfield  {author} {\bibinfo {author} {\bibfnamefont {I.}~\bibnamefont
  {Gupta}},\ }\href@noop {} {\  (\bibinfo {year} {2022})},\ \Eprint
  {http://arxiv.org/abs/2212.00163} {arXiv:2212.00163 [gr-qc]} \BibitemShut
  {NoStop}%
\bibitem [{\citenamefont {Sathyaprakash}\ \emph {et~al.}(2010)\citenamefont
  {Sathyaprakash}, \citenamefont {Schutz},\ and\ \citenamefont {Van
  Den~Broeck}}]{Sathyaprakash:2009xt}%
  \BibitemOpen
  \bibfield  {author} {\bibinfo {author} {\bibfnamefont {B.~S.}\ \bibnamefont
  {Sathyaprakash}}, \bibinfo {author} {\bibfnamefont {B.~F.}\ \bibnamefont
  {Schutz}}, \ and\ \bibinfo {author} {\bibfnamefont {C.}~\bibnamefont {Van
  Den~Broeck}},\ }\href {\doibase 10.1088/0264-9381/27/21/215006} {\bibfield
  {journal} {\bibinfo  {journal} {Class. Quant. Grav.}\ }\textbf {\bibinfo
  {volume} {27}},\ \bibinfo {pages} {215006} (\bibinfo {year} {2010})},\
  \Eprint {http://arxiv.org/abs/0906.4151} {arXiv:0906.4151 [astro-ph.CO]}
  \BibitemShut {NoStop}%
\bibitem [{\citenamefont {Zhao}\ \emph {et~al.}(2011)\citenamefont {Zhao},
  \citenamefont {Van Den~Broeck}, \citenamefont {Baskaran},\ and\ \citenamefont
  {Li}}]{Zhao:2010sz}%
  \BibitemOpen
  \bibfield  {author} {\bibinfo {author} {\bibfnamefont {W.}~\bibnamefont
  {Zhao}}, \bibinfo {author} {\bibfnamefont {C.}~\bibnamefont {Van
  Den~Broeck}}, \bibinfo {author} {\bibfnamefont {D.}~\bibnamefont {Baskaran}},
  \ and\ \bibinfo {author} {\bibfnamefont {T.~G.~F.}\ \bibnamefont {Li}},\
  }\href {\doibase 10.1103/PhysRevD.83.023005} {\bibfield  {journal} {\bibinfo
  {journal} {Phys. Rev. D}\ }\textbf {\bibinfo {volume} {83}},\ \bibinfo
  {pages} {023005} (\bibinfo {year} {2011})},\ \Eprint
  {http://arxiv.org/abs/1009.0206} {arXiv:1009.0206 [astro-ph.CO]} \BibitemShut
  {NoStop}%
\bibitem [{\citenamefont {Taylor}\ \emph
  {et~al.}(2012{\natexlab{b}})\citenamefont {Taylor}, \citenamefont {Gair},\
  and\ \citenamefont {Mandel}}]{Taylor:2011fs}%
  \BibitemOpen
  \bibfield  {author} {\bibinfo {author} {\bibfnamefont {S.~R.}\ \bibnamefont
  {Taylor}}, \bibinfo {author} {\bibfnamefont {J.~R.}\ \bibnamefont {Gair}}, \
  and\ \bibinfo {author} {\bibfnamefont {I.}~\bibnamefont {Mandel}},\ }\href
  {\doibase 10.1103/PhysRevD.85.023535} {\bibfield  {journal} {\bibinfo
  {journal} {Phys. Rev. D}\ }\textbf {\bibinfo {volume} {85}},\ \bibinfo
  {pages} {023535} (\bibinfo {year} {2012}{\natexlab{b}})},\ \Eprint
  {http://arxiv.org/abs/1108.5161} {arXiv:1108.5161 [gr-qc]} \BibitemShut
  {NoStop}%
\bibitem [{\citenamefont {Taylor}\ and\ \citenamefont
  {Gair}(2012)}]{Taylor:2012db}%
  \BibitemOpen
  \bibfield  {author} {\bibinfo {author} {\bibfnamefont {S.~R.}\ \bibnamefont
  {Taylor}}\ and\ \bibinfo {author} {\bibfnamefont {J.~R.}\ \bibnamefont
  {Gair}},\ }\href {\doibase 10.1103/PhysRevD.86.023502} {\bibfield  {journal}
  {\bibinfo  {journal} {Phys. Rev. D}\ }\textbf {\bibinfo {volume} {86}},\
  \bibinfo {pages} {023502} (\bibinfo {year} {2012})},\ \Eprint
  {http://arxiv.org/abs/1204.6739} {arXiv:1204.6739 [astro-ph.CO]} \BibitemShut
  {NoStop}%
\bibitem [{\citenamefont {Leandro}\ \emph
  {et~al.}(2022{\natexlab{b}})\citenamefont {Leandro}, \citenamefont {Marra},\
  and\ \citenamefont {Sturani}}]{Leandro:2021qlc}%
  \BibitemOpen
  \bibfield  {author} {\bibinfo {author} {\bibfnamefont {H.}~\bibnamefont
  {Leandro}}, \bibinfo {author} {\bibfnamefont {V.}~\bibnamefont {Marra}}, \
  and\ \bibinfo {author} {\bibfnamefont {R.}~\bibnamefont {Sturani}},\ }\href
  {\doibase 10.1103/PhysRevD.105.023523} {\bibfield  {journal} {\bibinfo
  {journal} {Phys. Rev. D}\ }\textbf {\bibinfo {volume} {105}},\ \bibinfo
  {pages} {023523} (\bibinfo {year} {2022}{\natexlab{b}})},\ \Eprint
  {http://arxiv.org/abs/2109.07537} {arXiv:2109.07537 [gr-qc]} \BibitemShut
  {NoStop}%
\bibitem [{\citenamefont {Ye}\ and\ \citenamefont
  {Fishbach}(2021)}]{Ye:2021klk}%
  \BibitemOpen
  \bibfield  {author} {\bibinfo {author} {\bibfnamefont {C.}~\bibnamefont
  {Ye}}\ and\ \bibinfo {author} {\bibfnamefont {M.}~\bibnamefont {Fishbach}},\
  }\href {\doibase 10.1103/PhysRevD.104.043507} {\bibfield  {journal} {\bibinfo
   {journal} {Phys. Rev. D}\ }\textbf {\bibinfo {volume} {104}},\ \bibinfo
  {pages} {043507} (\bibinfo {year} {2021})},\ \Eprint
  {http://arxiv.org/abs/2103.14038} {arXiv:2103.14038 [astro-ph.CO]}
  \BibitemShut {NoStop}%
\bibitem [{\citenamefont {Messenger}\ and\ \citenamefont
  {Read}(2012)}]{Messenger:2011gi}%
  \BibitemOpen
  \bibfield  {author} {\bibinfo {author} {\bibfnamefont {C.}~\bibnamefont
  {Messenger}}\ and\ \bibinfo {author} {\bibfnamefont {J.}~\bibnamefont
  {Read}},\ }\href {\doibase 10.1103/PhysRevLett.108.091101} {\bibfield
  {journal} {\bibinfo  {journal} {Phys. Rev. Lett.}\ }\textbf {\bibinfo
  {volume} {108}},\ \bibinfo {pages} {091101} (\bibinfo {year} {2012})},\
  \Eprint {http://arxiv.org/abs/1107.5725} {arXiv:1107.5725 [gr-qc]}
  \BibitemShut {NoStop}%
\bibitem [{\citenamefont {{Ghosh}}\ \emph {et~al.}(2022)\citenamefont
  {{Ghosh}}, \citenamefont {{Biswas}},\ and\ \citenamefont
  {{Bose}}}]{2022PhRvD.106l3529G}%
  \BibitemOpen
  \bibfield  {author} {\bibinfo {author} {\bibfnamefont {T.}~\bibnamefont
  {{Ghosh}}}, \bibinfo {author} {\bibfnamefont {B.}~\bibnamefont {{Biswas}}}, \
  and\ \bibinfo {author} {\bibfnamefont {S.}~\bibnamefont {{Bose}}},\ }\href
  {\doibase 10.1103/PhysRevD.106.123529} {\bibfield  {journal} {\bibinfo
  {journal} {\prd}\ }\textbf {\bibinfo {volume} {106}},\ \bibinfo {eid}
  {123529} (\bibinfo {year} {2022})},\ \Eprint
  {http://arxiv.org/abs/2203.11756} {arXiv:2203.11756 [astro-ph.CO]}
  \BibitemShut {NoStop}%
\bibitem [{\citenamefont {{Chatterjee}}\ \emph {et~al.}(2021)\citenamefont
  {{Chatterjee}}, \citenamefont {{Hegade K.~R.}}, \citenamefont {{Holder}},
  \citenamefont {{Holz}}, \citenamefont {{Perkins}}, \citenamefont {{Yagi}},\
  and\ \citenamefont {{Yunes}}}]{2021PhRvD.104h3528C}%
  \BibitemOpen
  \bibfield  {author} {\bibinfo {author} {\bibfnamefont {D.}~\bibnamefont
  {{Chatterjee}}}, \bibinfo {author} {\bibfnamefont {A.}~\bibnamefont {{Hegade
  K.~R.}}}, \bibinfo {author} {\bibfnamefont {G.}~\bibnamefont {{Holder}}},
  \bibinfo {author} {\bibfnamefont {D.~E.}\ \bibnamefont {{Holz}}}, \bibinfo
  {author} {\bibfnamefont {S.}~\bibnamefont {{Perkins}}}, \bibinfo {author}
  {\bibfnamefont {K.}~\bibnamefont {{Yagi}}}, \ and\ \bibinfo {author}
  {\bibfnamefont {N.}~\bibnamefont {{Yunes}}},\ }\href {\doibase
  10.1103/PhysRevD.104.083528} {\bibfield  {journal} {\bibinfo  {journal}
  {\prd}\ }\textbf {\bibinfo {volume} {104}},\ \bibinfo {eid} {083528}
  (\bibinfo {year} {2021})},\ \Eprint {http://arxiv.org/abs/2106.06589}
  {arXiv:2106.06589 [gr-qc]} \BibitemShut {NoStop}%
\bibitem [{\citenamefont {Jin}\ \emph {et~al.}(2022)\citenamefont {Jin},
  \citenamefont {Li}, \citenamefont {Zhang},\ and\ \citenamefont
  {Zhang}}]{Jin:2022qnj}%
  \BibitemOpen
  \bibfield  {author} {\bibinfo {author} {\bibfnamefont {S.-J.}\ \bibnamefont
  {Jin}}, \bibinfo {author} {\bibfnamefont {T.-N.}\ \bibnamefont {Li}},
  \bibinfo {author} {\bibfnamefont {J.-F.}\ \bibnamefont {Zhang}}, \ and\
  \bibinfo {author} {\bibfnamefont {X.}~\bibnamefont {Zhang}},\ }\href@noop {}
  {\  (\bibinfo {year} {2022})},\ \Eprint {http://arxiv.org/abs/2202.11882}
  {arXiv:2202.11882 [gr-qc]} \BibitemShut {NoStop}%
\bibitem [{\citenamefont {Yu}\ \emph {et~al.}(2020)\citenamefont {Yu},
  \citenamefont {Wang}, \citenamefont {Zhao},\ and\ \citenamefont
  {Lu}}]{Yu:2020vyy}%
  \BibitemOpen
  \bibfield  {author} {\bibinfo {author} {\bibfnamefont {J.}~\bibnamefont
  {Yu}}, \bibinfo {author} {\bibfnamefont {Y.}~\bibnamefont {Wang}}, \bibinfo
  {author} {\bibfnamefont {W.}~\bibnamefont {Zhao}}, \ and\ \bibinfo {author}
  {\bibfnamefont {Y.}~\bibnamefont {Lu}},\ }\href {\doibase
  10.1093/mnras/staa2465} {\bibfield  {journal} {\bibinfo  {journal} {Mon. Not.
  Roy. Astron. Soc.}\ }\textbf {\bibinfo {volume} {498}},\ \bibinfo {pages}
  {1786} (\bibinfo {year} {2020})},\ \Eprint {http://arxiv.org/abs/2003.06586}
  {arXiv:2003.06586 [astro-ph.CO]} \BibitemShut {NoStop}%
\bibitem [{\citenamefont {Song}\ \emph {et~al.}(2022)\citenamefont {Song},
  \citenamefont {Wang}, \citenamefont {Li}, \citenamefont {Zhao}, \citenamefont
  {Zhang}, \citenamefont {Zhao},\ and\ \citenamefont {Zhang}}]{Song:2022siz}%
  \BibitemOpen
  \bibfield  {author} {\bibinfo {author} {\bibfnamefont {J.-Y.}\ \bibnamefont
  {Song}}, \bibinfo {author} {\bibfnamefont {L.-F.}\ \bibnamefont {Wang}},
  \bibinfo {author} {\bibfnamefont {Y.}~\bibnamefont {Li}}, \bibinfo {author}
  {\bibfnamefont {Z.-W.}\ \bibnamefont {Zhao}}, \bibinfo {author}
  {\bibfnamefont {J.-F.}\ \bibnamefont {Zhang}}, \bibinfo {author}
  {\bibfnamefont {W.}~\bibnamefont {Zhao}}, \ and\ \bibinfo {author}
  {\bibfnamefont {X.}~\bibnamefont {Zhang}},\ }\href@noop {} {\  (\bibinfo
  {year} {2022})},\ \Eprint {http://arxiv.org/abs/2212.00531} {arXiv:2212.00531
  [astro-ph.CO]} \BibitemShut {NoStop}%
\bibitem [{\citenamefont {Borhanian}\ \emph {et~al.}(2020)\citenamefont
  {Borhanian}, \citenamefont {Dhani}, \citenamefont {Gupta}, \citenamefont
  {Arun},\ and\ \citenamefont {Sathyaprakash}}]{Borhanian:2020vyr}%
  \BibitemOpen
  \bibfield  {author} {\bibinfo {author} {\bibfnamefont {S.}~\bibnamefont
  {Borhanian}}, \bibinfo {author} {\bibfnamefont {A.}~\bibnamefont {Dhani}},
  \bibinfo {author} {\bibfnamefont {A.}~\bibnamefont {Gupta}}, \bibinfo
  {author} {\bibfnamefont {K.~G.}\ \bibnamefont {Arun}}, \ and\ \bibinfo
  {author} {\bibfnamefont {B.~S.}\ \bibnamefont {Sathyaprakash}},\ }\href
  {\doibase 10.3847/2041-8213/abcaf5} {\bibfield  {journal} {\bibinfo
  {journal} {Astrophys. J. Lett.}\ }\textbf {\bibinfo {volume} {905}},\
  \bibinfo {pages} {L28} (\bibinfo {year} {2020})},\ \Eprint
  {http://arxiv.org/abs/2007.02883} {arXiv:2007.02883 [astro-ph.CO]}
  \BibitemShut {NoStop}%
\bibitem [{\citenamefont {Zhu}\ and\ \citenamefont {Chen}(2023)}]{Zhu:2023jti}%
  \BibitemOpen
  \bibfield  {author} {\bibinfo {author} {\bibfnamefont {L.-G.}\ \bibnamefont
  {Zhu}}\ and\ \bibinfo {author} {\bibfnamefont {X.}~\bibnamefont {Chen}},\
  }\href@noop {} {\  (\bibinfo {year} {2023})},\ \Eprint
  {http://arxiv.org/abs/2302.10621} {arXiv:2302.10621 [astro-ph.CO]}
  \BibitemShut {NoStop}%
\bibitem [{\citenamefont {Auclair}\ \emph {et~al.}(2022)\citenamefont {Auclair}
  \emph {et~al.}}]{LISACosmologyWorkingGroup:2022jok}%
  \BibitemOpen
  \bibfield  {author} {\bibinfo {author} {\bibfnamefont {P.}~\bibnamefont
  {Auclair}} \emph {et~al.} (\bibinfo {collaboration} {LISA Cosmology Working
  Group}),\ }\href@noop {} {\  (\bibinfo {year} {2022})},\ \Eprint
  {http://arxiv.org/abs/2204.05434} {arXiv:2204.05434 [astro-ph.CO]}
  \BibitemShut {NoStop}%
\bibitem [{\citenamefont {Tamanini}\ \emph {et~al.}(2016)\citenamefont
  {Tamanini}, \citenamefont {Caprini}, \citenamefont {Barausse}, \citenamefont
  {Sesana}, \citenamefont {Klein},\ and\ \citenamefont
  {Petiteau}}]{Tamanini:2016zlh}%
  \BibitemOpen
  \bibfield  {author} {\bibinfo {author} {\bibfnamefont {N.}~\bibnamefont
  {Tamanini}}, \bibinfo {author} {\bibfnamefont {C.}~\bibnamefont {Caprini}},
  \bibinfo {author} {\bibfnamefont {E.}~\bibnamefont {Barausse}}, \bibinfo
  {author} {\bibfnamefont {A.}~\bibnamefont {Sesana}}, \bibinfo {author}
  {\bibfnamefont {A.}~\bibnamefont {Klein}}, \ and\ \bibinfo {author}
  {\bibfnamefont {A.}~\bibnamefont {Petiteau}},\ }\href {\doibase
  10.1088/1475-7516/2016/04/002} {\bibfield  {journal} {\bibinfo  {journal}
  {JCAP}\ }\textbf {\bibinfo {volume} {04}},\ \bibinfo {pages} {002} (\bibinfo
  {year} {2016})},\ \Eprint {http://arxiv.org/abs/1601.07112} {arXiv:1601.07112
  [astro-ph.CO]} \BibitemShut {NoStop}%
\bibitem [{\citenamefont {Caprini}\ and\ \citenamefont
  {Tamanini}(2016)}]{Caprini:2016qxs}%
  \BibitemOpen
  \bibfield  {author} {\bibinfo {author} {\bibfnamefont {C.}~\bibnamefont
  {Caprini}}\ and\ \bibinfo {author} {\bibfnamefont {N.}~\bibnamefont
  {Tamanini}},\ }\href {\doibase 10.1088/1475-7516/2016/10/006} {\bibfield
  {journal} {\bibinfo  {journal} {JCAP}\ }\textbf {\bibinfo {volume} {10}},\
  \bibinfo {pages} {006} (\bibinfo {year} {2016})},\ \Eprint
  {http://arxiv.org/abs/1607.08755} {arXiv:1607.08755 [astro-ph.CO]}
  \BibitemShut {NoStop}%
\bibitem [{\citenamefont {Cai}\ \emph {et~al.}(2017)\citenamefont {Cai},
  \citenamefont {Tamanini},\ and\ \citenamefont {Yang}}]{Cai:2017yww}%
  \BibitemOpen
  \bibfield  {author} {\bibinfo {author} {\bibfnamefont {R.-G.}\ \bibnamefont
  {Cai}}, \bibinfo {author} {\bibfnamefont {N.}~\bibnamefont {Tamanini}}, \
  and\ \bibinfo {author} {\bibfnamefont {T.}~\bibnamefont {Yang}},\ }\href
  {\doibase 10.1088/1475-7516/2017/05/031} {\bibfield  {journal} {\bibinfo
  {journal} {JCAP}\ }\textbf {\bibinfo {volume} {05}},\ \bibinfo {pages} {031}
  (\bibinfo {year} {2017})},\ \Eprint {http://arxiv.org/abs/1703.07323}
  {arXiv:1703.07323 [astro-ph.CO]} \BibitemShut {NoStop}%
\bibitem [{\citenamefont {Del~Pozzo}\ \emph
  {et~al.}(2018{\natexlab{a}})\citenamefont {Del~Pozzo}, \citenamefont
  {Sesana},\ and\ \citenamefont {Klein}}]{DelPozzo:2017kme}%
  \BibitemOpen
  \bibfield  {author} {\bibinfo {author} {\bibfnamefont {W.}~\bibnamefont
  {Del~Pozzo}}, \bibinfo {author} {\bibfnamefont {A.}~\bibnamefont {Sesana}}, \
  and\ \bibinfo {author} {\bibfnamefont {A.}~\bibnamefont {Klein}},\ }\href
  {\doibase 10.1093/mnras/sty057} {\bibfield  {journal} {\bibinfo  {journal}
  {Mon. Not. Roy. Astron. Soc.}\ }\textbf {\bibinfo {volume} {475}},\ \bibinfo
  {pages} {3485} (\bibinfo {year} {2018}{\natexlab{a}})},\ \Eprint
  {http://arxiv.org/abs/1703.01300} {arXiv:1703.01300 [astro-ph.CO]}
  \BibitemShut {NoStop}%
\bibitem [{\citenamefont {Belgacem}\ \emph
  {et~al.}(2019{\natexlab{b}})\citenamefont {Belgacem} \emph
  {et~al.}}]{LISACosmologyWorkingGroup:2019mwx}%
  \BibitemOpen
  \bibfield  {author} {\bibinfo {author} {\bibfnamefont {E.}~\bibnamefont
  {Belgacem}} \emph {et~al.} (\bibinfo {collaboration} {LISA Cosmology Working
  Group}),\ }\href {\doibase 10.1088/1475-7516/2019/07/024} {\bibfield
  {journal} {\bibinfo  {journal} {JCAP}\ }\textbf {\bibinfo {volume} {07}},\
  \bibinfo {pages} {024} (\bibinfo {year} {2019}{\natexlab{b}})},\ \Eprint
  {http://arxiv.org/abs/1906.01593} {arXiv:1906.01593 [astro-ph.CO]}
  \BibitemShut {NoStop}%
\bibitem [{\citenamefont {Speri}\ \emph {et~al.}(2021)\citenamefont {Speri},
  \citenamefont {Tamanini}, \citenamefont {Caldwell}, \citenamefont {Gair},\
  and\ \citenamefont {Wang}}]{Speri:2020hwc}%
  \BibitemOpen
  \bibfield  {author} {\bibinfo {author} {\bibfnamefont {L.}~\bibnamefont
  {Speri}}, \bibinfo {author} {\bibfnamefont {N.}~\bibnamefont {Tamanini}},
  \bibinfo {author} {\bibfnamefont {R.~R.}\ \bibnamefont {Caldwell}}, \bibinfo
  {author} {\bibfnamefont {J.~R.}\ \bibnamefont {Gair}}, \ and\ \bibinfo
  {author} {\bibfnamefont {B.}~\bibnamefont {Wang}},\ }\href {\doibase
  10.1103/PhysRevD.103.083526} {\bibfield  {journal} {\bibinfo  {journal}
  {Phys. Rev. D}\ }\textbf {\bibinfo {volume} {103}},\ \bibinfo {pages}
  {083526} (\bibinfo {year} {2021})},\ \Eprint
  {http://arxiv.org/abs/2010.09049} {arXiv:2010.09049 [astro-ph.CO]}
  \BibitemShut {NoStop}%
\bibitem [{\citenamefont {{Laghi}}\ \emph {et~al.}(2021)\citenamefont
  {{Laghi}}, \citenamefont {{Tamanini}}, \citenamefont {{Del Pozzo}},
  \citenamefont {{Sesana}}, \citenamefont {{Gair}}, \citenamefont {{Babak}},\
  and\ \citenamefont {{Izquierdo-Villalba}}}]{2021MNRAS.508.4512L}%
  \BibitemOpen
  \bibfield  {author} {\bibinfo {author} {\bibfnamefont {D.}~\bibnamefont
  {{Laghi}}}, \bibinfo {author} {\bibfnamefont {N.}~\bibnamefont {{Tamanini}}},
  \bibinfo {author} {\bibfnamefont {W.}~\bibnamefont {{Del Pozzo}}}, \bibinfo
  {author} {\bibfnamefont {A.}~\bibnamefont {{Sesana}}}, \bibinfo {author}
  {\bibfnamefont {J.}~\bibnamefont {{Gair}}}, \bibinfo {author} {\bibfnamefont
  {S.}~\bibnamefont {{Babak}}}, \ and\ \bibinfo {author} {\bibfnamefont
  {D.}~\bibnamefont {{Izquierdo-Villalba}}},\ }\href {\doibase
  10.1093/mnras/stab2741} {\bibfield  {journal} {\bibinfo  {journal} {\mnras}\
  }\textbf {\bibinfo {volume} {508}},\ \bibinfo {pages} {4512} (\bibinfo {year}
  {2021})},\ \Eprint {http://arxiv.org/abs/2102.01708} {arXiv:2102.01708
  [astro-ph.CO]} \BibitemShut {NoStop}%
\bibitem [{\citenamefont {Muttoni}\ \emph {et~al.}(2022)\citenamefont
  {Muttoni}, \citenamefont {Mangiagli}, \citenamefont {Sesana}, \citenamefont
  {Laghi}, \citenamefont {Del~Pozzo}, \citenamefont {Izquierdo-Villalba},\ and\
  \citenamefont {Rosati}}]{Muttoni:2021veo}%
  \BibitemOpen
  \bibfield  {author} {\bibinfo {author} {\bibfnamefont {N.}~\bibnamefont
  {Muttoni}}, \bibinfo {author} {\bibfnamefont {A.}~\bibnamefont {Mangiagli}},
  \bibinfo {author} {\bibfnamefont {A.}~\bibnamefont {Sesana}}, \bibinfo
  {author} {\bibfnamefont {D.}~\bibnamefont {Laghi}}, \bibinfo {author}
  {\bibfnamefont {W.}~\bibnamefont {Del~Pozzo}}, \bibinfo {author}
  {\bibfnamefont {D.}~\bibnamefont {Izquierdo-Villalba}}, \ and\ \bibinfo
  {author} {\bibfnamefont {M.}~\bibnamefont {Rosati}},\ }\href {\doibase
  10.1103/PhysRevD.105.043509} {\bibfield  {journal} {\bibinfo  {journal}
  {Phys. Rev. D}\ }\textbf {\bibinfo {volume} {105}},\ \bibinfo {pages}
  {043509} (\bibinfo {year} {2022})},\ \Eprint
  {http://arxiv.org/abs/2109.13934} {arXiv:2109.13934 [astro-ph.CO]}
  \BibitemShut {NoStop}%
\bibitem [{\citenamefont {Yang}(2021)}]{Yang:2021qge}%
  \BibitemOpen
  \bibfield  {author} {\bibinfo {author} {\bibfnamefont {T.}~\bibnamefont
  {Yang}},\ }\href {\doibase 10.1088/1475-7516/2021/05/044} {\bibfield
  {journal} {\bibinfo  {journal} {JCAP}\ }\textbf {\bibinfo {volume} {05}},\
  \bibinfo {pages} {044} (\bibinfo {year} {2021})},\ \Eprint
  {http://arxiv.org/abs/2103.01923} {arXiv:2103.01923 [astro-ph.CO]}
  \BibitemShut {NoStop}%
\bibitem [{\citenamefont {Abbott}\ \emph {et~al.}(2019)\citenamefont {Abbott},
  \citenamefont {Abbott}, \citenamefont {Abbott}, \citenamefont {Abraham},
  \citenamefont {Acernese}, \citenamefont {Ackley}, \citenamefont {Adams},
  \citenamefont {Adhikari}, \citenamefont {Adya}, \citenamefont {Affeldt} \emph
  {et~al.}}]{abbott2019gwtc1}%
  \BibitemOpen
  \bibfield  {author} {\bibinfo {author} {\bibfnamefont {B.}~\bibnamefont
  {Abbott}}, \bibinfo {author} {\bibfnamefont {R.}~\bibnamefont {Abbott}},
  \bibinfo {author} {\bibfnamefont {T.}~\bibnamefont {Abbott}}, \bibinfo
  {author} {\bibfnamefont {S.}~\bibnamefont {Abraham}}, \bibinfo {author}
  {\bibfnamefont {F.}~\bibnamefont {Acernese}}, \bibinfo {author}
  {\bibfnamefont {K.}~\bibnamefont {Ackley}}, \bibinfo {author} {\bibfnamefont
  {C.}~\bibnamefont {Adams}}, \bibinfo {author} {\bibfnamefont
  {R.}~\bibnamefont {Adhikari}}, \bibinfo {author} {\bibfnamefont
  {V.}~\bibnamefont {Adya}}, \bibinfo {author} {\bibfnamefont {C.}~\bibnamefont
  {Affeldt}},  \emph {et~al.},\ }\href
  {https://journals.aps.org/prx/abstract/10.1103/PhysRevX.9.031040} {\bibfield
  {journal} {\bibinfo  {journal} {Physical Review X}\ }\textbf {\bibinfo
  {volume} {9}},\ \bibinfo {pages} {031040} (\bibinfo {year}
  {2019})}\BibitemShut {NoStop}%
\bibitem [{\citenamefont {Abbott}\ \emph
  {et~al.}(2021{\natexlab{c}})\citenamefont {Abbott}, \citenamefont {Abbott},
  \citenamefont {Abraham}, \citenamefont {Acernese}, \citenamefont {Ackley},
  \citenamefont {Adams}, \citenamefont {Adams}, \citenamefont {Adhikari},
  \citenamefont {Adya}, \citenamefont {Affeldt} \emph
  {et~al.}}]{abbott2021gwtc2}%
  \BibitemOpen
  \bibfield  {author} {\bibinfo {author} {\bibfnamefont {R.}~\bibnamefont
  {Abbott}}, \bibinfo {author} {\bibfnamefont {T.}~\bibnamefont {Abbott}},
  \bibinfo {author} {\bibfnamefont {S.}~\bibnamefont {Abraham}}, \bibinfo
  {author} {\bibfnamefont {F.}~\bibnamefont {Acernese}}, \bibinfo {author}
  {\bibfnamefont {K.}~\bibnamefont {Ackley}}, \bibinfo {author} {\bibfnamefont
  {A.}~\bibnamefont {Adams}}, \bibinfo {author} {\bibfnamefont
  {C.}~\bibnamefont {Adams}}, \bibinfo {author} {\bibfnamefont
  {R.}~\bibnamefont {Adhikari}}, \bibinfo {author} {\bibfnamefont
  {V.}~\bibnamefont {Adya}}, \bibinfo {author} {\bibfnamefont {C.}~\bibnamefont
  {Affeldt}},  \emph {et~al.},\ }\href {https://arxiv.org/abs/2010.14527}
  {\bibfield  {journal} {\bibinfo  {journal} {Physical Review X}\ }\textbf
  {\bibinfo {volume} {11}},\ \bibinfo {pages} {021053} (\bibinfo {year}
  {2021}{\natexlab{c}})}\BibitemShut {NoStop}%
\bibitem [{\citenamefont {Abbott}\ \emph
  {et~al.}(2021{\natexlab{d}})\citenamefont {Abbott}, \citenamefont {Abbott},
  \citenamefont {Acernese}, \citenamefont {Ackley}, \citenamefont {Adams},
  \citenamefont {Adhikari}, \citenamefont {Adhikari}, \citenamefont {Adya},
  \citenamefont {Affeldt}, \citenamefont {Agarwal} \emph
  {et~al.}}]{abbott2021gwtc3}%
  \BibitemOpen
  \bibfield  {author} {\bibinfo {author} {\bibfnamefont {R.}~\bibnamefont
  {Abbott}}, \bibinfo {author} {\bibfnamefont {T.}~\bibnamefont {Abbott}},
  \bibinfo {author} {\bibfnamefont {F.}~\bibnamefont {Acernese}}, \bibinfo
  {author} {\bibfnamefont {K.}~\bibnamefont {Ackley}}, \bibinfo {author}
  {\bibfnamefont {C.}~\bibnamefont {Adams}}, \bibinfo {author} {\bibfnamefont
  {N.}~\bibnamefont {Adhikari}}, \bibinfo {author} {\bibfnamefont
  {R.}~\bibnamefont {Adhikari}}, \bibinfo {author} {\bibfnamefont
  {V.}~\bibnamefont {Adya}}, \bibinfo {author} {\bibfnamefont {C.}~\bibnamefont
  {Affeldt}}, \bibinfo {author} {\bibfnamefont {D.}~\bibnamefont {Agarwal}},
  \emph {et~al.},\ }\href {https://arxiv.org/abs/2111.03606} {\bibfield
  {journal} {\bibinfo  {journal} {arXiv preprint arXiv:2111.03606}\ } (\bibinfo
  {year} {2021}{\natexlab{d}})}\BibitemShut {NoStop}%
\bibitem [{\citenamefont {Collaboration}\ \emph {et~al.}(2021)\citenamefont
  {Collaboration}, \citenamefont {Collaboration}, \citenamefont {Collaboration}
  \emph {et~al.}}]{ligo2021population}%
  \BibitemOpen
  \bibfield  {author} {\bibinfo {author} {\bibfnamefont {L.~S.}\ \bibnamefont
  {Collaboration}}, \bibinfo {author} {\bibfnamefont {V.}~\bibnamefont
  {Collaboration}}, \bibinfo {author} {\bibfnamefont {K.~S.}\ \bibnamefont
  {Collaboration}},  \emph {et~al.},\ }\href {https://arxiv.org/abs/2111.03634}
  {\bibfield  {journal} {\bibinfo  {journal} {arXiv preprint arXiv:2111.03634}\
  } (\bibinfo {year} {2021})}\BibitemShut {NoStop}%
\bibitem [{\citenamefont {Borhanian}\ and\ \citenamefont
  {Sathyaprakash}(2022)}]{borhanian2022listening}%
  \BibitemOpen
  \bibfield  {author} {\bibinfo {author} {\bibfnamefont {S.}~\bibnamefont
  {Borhanian}}\ and\ \bibinfo {author} {\bibfnamefont {B.}~\bibnamefont
  {Sathyaprakash}},\ }\href {\doibase 10.48550/arXiv.2202.11048} {\  (\bibinfo
  {year} {2022}),\ 10.48550/arXiv.2202.11048}\BibitemShut {NoStop}%
\bibitem [{\citenamefont {Madau}\ and\ \citenamefont
  {Fragos}(2017)}]{madau2017radiation}%
  \BibitemOpen
  \bibfield  {author} {\bibinfo {author} {\bibfnamefont {P.}~\bibnamefont
  {Madau}}\ and\ \bibinfo {author} {\bibfnamefont {T.}~\bibnamefont {Fragos}},\
  }\href {\doibase 10.3847/1538-4357/aa6af9} {\bibfield  {journal} {\bibinfo
  {journal} {The Astrophysical Journal}\ }\textbf {\bibinfo {volume} {840}},\
  \bibinfo {pages} {39} (\bibinfo {year} {2017})}\BibitemShut {NoStop}%
\bibitem [{\citenamefont {Iacovelli}\ \emph
  {et~al.}(2022{\natexlab{a}})\citenamefont {Iacovelli}, \citenamefont
  {Mancarella}, \citenamefont {Foffa},\ and\ \citenamefont
  {Maggiore}}]{iacovelli2022forecasting}%
  \BibitemOpen
  \bibfield  {author} {\bibinfo {author} {\bibfnamefont {F.}~\bibnamefont
  {Iacovelli}}, \bibinfo {author} {\bibfnamefont {M.}~\bibnamefont
  {Mancarella}}, \bibinfo {author} {\bibfnamefont {S.}~\bibnamefont {Foffa}}, \
  and\ \bibinfo {author} {\bibfnamefont {M.}~\bibnamefont {Maggiore}},\ }\href
  {\doibase 10.3847/1538-4357/ac9cd4} {\bibfield  {journal} {\bibinfo
  {journal} {The Astrophysical Journal}\ }\textbf {\bibinfo {volume} {941}},\
  \bibinfo {pages} {208} (\bibinfo {year} {2022}{\natexlab{a}})}\BibitemShut
  {NoStop}%
\bibitem [{\citenamefont {{Planck Collaboration}}\ \emph
  {et~al.}(2014)\citenamefont {{Planck Collaboration}}, \citenamefont {{Ade}},
  \citenamefont {{Aghanim}}, \citenamefont {{Armitage-Caplan}}, \citenamefont
  {{Arnaud}}, \citenamefont {{Ashdown}}, \citenamefont {{Atrio-Barandela}},
  \citenamefont {{Aumont}}, \citenamefont {{Baccigalupi}}, \citenamefont
  {{Banday}},\ and\ \citenamefont {et~al.}}]{PlanckCollaboration2014}%
  \BibitemOpen
  \bibfield  {author} {\bibinfo {author} {\bibnamefont {{Planck
  Collaboration}}}, \bibinfo {author} {\bibfnamefont {P.~A.~R.}\ \bibnamefont
  {{Ade}}}, \bibinfo {author} {\bibfnamefont {N.}~\bibnamefont {{Aghanim}}},
  \bibinfo {author} {\bibfnamefont {C.}~\bibnamefont {{Armitage-Caplan}}},
  \bibinfo {author} {\bibfnamefont {M.}~\bibnamefont {{Arnaud}}}, \bibinfo
  {author} {\bibfnamefont {M.}~\bibnamefont {{Ashdown}}}, \bibinfo {author}
  {\bibfnamefont {F.}~\bibnamefont {{Atrio-Barandela}}}, \bibinfo {author}
  {\bibfnamefont {J.}~\bibnamefont {{Aumont}}}, \bibinfo {author}
  {\bibfnamefont {C.}~\bibnamefont {{Baccigalupi}}}, \bibinfo {author}
  {\bibfnamefont {A.~J.}\ \bibnamefont {{Banday}}}, \ and\ \bibinfo {author}
  {\bibnamefont {et~al.}},\ }\href {\doibase 10.1051/0004-6361/201321591}
  {\bibfield  {journal} {\bibinfo  {journal} {\aap}\ }\textbf {\bibinfo
  {volume} {571}},\ \bibinfo {eid} {A16} (\bibinfo {year} {2014})},\ \Eprint
  {http://arxiv.org/abs/1303.5076} {arXiv:1303.5076} \BibitemShut {NoStop}%
\bibitem [{\citenamefont {Samajdar}\ \emph {et~al.}(2021)\citenamefont
  {Samajdar}, \citenamefont {Janquart}, \citenamefont {Van Den~Broeck},\ and\
  \citenamefont {Dietrich}}]{samajdar2021biases}%
  \BibitemOpen
  \bibfield  {author} {\bibinfo {author} {\bibfnamefont {A.}~\bibnamefont
  {Samajdar}}, \bibinfo {author} {\bibfnamefont {J.}~\bibnamefont {Janquart}},
  \bibinfo {author} {\bibfnamefont {C.}~\bibnamefont {Van Den~Broeck}}, \ and\
  \bibinfo {author} {\bibfnamefont {T.}~\bibnamefont {Dietrich}},\ }\href
  {\doibase 10.1103/PhysRevD.104.044003} {\bibfield  {journal} {\bibinfo
  {journal} {Physical Review D}\ }\textbf {\bibinfo {volume} {104}},\ \bibinfo
  {pages} {044003} (\bibinfo {year} {2021})}\BibitemShut {NoStop}%
\bibitem [{\citenamefont {Himemoto}\ \emph {et~al.}(2021)\citenamefont
  {Himemoto}, \citenamefont {Nishizawa},\ and\ \citenamefont
  {Taruya}}]{himemoto2021impacts}%
  \BibitemOpen
  \bibfield  {author} {\bibinfo {author} {\bibfnamefont {Y.}~\bibnamefont
  {Himemoto}}, \bibinfo {author} {\bibfnamefont {A.}~\bibnamefont {Nishizawa}},
  \ and\ \bibinfo {author} {\bibfnamefont {A.}~\bibnamefont {Taruya}},\ }\href
  {\doibase 10.1103/PhysRevD.104.044010} {\bibfield  {journal} {\bibinfo
  {journal} {Physical Review D}\ }\textbf {\bibinfo {volume} {104}},\ \bibinfo
  {pages} {044010} (\bibinfo {year} {2021})}\BibitemShut {NoStop}%
\bibitem [{\citenamefont {Pizzati}\ \emph {et~al.}(2022)\citenamefont
  {Pizzati}, \citenamefont {Sachdev}, \citenamefont {Gupta},\ and\
  \citenamefont {Sathyaprakash}}]{pizzati2022toward}%
  \BibitemOpen
  \bibfield  {author} {\bibinfo {author} {\bibfnamefont {E.}~\bibnamefont
  {Pizzati}}, \bibinfo {author} {\bibfnamefont {S.}~\bibnamefont {Sachdev}},
  \bibinfo {author} {\bibfnamefont {A.}~\bibnamefont {Gupta}}, \ and\ \bibinfo
  {author} {\bibfnamefont {B.}~\bibnamefont {Sathyaprakash}},\ }\href {\doibase
  10.1103/PhysRevD.105.104016} {\bibfield  {journal} {\bibinfo  {journal}
  {Physical Review D}\ }\textbf {\bibinfo {volume} {105}},\ \bibinfo {pages}
  {104016} (\bibinfo {year} {2022})}\BibitemShut {NoStop}%
\bibitem [{\citenamefont {Garc\'{\i}a-Quir\'os}\ \emph
  {et~al.}(2020)\citenamefont {Garc\'{\i}a-Quir\'os}, \citenamefont {Colleoni},
  \citenamefont {Husa}, \citenamefont {Estell\'es}, \citenamefont {Pratten},
  \citenamefont {Ramos-Buades}, \citenamefont {Mateu-Lucena},\ and\
  \citenamefont {Jaume}}]{PhysRevD.102.064002}%
  \BibitemOpen
  \bibfield  {author} {\bibinfo {author} {\bibfnamefont {C.}~\bibnamefont
  {Garc\'{\i}a-Quir\'os}}, \bibinfo {author} {\bibfnamefont {M.}~\bibnamefont
  {Colleoni}}, \bibinfo {author} {\bibfnamefont {S.}~\bibnamefont {Husa}},
  \bibinfo {author} {\bibfnamefont {H.}~\bibnamefont {Estell\'es}}, \bibinfo
  {author} {\bibfnamefont {G.}~\bibnamefont {Pratten}}, \bibinfo {author}
  {\bibfnamefont {A.}~\bibnamefont {Ramos-Buades}}, \bibinfo {author}
  {\bibfnamefont {M.}~\bibnamefont {Mateu-Lucena}}, \ and\ \bibinfo {author}
  {\bibfnamefont {R.}~\bibnamefont {Jaume}},\ }\href {\doibase
  10.1103/PhysRevD.102.064002} {\bibfield  {journal} {\bibinfo  {journal}
  {Phys. Rev. D}\ }\textbf {\bibinfo {volume} {102}},\ \bibinfo {pages}
  {064002} (\bibinfo {year} {2020})}\BibitemShut {NoStop}%
\bibitem [{\citenamefont {Littenberg}\ \emph {et~al.}(2013)\citenamefont
  {Littenberg}, \citenamefont {Baker}, \citenamefont {Buonanno},\ and\
  \citenamefont {Kelly}}]{Littenberg2012}%
  \BibitemOpen
  \bibfield  {author} {\bibinfo {author} {\bibfnamefont {T.~B.}\ \bibnamefont
  {Littenberg}}, \bibinfo {author} {\bibfnamefont {J.~G.}\ \bibnamefont
  {Baker}}, \bibinfo {author} {\bibfnamefont {A.}~\bibnamefont {Buonanno}}, \
  and\ \bibinfo {author} {\bibfnamefont {B.~J.}\ \bibnamefont {Kelly}},\ }\href
  {\doibase 10.1103/PhysRevD.87.104003} {\bibfield  {journal} {\bibinfo
  {journal} {Phys. Rev. D}\ }\textbf {\bibinfo {volume} {87}},\ \bibinfo
  {pages} {104003} (\bibinfo {year} {2013})},\ \Eprint
  {http://arxiv.org/abs/1210.0893} {arXiv:1210.0893 [gr-qc]} \BibitemShut
  {NoStop}%
\bibitem [{\citenamefont {Varma}\ \emph {et~al.}(2014)\citenamefont {Varma},
  \citenamefont {Ajith}, \citenamefont {Husa}, \citenamefont {Bustillo},
  \citenamefont {Hannam},\ and\ \citenamefont {P\"urrer}}]{Varma2014}%
  \BibitemOpen
  \bibfield  {author} {\bibinfo {author} {\bibfnamefont {V.}~\bibnamefont
  {Varma}}, \bibinfo {author} {\bibfnamefont {P.}~\bibnamefont {Ajith}},
  \bibinfo {author} {\bibfnamefont {S.}~\bibnamefont {Husa}}, \bibinfo {author}
  {\bibfnamefont {J.~C.}\ \bibnamefont {Bustillo}}, \bibinfo {author}
  {\bibfnamefont {M.}~\bibnamefont {Hannam}}, \ and\ \bibinfo {author}
  {\bibfnamefont {M.}~\bibnamefont {P\"urrer}},\ }\href {\doibase
  10.1103/PhysRevD.90.124004} {\bibfield  {journal} {\bibinfo  {journal} {Phys.
  Rev. D}\ }\textbf {\bibinfo {volume} {90}},\ \bibinfo {pages} {124004}
  (\bibinfo {year} {2014})},\ \Eprint {http://arxiv.org/abs/1409.2349}
  {arXiv:1409.2349 [gr-qc]} \BibitemShut {NoStop}%
\bibitem [{\citenamefont {Shaik}\ \emph {et~al.}(2020)\citenamefont {Shaik},
  \citenamefont {Lange}, \citenamefont {Field}, \citenamefont {O'Shaughnessy},
  \citenamefont {Varma}, \citenamefont {Kidder}, \citenamefont {Pfeiffer},\
  and\ \citenamefont {Wysocki}}]{Shaik2019}%
  \BibitemOpen
  \bibfield  {author} {\bibinfo {author} {\bibfnamefont {F.~H.}\ \bibnamefont
  {Shaik}}, \bibinfo {author} {\bibfnamefont {J.}~\bibnamefont {Lange}},
  \bibinfo {author} {\bibfnamefont {S.~E.}\ \bibnamefont {Field}}, \bibinfo
  {author} {\bibfnamefont {R.}~\bibnamefont {O'Shaughnessy}}, \bibinfo {author}
  {\bibfnamefont {V.}~\bibnamefont {Varma}}, \bibinfo {author} {\bibfnamefont
  {L.~E.}\ \bibnamefont {Kidder}}, \bibinfo {author} {\bibfnamefont {H.~P.}\
  \bibnamefont {Pfeiffer}}, \ and\ \bibinfo {author} {\bibfnamefont
  {D.}~\bibnamefont {Wysocki}},\ }\href {\doibase 10.1103/PhysRevD.101.124054}
  {\bibfield  {journal} {\bibinfo  {journal} {Phys. Rev. D}\ }\textbf {\bibinfo
  {volume} {101}},\ \bibinfo {pages} {124054} (\bibinfo {year} {2020})},\
  \Eprint {http://arxiv.org/abs/1911.02693} {arXiv:1911.02693 [gr-qc]}
  \BibitemShut {NoStop}%
\bibitem [{\citenamefont {Nitz}\ \emph {et~al.}(2021)\citenamefont {Nitz},
  \citenamefont {Harry}, \citenamefont {Brown}, \citenamefont {Biwer},
  \citenamefont {Willis}, \citenamefont {Canton}, \citenamefont {Capano},
  \citenamefont {Dent}, \citenamefont {Pekowsky}, \citenamefont {Williamson},
  \citenamefont {Davies}, \citenamefont {De}, \citenamefont {Cabero},
  \citenamefont {Machenschalk}, \citenamefont {Kumar}, \citenamefont {Macleod},
  \citenamefont {Reyes}, \citenamefont {dfinstad}, \citenamefont {Pannarale},
  \citenamefont {Massinger}, \citenamefont {Kumar}, \citenamefont {Tápai},
  \citenamefont {Singer}, \citenamefont {Khan}, \citenamefont {Fairhurst},
  \citenamefont {Nielsen}, \citenamefont {Singh}, \citenamefont {Chandra},
  \citenamefont {shasvath},\ and\ \citenamefont
  {Gadre}}]{alex_nitz_2021_5256134}%
  \BibitemOpen
  \bibfield  {author} {\bibinfo {author} {\bibfnamefont {A.}~\bibnamefont
  {Nitz}}, \bibinfo {author} {\bibfnamefont {I.}~\bibnamefont {Harry}},
  \bibinfo {author} {\bibfnamefont {D.}~\bibnamefont {Brown}}, \bibinfo
  {author} {\bibfnamefont {C.~M.}\ \bibnamefont {Biwer}}, \bibinfo {author}
  {\bibfnamefont {J.}~\bibnamefont {Willis}}, \bibinfo {author} {\bibfnamefont
  {T.~D.}\ \bibnamefont {Canton}}, \bibinfo {author} {\bibfnamefont
  {C.}~\bibnamefont {Capano}}, \bibinfo {author} {\bibfnamefont
  {T.}~\bibnamefont {Dent}}, \bibinfo {author} {\bibfnamefont {L.}~\bibnamefont
  {Pekowsky}}, \bibinfo {author} {\bibfnamefont {A.~R.}\ \bibnamefont
  {Williamson}}, \bibinfo {author} {\bibfnamefont {G.~S.~C.}\ \bibnamefont
  {Davies}}, \bibinfo {author} {\bibfnamefont {S.}~\bibnamefont {De}}, \bibinfo
  {author} {\bibfnamefont {M.}~\bibnamefont {Cabero}}, \bibinfo {author}
  {\bibfnamefont {B.}~\bibnamefont {Machenschalk}}, \bibinfo {author}
  {\bibfnamefont {P.}~\bibnamefont {Kumar}}, \bibinfo {author} {\bibfnamefont
  {D.}~\bibnamefont {Macleod}}, \bibinfo {author} {\bibfnamefont
  {S.}~\bibnamefont {Reyes}}, \bibinfo {author} {\bibnamefont {dfinstad}},
  \bibinfo {author} {\bibfnamefont {F.}~\bibnamefont {Pannarale}}, \bibinfo
  {author} {\bibfnamefont {T.}~\bibnamefont {Massinger}}, \bibinfo {author}
  {\bibfnamefont {S.}~\bibnamefont {Kumar}}, \bibinfo {author} {\bibfnamefont
  {M.}~\bibnamefont {Tápai}}, \bibinfo {author} {\bibfnamefont
  {L.}~\bibnamefont {Singer}}, \bibinfo {author} {\bibfnamefont
  {S.}~\bibnamefont {Khan}}, \bibinfo {author} {\bibfnamefont {S.}~\bibnamefont
  {Fairhurst}}, \bibinfo {author} {\bibfnamefont {A.}~\bibnamefont {Nielsen}},
  \bibinfo {author} {\bibfnamefont {S.}~\bibnamefont {Singh}}, \bibinfo
  {author} {\bibfnamefont {K.}~\bibnamefont {Chandra}}, \bibinfo {author}
  {\bibnamefont {shasvath}}, \ and\ \bibinfo {author} {\bibfnamefont
  {B.~U.~V.}\ \bibnamefont {Gadre}},\ }\href {\doibase 10.5281/zenodo.5256134}
  {\enquote {\bibinfo {title} {gwastro/pycbc: Release v1.18.3 of pycbc},}\ }
  (\bibinfo {year} {2021})\BibitemShut {NoStop}%
\bibitem [{\citenamefont {Cutler}\ and\ \citenamefont
  {Flanagan}(1994)}]{cutler1994gravitational}%
  \BibitemOpen
  \bibfield  {author} {\bibinfo {author} {\bibfnamefont {C.}~\bibnamefont
  {Cutler}}\ and\ \bibinfo {author} {\bibfnamefont {E.~E.}\ \bibnamefont
  {Flanagan}},\ }\href {\doibase 10.1103/PhysRevD.49.2658} {\bibfield
  {journal} {\bibinfo  {journal} {Phys. Rev. D}\ }\textbf {\bibinfo {volume}
  {49}},\ \bibinfo {pages} {2658} (\bibinfo {year} {1994})}\BibitemShut
  {NoStop}%
\bibitem [{\citenamefont {Press}\ \emph {et~al.}(1992)\citenamefont {Press},
  \citenamefont {Teukolsky}, \citenamefont {Vetterling},\ and\ \citenamefont
  {Flannery}}]{PresTeukVettFlan92}%
  \BibitemOpen
  \bibfield  {author} {\bibinfo {author} {\bibfnamefont {W.~H.}\ \bibnamefont
  {Press}}, \bibinfo {author} {\bibfnamefont {S.~A.}\ \bibnamefont
  {Teukolsky}}, \bibinfo {author} {\bibfnamefont {W.~T.}\ \bibnamefont
  {Vetterling}}, \ and\ \bibinfo {author} {\bibfnamefont {B.~P.}\ \bibnamefont
  {Flannery}},\ }\href@noop {} {\emph {\bibinfo {title} {Numerical Recipes in
  C}}},\ \bibinfo {edition} {2nd}\ ed.\ (\bibinfo  {publisher} {Cambridge
  University Press},\ \bibinfo {address} {Cambridge, USA},\ \bibinfo {year}
  {1992})\BibitemShut {NoStop}%
\bibitem [{\citenamefont {Iacovelli}\ \emph
  {et~al.}(2022{\natexlab{b}})\citenamefont {Iacovelli}, \citenamefont
  {Mancarella}, \citenamefont {Foffa},\ and\ \citenamefont
  {Maggiore}}]{Iacovelli_2022}%
  \BibitemOpen
  \bibfield  {author} {\bibinfo {author} {\bibfnamefont {F.}~\bibnamefont
  {Iacovelli}}, \bibinfo {author} {\bibfnamefont {M.}~\bibnamefont
  {Mancarella}}, \bibinfo {author} {\bibfnamefont {S.}~\bibnamefont {Foffa}}, \
  and\ \bibinfo {author} {\bibfnamefont {M.}~\bibnamefont {Maggiore}},\ }\href
  {\doibase 10.3847/1538-4365/ac9129} {\bibfield  {journal} {\bibinfo
  {journal} {The Astrophysical Journal Supplement Series}\ }\textbf {\bibinfo
  {volume} {263}},\ \bibinfo {pages} {2} (\bibinfo {year}
  {2022}{\natexlab{b}})}\BibitemShut {NoStop}%
\bibitem [{\citenamefont {Dupletsa}\ \emph {et~al.}(2023)\citenamefont
  {Dupletsa}, \citenamefont {Harms}, \citenamefont {Banerjee}, \citenamefont
  {Branchesi}, \citenamefont {Goncharov}, \citenamefont {Maselli},
  \citenamefont {Oliveira}, \citenamefont {Ronchini},\ and\ \citenamefont
  {Tissino}}]{dupletsa2023gwfish}%
  \BibitemOpen
  \bibfield  {author} {\bibinfo {author} {\bibfnamefont {U.}~\bibnamefont
  {Dupletsa}}, \bibinfo {author} {\bibfnamefont {J.}~\bibnamefont {Harms}},
  \bibinfo {author} {\bibfnamefont {B.}~\bibnamefont {Banerjee}}, \bibinfo
  {author} {\bibfnamefont {M.}~\bibnamefont {Branchesi}}, \bibinfo {author}
  {\bibfnamefont {B.}~\bibnamefont {Goncharov}}, \bibinfo {author}
  {\bibfnamefont {A.}~\bibnamefont {Maselli}}, \bibinfo {author} {\bibfnamefont
  {A.}~\bibnamefont {Oliveira}}, \bibinfo {author} {\bibfnamefont
  {S.}~\bibnamefont {Ronchini}}, \ and\ \bibinfo {author} {\bibfnamefont
  {J.}~\bibnamefont {Tissino}},\ }\href {\doibase 10.1016/j.ascom.2022.100671}
  {\bibfield  {journal} {\bibinfo  {journal} {Astronomy and Computing}\
  }\textbf {\bibinfo {volume} {42}},\ \bibinfo {pages} {100671} (\bibinfo
  {year} {2023})}\BibitemShut {NoStop}%
\bibitem [{\citenamefont {Taylor}\ and\ \citenamefont
  {Gerosa}(2018)}]{taylor2018mining}%
  \BibitemOpen
  \bibfield  {author} {\bibinfo {author} {\bibfnamefont {S.~R.}\ \bibnamefont
  {Taylor}}\ and\ \bibinfo {author} {\bibfnamefont {D.}~\bibnamefont
  {Gerosa}},\ }\href {\doibase 10.1103/PhysRevD.98.083017} {\bibfield
  {journal} {\bibinfo  {journal} {Physical Review D}\ }\textbf {\bibinfo
  {volume} {98}},\ \bibinfo {pages} {083017} (\bibinfo {year}
  {2018})}\BibitemShut {NoStop}%
\bibitem [{\citenamefont {Mould}\ \emph {et~al.}(2022)\citenamefont {Mould},
  \citenamefont {Gerosa},\ and\ \citenamefont {Taylor}}]{mould2022deep}%
  \BibitemOpen
  \bibfield  {author} {\bibinfo {author} {\bibfnamefont {M.}~\bibnamefont
  {Mould}}, \bibinfo {author} {\bibfnamefont {D.}~\bibnamefont {Gerosa}}, \
  and\ \bibinfo {author} {\bibfnamefont {S.~R.}\ \bibnamefont {Taylor}},\
  }\href {\doibase 10.1103/PhysRevD.106.103013} {\bibfield  {journal} {\bibinfo
   {journal} {Physical Review D}\ }\textbf {\bibinfo {volume} {106}},\ \bibinfo
  {pages} {103013} (\bibinfo {year} {2022})}\BibitemShut {NoStop}%
\bibitem [{\citenamefont {Pieroni}\ \emph {et~al.}(2022)\citenamefont
  {Pieroni}, \citenamefont {Ricciardone},\ and\ \citenamefont
  {Barausse}}]{pieroni2022detectability}%
  \BibitemOpen
  \bibfield  {author} {\bibinfo {author} {\bibfnamefont {M.}~\bibnamefont
  {Pieroni}}, \bibinfo {author} {\bibfnamefont {A.}~\bibnamefont
  {Ricciardone}}, \ and\ \bibinfo {author} {\bibfnamefont {E.}~\bibnamefont
  {Barausse}},\ }\href {\doibase 10.1038/s41598-022-19540-7} {\bibfield
  {journal} {\bibinfo  {journal} {Scientific Reports}\ }\textbf {\bibinfo
  {volume} {12}},\ \bibinfo {pages} {17940} (\bibinfo {year}
  {2022})}\BibitemShut {NoStop}%
\bibitem [{\citenamefont {{Henriques}}\ \emph {et~al.}(2015)\citenamefont
  {{Henriques}}, \citenamefont {{White}}, \citenamefont {{Thomas}},
  \citenamefont {{Angulo}}, \citenamefont {{Guo}}, \citenamefont {{Lemson}},
  \citenamefont {{Springel}},\ and\ \citenamefont
  {{Overzier}}}]{Henriques2015}%
  \BibitemOpen
  \bibfield  {author} {\bibinfo {author} {\bibfnamefont {B.~M.~B.}\
  \bibnamefont {{Henriques}}}, \bibinfo {author} {\bibfnamefont {S.~D.~M.}\
  \bibnamefont {{White}}}, \bibinfo {author} {\bibfnamefont {P.~A.}\
  \bibnamefont {{Thomas}}}, \bibinfo {author} {\bibfnamefont {R.}~\bibnamefont
  {{Angulo}}}, \bibinfo {author} {\bibfnamefont {Q.}~\bibnamefont {{Guo}}},
  \bibinfo {author} {\bibfnamefont {G.}~\bibnamefont {{Lemson}}}, \bibinfo
  {author} {\bibfnamefont {V.}~\bibnamefont {{Springel}}}, \ and\ \bibinfo
  {author} {\bibfnamefont {R.}~\bibnamefont {{Overzier}}},\ }\href {\doibase
  10.1093/mnras/stv705} {\bibfield  {journal} {\bibinfo  {journal} {\mnras}\
  }\textbf {\bibinfo {volume} {451}},\ \bibinfo {pages} {2663} (\bibinfo {year}
  {2015})},\ \Eprint {http://arxiv.org/abs/1410.0365} {arXiv:1410.0365
  [astro-ph.GA]} \BibitemShut {NoStop}%
\bibitem [{\citenamefont {{Springel}}(2005)}]{Springel2005}%
  \BibitemOpen
  \bibfield  {author} {\bibinfo {author} {\bibfnamefont {V.}~\bibnamefont
  {{Springel}}},\ }\href {\doibase 10.1111/j.1365-2966.2005.09655.x} {\bibfield
   {journal} {\bibinfo  {journal} {\mnras}\ }\textbf {\bibinfo {volume}
  {364}},\ \bibinfo {pages} {1105} (\bibinfo {year} {2005})},\ \Eprint
  {http://arxiv.org/abs/astro-ph/0505010} {astro-ph/0505010} \BibitemShut
  {NoStop}%
\bibitem [{\citenamefont {Izquierdo-Villalba}\ \emph
  {et~al.}(2019)\citenamefont {Izquierdo-Villalba}, \citenamefont {Angulo},
  \citenamefont {Orsi}, \citenamefont {Hurier}, \citenamefont {Vilella-Rojo},
  \citenamefont {Bonoli}, \citenamefont {L{\'o}pez-Sanjuan}, \citenamefont
  {Alcaniz}, \citenamefont {Cenarro}, \citenamefont {Crist{\'o}bal-Hornillos}
  \emph {et~al.}}]{IzquierdoVillalba2019}%
  \BibitemOpen
  \bibfield  {author} {\bibinfo {author} {\bibfnamefont {D.}~\bibnamefont
  {Izquierdo-Villalba}}, \bibinfo {author} {\bibfnamefont {R.~E.}\ \bibnamefont
  {Angulo}}, \bibinfo {author} {\bibfnamefont {A.}~\bibnamefont {Orsi}},
  \bibinfo {author} {\bibfnamefont {G.}~\bibnamefont {Hurier}}, \bibinfo
  {author} {\bibfnamefont {G.}~\bibnamefont {Vilella-Rojo}}, \bibinfo {author}
  {\bibfnamefont {S.}~\bibnamefont {Bonoli}}, \bibinfo {author} {\bibfnamefont
  {C.}~\bibnamefont {L{\'o}pez-Sanjuan}}, \bibinfo {author} {\bibfnamefont
  {J.}~\bibnamefont {Alcaniz}}, \bibinfo {author} {\bibfnamefont
  {J.}~\bibnamefont {Cenarro}}, \bibinfo {author} {\bibfnamefont
  {D.}~\bibnamefont {Crist{\'o}bal-Hornillos}},  \emph {et~al.},\ }\href
  {\doibase 10.1051/0004-6361/201936232} {\bibfield  {journal} {\bibinfo
  {journal} {Astronomy \& Astrophysics}\ }\textbf {\bibinfo {volume} {631}},\
  \bibinfo {pages} {A82} (\bibinfo {year} {2019})}\BibitemShut {NoStop}%
\bibitem [{\citenamefont {York}\ \emph {et~al.}(2000)\citenamefont {York},
  \citenamefont {Adelman}, \citenamefont {Anderson~Jr}, \citenamefont
  {Anderson}, \citenamefont {Annis}, \citenamefont {Bahcall}, \citenamefont
  {Bakken}, \citenamefont {Barkhouser}, \citenamefont {Bastian}, \citenamefont
  {Berman} \emph {et~al.}}]{SDSS2000}%
  \BibitemOpen
  \bibfield  {author} {\bibinfo {author} {\bibfnamefont {D.~G.}\ \bibnamefont
  {York}}, \bibinfo {author} {\bibfnamefont {J.}~\bibnamefont {Adelman}},
  \bibinfo {author} {\bibfnamefont {J.~E.}\ \bibnamefont {Anderson~Jr}},
  \bibinfo {author} {\bibfnamefont {S.~F.}\ \bibnamefont {Anderson}}, \bibinfo
  {author} {\bibfnamefont {J.}~\bibnamefont {Annis}}, \bibinfo {author}
  {\bibfnamefont {N.~A.}\ \bibnamefont {Bahcall}}, \bibinfo {author}
  {\bibfnamefont {J.}~\bibnamefont {Bakken}}, \bibinfo {author} {\bibfnamefont
  {R.}~\bibnamefont {Barkhouser}}, \bibinfo {author} {\bibfnamefont
  {S.}~\bibnamefont {Bastian}}, \bibinfo {author} {\bibfnamefont
  {E.}~\bibnamefont {Berman}},  \emph {et~al.},\ }\href {\doibase
  10.1086/301513} {\bibfield  {journal} {\bibinfo  {journal} {The Astronomical
  Journal}\ }\textbf {\bibinfo {volume} {120}},\ \bibinfo {pages} {1579}
  (\bibinfo {year} {2000})}\BibitemShut {NoStop}%
\bibitem [{\citenamefont {Amiaux}\ \emph {et~al.}(2012)\citenamefont {Amiaux},
  \citenamefont {Scaramella}, \citenamefont {Mellier}, \citenamefont {Altieri},
  \citenamefont {Burigana}, \citenamefont {Silva}, \citenamefont {G{\'o}mez},
  \citenamefont {Hoar}, \citenamefont {Laureijs}, \citenamefont {Maiorano},
  \citenamefont {Oliveira}, \citenamefont {Renk}, \citenamefont {Criado},
  \citenamefont {Tereno}, \citenamefont {Augu{\`e}res}, \citenamefont
  {Brinchmann}, \citenamefont {Cropper}, \citenamefont {Duvet}, \citenamefont
  {Ealet}, \citenamefont {Franzetti}, \citenamefont {Garilli}, \citenamefont
  {Gondoin}, \citenamefont {Guzzo}, \citenamefont {Hoekstra}, \citenamefont
  {Holmes}, \citenamefont {Jahnke}, \citenamefont {Kitching}, \citenamefont
  {Meneghetti}, \citenamefont {Percival},\ and\ \citenamefont
  {Warren}}]{EUCLID2012}%
  \BibitemOpen
  \bibfield  {author} {\bibinfo {author} {\bibfnamefont {J.}~\bibnamefont
  {Amiaux}}, \bibinfo {author} {\bibfnamefont {R.}~\bibnamefont {Scaramella}},
  \bibinfo {author} {\bibfnamefont {Y.}~\bibnamefont {Mellier}}, \bibinfo
  {author} {\bibfnamefont {B.}~\bibnamefont {Altieri}}, \bibinfo {author}
  {\bibfnamefont {C.}~\bibnamefont {Burigana}}, \bibinfo {author}
  {\bibfnamefont {A.~D.}\ \bibnamefont {Silva}}, \bibinfo {author}
  {\bibfnamefont {P.~M.}\ \bibnamefont {G{\'o}mez}}, \bibinfo {author}
  {\bibfnamefont {J.}~\bibnamefont {Hoar}}, \bibinfo {author} {\bibfnamefont
  {R.~J.}\ \bibnamefont {Laureijs}}, \bibinfo {author} {\bibfnamefont
  {E.}~\bibnamefont {Maiorano}}, \bibinfo {author} {\bibfnamefont {D.~M.}\
  \bibnamefont {Oliveira}}, \bibinfo {author} {\bibfnamefont {F.}~\bibnamefont
  {Renk}}, \bibinfo {author} {\bibfnamefont {G.~S.}\ \bibnamefont {Criado}},
  \bibinfo {author} {\bibfnamefont {I.}~\bibnamefont {Tereno}}, \bibinfo
  {author} {\bibfnamefont {J.-L.}\ \bibnamefont {Augu{\`e}res}}, \bibinfo
  {author} {\bibfnamefont {J.}~\bibnamefont {Brinchmann}}, \bibinfo {author}
  {\bibfnamefont {M.}~\bibnamefont {Cropper}}, \bibinfo {author} {\bibfnamefont
  {L.}~\bibnamefont {Duvet}}, \bibinfo {author} {\bibfnamefont
  {A.}~\bibnamefont {Ealet}}, \bibinfo {author} {\bibfnamefont
  {P.}~\bibnamefont {Franzetti}}, \bibinfo {author} {\bibfnamefont
  {B.}~\bibnamefont {Garilli}}, \bibinfo {author} {\bibfnamefont
  {P.}~\bibnamefont {Gondoin}}, \bibinfo {author} {\bibfnamefont
  {L.}~\bibnamefont {Guzzo}}, \bibinfo {author} {\bibfnamefont
  {H.}~\bibnamefont {Hoekstra}}, \bibinfo {author} {\bibfnamefont
  {R.}~\bibnamefont {Holmes}}, \bibinfo {author} {\bibfnamefont
  {K.}~\bibnamefont {Jahnke}}, \bibinfo {author} {\bibfnamefont
  {T.}~\bibnamefont {Kitching}}, \bibinfo {author} {\bibfnamefont
  {M.}~\bibnamefont {Meneghetti}}, \bibinfo {author} {\bibfnamefont {W.~J.}\
  \bibnamefont {Percival}}, \ and\ \bibinfo {author} {\bibfnamefont {S.~J.}\
  \bibnamefont {Warren}},\ }\href {\doibase 10.1117/12.926513} {\bibfield
  {journal} {\bibinfo  {journal} {Space Telescopes and Instrumentation 2012:
  Optical, Infrared, and Millimeter Wave}\ }\textbf {\bibinfo {volume}
  {8442}},\ \bibinfo {pages} {380} (\bibinfo {year} {2012})},\ \bibinfo {note}
  {\url{https://arxiv.org/pdf/1209.2228.pdf} ;
  \url{https://arxiv.org/abs/1209.2228} ;
  \url{http://arxiv.org/abs/1209.2228}}\BibitemShut {NoStop}%
\bibitem [{\citenamefont {Spergel}\ \emph {et~al.}(2015)\citenamefont
  {Spergel}, \citenamefont {Gehrels}, \citenamefont {Baltay}, \citenamefont
  {Bennett}, \citenamefont {Breckinridge}, \citenamefont {Donahue},
  \citenamefont {Dressler}, \citenamefont {Gaudi}, \citenamefont {Greene},
  \citenamefont {Guyon} \emph {et~al.}}]{WFIRST2015}%
  \BibitemOpen
  \bibfield  {author} {\bibinfo {author} {\bibfnamefont {D.}~\bibnamefont
  {Spergel}}, \bibinfo {author} {\bibfnamefont {N.}~\bibnamefont {Gehrels}},
  \bibinfo {author} {\bibfnamefont {C.}~\bibnamefont {Baltay}}, \bibinfo
  {author} {\bibfnamefont {D.}~\bibnamefont {Bennett}}, \bibinfo {author}
  {\bibfnamefont {J.}~\bibnamefont {Breckinridge}}, \bibinfo {author}
  {\bibfnamefont {M.}~\bibnamefont {Donahue}}, \bibinfo {author} {\bibfnamefont
  {A.}~\bibnamefont {Dressler}}, \bibinfo {author} {\bibfnamefont
  {B.}~\bibnamefont {Gaudi}}, \bibinfo {author} {\bibfnamefont
  {T.}~\bibnamefont {Greene}}, \bibinfo {author} {\bibfnamefont
  {O.}~\bibnamefont {Guyon}},  \emph {et~al.},\ }\href {\doibase
  10.48550/arXiv.1503.03757} {\bibfield  {journal} {\bibinfo  {journal} {arXiv
  preprint arXiv:1503.03757}\ } (\bibinfo {year} {2015}),\
  10.48550/arXiv.1503.03757}\BibitemShut {NoStop}%
\bibitem [{\citenamefont {Ivezi{\'c}}\ \emph {et~al.}(2019)\citenamefont
  {Ivezi{\'c}}, \citenamefont {Kahn}, \citenamefont {Tyson}, \citenamefont
  {Abel}, \citenamefont {Acosta}, \citenamefont {Allsman}, \citenamefont
  {Alonso}, \citenamefont {AlSayyad}, \citenamefont {Anderson}, \citenamefont
  {Andrew} \emph {et~al.}}]{LSST2019}%
  \BibitemOpen
  \bibfield  {author} {\bibinfo {author} {\bibfnamefont {{\v{Z}}.}~\bibnamefont
  {Ivezi{\'c}}}, \bibinfo {author} {\bibfnamefont {S.~M.}\ \bibnamefont
  {Kahn}}, \bibinfo {author} {\bibfnamefont {J.~A.}\ \bibnamefont {Tyson}},
  \bibinfo {author} {\bibfnamefont {B.}~\bibnamefont {Abel}}, \bibinfo {author}
  {\bibfnamefont {E.}~\bibnamefont {Acosta}}, \bibinfo {author} {\bibfnamefont
  {R.}~\bibnamefont {Allsman}}, \bibinfo {author} {\bibfnamefont
  {D.}~\bibnamefont {Alonso}}, \bibinfo {author} {\bibfnamefont
  {Y.}~\bibnamefont {AlSayyad}}, \bibinfo {author} {\bibfnamefont {S.~F.}\
  \bibnamefont {Anderson}}, \bibinfo {author} {\bibfnamefont {J.}~\bibnamefont
  {Andrew}},  \emph {et~al.},\ }\href {\doibase 10.3847/1538-4357/ab042c}
  {\bibfield  {journal} {\bibinfo  {journal} {The Astrophysical Journal}\
  }\textbf {\bibinfo {volume} {873}},\ \bibinfo {pages} {111} (\bibinfo {year}
  {2019})}\BibitemShut {NoStop}%
\bibitem [{\citenamefont {{Baldry}}\ \emph {et~al.}(2008)\citenamefont
  {{Baldry}}, \citenamefont {{Glazebrook}},\ and\ \citenamefont
  {{Driver}}}]{Baldry2008}%
  \BibitemOpen
  \bibfield  {author} {\bibinfo {author} {\bibfnamefont {I.~K.}\ \bibnamefont
  {{Baldry}}}, \bibinfo {author} {\bibfnamefont {K.}~\bibnamefont
  {{Glazebrook}}}, \ and\ \bibinfo {author} {\bibfnamefont {S.~P.}\
  \bibnamefont {{Driver}}},\ }\href {\doibase 10.1111/j.1365-2966.2008.13348.x}
  {\bibfield  {journal} {\bibinfo  {journal} {\mnras}\ }\textbf {\bibinfo
  {volume} {388}},\ \bibinfo {pages} {945} (\bibinfo {year} {2008})},\ \Eprint
  {http://arxiv.org/abs/0804.2892} {arXiv:0804.2892 [astro-ph]} \BibitemShut
  {NoStop}%
\bibitem [{\citenamefont {Yasuda}\ \emph {et~al.}(2001)\citenamefont {Yasuda},
  \citenamefont {Fukugita}, \citenamefont {Narayanan}, \citenamefont {Lupton},
  \citenamefont {Strateva}, \citenamefont {Strauss}, \citenamefont
  {Ivezi{\'c}}, \citenamefont {Kim}, \citenamefont {Hogg}, \citenamefont
  {Weinberg} \emph {et~al.}}]{Yasuda2001}%
  \BibitemOpen
  \bibfield  {author} {\bibinfo {author} {\bibfnamefont {N.}~\bibnamefont
  {Yasuda}}, \bibinfo {author} {\bibfnamefont {M.}~\bibnamefont {Fukugita}},
  \bibinfo {author} {\bibfnamefont {V.~K.}\ \bibnamefont {Narayanan}}, \bibinfo
  {author} {\bibfnamefont {R.~H.}\ \bibnamefont {Lupton}}, \bibinfo {author}
  {\bibfnamefont {I.}~\bibnamefont {Strateva}}, \bibinfo {author}
  {\bibfnamefont {M.~A.}\ \bibnamefont {Strauss}}, \bibinfo {author}
  {\bibfnamefont {{\v{Z}}.}~\bibnamefont {Ivezi{\'c}}}, \bibinfo {author}
  {\bibfnamefont {R.~S.}\ \bibnamefont {Kim}}, \bibinfo {author} {\bibfnamefont
  {D.~W.}\ \bibnamefont {Hogg}}, \bibinfo {author} {\bibfnamefont {D.~H.}\
  \bibnamefont {Weinberg}},  \emph {et~al.},\ }\href {\doibase 10.1086/322093}
  {\bibfield  {journal} {\bibinfo  {journal} {The Astronomical Journal}\
  }\textbf {\bibinfo {volume} {122}},\ \bibinfo {pages} {1104} (\bibinfo {year}
  {2001})}\BibitemShut {NoStop}%
\bibitem [{\citenamefont {{Rovilos}}\ \emph {et~al.}(2009)\citenamefont
  {{Rovilos}}, \citenamefont {{Burwitz}}, \citenamefont {{Szokoly}},
  \citenamefont {{Hasinger}}, \citenamefont {{Egami}}, \citenamefont
  {{Bouch{\'e}}}, \citenamefont {{Berta}}, \citenamefont {{Salvato}},
  \citenamefont {{Lutz}},\ and\ \citenamefont {{Genzel}}}]{Rovilos2009}%
  \BibitemOpen
  \bibfield  {author} {\bibinfo {author} {\bibfnamefont {E.}~\bibnamefont
  {{Rovilos}}}, \bibinfo {author} {\bibfnamefont {V.}~\bibnamefont
  {{Burwitz}}}, \bibinfo {author} {\bibfnamefont {G.}~\bibnamefont
  {{Szokoly}}}, \bibinfo {author} {\bibfnamefont {G.}~\bibnamefont
  {{Hasinger}}}, \bibinfo {author} {\bibfnamefont {E.}~\bibnamefont {{Egami}}},
  \bibinfo {author} {\bibfnamefont {N.}~\bibnamefont {{Bouch{\'e}}}}, \bibinfo
  {author} {\bibfnamefont {S.}~\bibnamefont {{Berta}}}, \bibinfo {author}
  {\bibfnamefont {M.}~\bibnamefont {{Salvato}}}, \bibinfo {author}
  {\bibfnamefont {D.}~\bibnamefont {{Lutz}}}, \ and\ \bibinfo {author}
  {\bibfnamefont {R.}~\bibnamefont {{Genzel}}},\ }\href {\doibase
  10.1051/0004-6361/200912626} {\bibfield  {journal} {\bibinfo  {journal}
  {\aap}\ }\textbf {\bibinfo {volume} {507}},\ \bibinfo {pages} {195} (\bibinfo
  {year} {2009})},\ \Eprint {http://arxiv.org/abs/0909.0661} {arXiv:0909.0661
  [astro-ph.CO]} \BibitemShut {NoStop}%
\bibitem [{\citenamefont {Dom{\'\i}nguez~S{\'a}nchez}\ \emph
  {et~al.}(2011)\citenamefont {Dom{\'\i}nguez~S{\'a}nchez}, \citenamefont
  {Pozzi}, \citenamefont {Gruppioni}, \citenamefont {Cimatti}, \citenamefont
  {Ilbert}, \citenamefont {Pozzetti}, \citenamefont {McCracken}, \citenamefont
  {Capak}, \citenamefont {Le~Floch}, \citenamefont {Salvato} \emph
  {et~al.}}]{DominguezSanchez2011}%
  \BibitemOpen
  \bibfield  {author} {\bibinfo {author} {\bibfnamefont {H.}~\bibnamefont
  {Dom{\'\i}nguez~S{\'a}nchez}}, \bibinfo {author} {\bibfnamefont
  {F.}~\bibnamefont {Pozzi}}, \bibinfo {author} {\bibfnamefont
  {C.}~\bibnamefont {Gruppioni}}, \bibinfo {author} {\bibfnamefont
  {A.}~\bibnamefont {Cimatti}}, \bibinfo {author} {\bibfnamefont
  {O.}~\bibnamefont {Ilbert}}, \bibinfo {author} {\bibfnamefont
  {L.}~\bibnamefont {Pozzetti}}, \bibinfo {author} {\bibfnamefont
  {H.}~\bibnamefont {McCracken}}, \bibinfo {author} {\bibfnamefont
  {P.}~\bibnamefont {Capak}}, \bibinfo {author} {\bibfnamefont
  {E.}~\bibnamefont {Le~Floch}}, \bibinfo {author} {\bibfnamefont
  {M.}~\bibnamefont {Salvato}},  \emph {et~al.},\ }\href {\doibase
  10.1111/j.1365-2966.2011.19263.x} {\bibfield  {journal} {\bibinfo  {journal}
  {Monthly Notices of the Royal Astronomical Society}\ }\textbf {\bibinfo
  {volume} {417}},\ \bibinfo {pages} {900} (\bibinfo {year}
  {2011})}\BibitemShut {NoStop}%
\bibitem [{\citenamefont {Cusin}\ and\ \citenamefont
  {Tamanini}(2021)}]{Cusin:2020ezb}%
  \BibitemOpen
  \bibfield  {author} {\bibinfo {author} {\bibfnamefont {G.}~\bibnamefont
  {Cusin}}\ and\ \bibinfo {author} {\bibfnamefont {N.}~\bibnamefont
  {Tamanini}},\ }\href {\doibase 10.1093/mnras/stab1130} {\bibfield  {journal}
  {\bibinfo  {journal} {Mon. Not. Roy. Astron. Soc.}\ }\textbf {\bibinfo
  {volume} {504}},\ \bibinfo {pages} {3610} (\bibinfo {year} {2021})},\ \Eprint
  {http://arxiv.org/abs/2011.15109} {arXiv:2011.15109 [astro-ph.CO]}
  \BibitemShut {NoStop}%
\bibitem [{\citenamefont {Wen}\ and\ \citenamefont
  {Chen}(2010)}]{PhysRevD.81.082001}%
  \BibitemOpen
  \bibfield  {author} {\bibinfo {author} {\bibfnamefont {L.}~\bibnamefont
  {Wen}}\ and\ \bibinfo {author} {\bibfnamefont {Y.}~\bibnamefont {Chen}},\
  }\href {\doibase 10.1103/PhysRevD.81.082001} {\bibfield  {journal} {\bibinfo
  {journal} {Phys. Rev. D}\ }\textbf {\bibinfo {volume} {81}},\ \bibinfo
  {pages} {082001} (\bibinfo {year} {2010})}\BibitemShut {NoStop}%
\bibitem [{\citenamefont {Laghi}\ \emph {et~al.}(2021)\citenamefont {Laghi},
  \citenamefont {Tamanini}, \citenamefont {Del Pozzo}, \citenamefont {Sesana},
  \citenamefont {Gair}, \citenamefont {Babak},\ and\ \citenamefont
  {Izquierdo-Villalba}}]{Laghi_2021}%
  \BibitemOpen
  \bibfield  {author} {\bibinfo {author} {\bibfnamefont {D.}~\bibnamefont
  {Laghi}}, \bibinfo {author} {\bibfnamefont {N.}~\bibnamefont {Tamanini}},
  \bibinfo {author} {\bibfnamefont {W.}~\bibnamefont {Del Pozzo}}, \bibinfo
  {author} {\bibfnamefont {A.}~\bibnamefont {Sesana}}, \bibinfo {author}
  {\bibfnamefont {J.}~\bibnamefont {Gair}}, \bibinfo {author} {\bibfnamefont
  {S.}~\bibnamefont {Babak}}, \ and\ \bibinfo {author} {\bibfnamefont
  {D.}~\bibnamefont {Izquierdo-Villalba}},\ }\href {\doibase
  10.1093/mnras/stab2741} {\bibfield  {journal} {\bibinfo  {journal} {Monthly
  Notices of the Royal Astronomical Society}\ }\textbf {\bibinfo {volume}
  {508}},\ \bibinfo {pages} {4512} (\bibinfo {year} {2021})}\BibitemShut
  {NoStop}%
\bibitem [{\citenamefont {Hogg}(1999)}]{hogg1999}%
  \BibitemOpen
  \bibfield  {author} {\bibinfo {author} {\bibfnamefont {D.~W.}\ \bibnamefont
  {Hogg}},\ }\href@noop {} {\  (\bibinfo {year} {1999})},\ \Eprint
  {http://arxiv.org/abs/9905116} {9905116 [astro-ph.CO]} \BibitemShut {NoStop}%
\bibitem [{\citenamefont {{Del Pozzo}}(2012)}]{DelPozzo:2011yh}%
  \BibitemOpen
  \bibfield  {author} {\bibinfo {author} {\bibfnamefont {W.}~\bibnamefont {{Del
  Pozzo}}},\ }\href {\doibase 10.1103/PhysRevD.86.043011} {\bibfield  {journal}
  {\bibinfo  {journal} {Phys. Rev. D}\ }\textbf {\bibinfo {volume} {86}},\
  \bibinfo {pages} {043011} (\bibinfo {year} {2012})},\ \Eprint
  {http://arxiv.org/abs/1108.1317} {arXiv:1108.1317 [astro-ph.CO]} \BibitemShut
  {NoStop}%
%%CITATION = ARXIV:1108.1317;%%
\bibitem [{\citenamefont {Del~Pozzo}\ \emph {et~al.}(2017)\citenamefont
  {Del~Pozzo}, \citenamefont {Li},\ and\ \citenamefont
  {Messenger}}]{DelPozzo:2015bna}%
  \BibitemOpen
  \bibfield  {author} {\bibinfo {author} {\bibfnamefont {W.}~\bibnamefont
  {Del~Pozzo}}, \bibinfo {author} {\bibfnamefont {T.~G.~F.}\ \bibnamefont
  {Li}}, \ and\ \bibinfo {author} {\bibfnamefont {C.}~\bibnamefont
  {Messenger}},\ }\href {\doibase 10.1103/PhysRevD.95.043502} {\bibfield
  {journal} {\bibinfo  {journal} {Phys. Rev. D}\ }\textbf {\bibinfo {volume}
  {95}},\ \bibinfo {pages} {043502} (\bibinfo {year} {2017})},\ \Eprint
  {http://arxiv.org/abs/1506.06590} {arXiv:1506.06590 [gr-qc]} \BibitemShut
  {NoStop}%
\bibitem [{\citenamefont {Mandel}\ \emph {et~al.}(2019)\citenamefont {Mandel},
  \citenamefont {Farr},\ and\ \citenamefont {Gair}}]{mandel2019extracting}%
  \BibitemOpen
  \bibfield  {author} {\bibinfo {author} {\bibfnamefont {I.}~\bibnamefont
  {Mandel}}, \bibinfo {author} {\bibfnamefont {W.~M.}\ \bibnamefont {Farr}}, \
  and\ \bibinfo {author} {\bibfnamefont {J.~R.}\ \bibnamefont {Gair}},\ }\href
  {\doibase 10.1093/mnras/stz896} {\bibfield  {journal} {\bibinfo  {journal}
  {Monthly Notices of the Royal Astronomical Society}\ }\textbf {\bibinfo
  {volume} {486}},\ \bibinfo {pages} {1086} (\bibinfo {year}
  {2019})}\BibitemShut {NoStop}%
\bibitem [{\citenamefont {Vitale}\ \emph {et~al.}(2022)\citenamefont {Vitale},
  \citenamefont {Gerosa}, \citenamefont {Farr},\ and\ \citenamefont
  {Taylor}}]{vitale2022inferring}%
  \BibitemOpen
  \bibfield  {author} {\bibinfo {author} {\bibfnamefont {S.}~\bibnamefont
  {Vitale}}, \bibinfo {author} {\bibfnamefont {D.}~\bibnamefont {Gerosa}},
  \bibinfo {author} {\bibfnamefont {W.~M.}\ \bibnamefont {Farr}}, \ and\
  \bibinfo {author} {\bibfnamefont {S.~R.}\ \bibnamefont {Taylor}},\ }in\ \href
  {\doibase 10.1007/978-981-15-4702-7_45-1} {\emph {\bibinfo {booktitle}
  {Handbook of Gravitational Wave Astronomy}}}\ (\bibinfo  {publisher}
  {Springer},\ \bibinfo {year} {2022})\ pp.\ \bibinfo {pages}
  {1--60}\BibitemShut {NoStop}%
\bibitem [{\citenamefont {Jaynes}(2003)}]{jaynes2003}%
  \BibitemOpen
  \bibfield  {author} {\bibinfo {author} {\bibfnamefont {E.~T.}\ \bibnamefont
  {Jaynes}},\ }\href@noop {} {\emph {\bibinfo {title} {Probability theory: The
  logic of science}}}\ (\bibinfo  {publisher} {Cambridge university press},\
  \bibinfo {year} {2003})\BibitemShut {NoStop}%
\bibitem [{\citenamefont {{Del Pozzo}}\ and\ \citenamefont
  {Laghi}(2022)}]{cosmoLISA}%
  \BibitemOpen
  \bibfield  {author} {\bibinfo {author} {\bibfnamefont {W.}~\bibnamefont {{Del
  Pozzo}}}\ and\ \bibinfo {author} {\bibfnamefont {D.}~\bibnamefont {Laghi}},\
  }\href {https://github.com/wdpozzo/cosmolisa} {\enquote {\bibinfo {title}
  {wdpozzo/cosmolisa},}\ } (\bibinfo {year} {2022})\BibitemShut {NoStop}%
\bibitem [{\citenamefont {Veitch}\ \emph {et~al.}(2022)\citenamefont {Veitch},
  \citenamefont {Pozzo}, \citenamefont {Lyttle}, \citenamefont {Williams},
  \citenamefont {Talbot}, \citenamefont {Pitkin}, \citenamefont {Ashton},
  \citenamefont {Cody}, \citenamefont {Hübner}, \citenamefont {Nitz},
  \citenamefont {Mihaylov}, \citenamefont {Macleod}, \citenamefont {Carullo},
  \citenamefont {Davies},\ and\ \citenamefont {ttw}}]{CPNest}%
  \BibitemOpen
  \bibfield  {author} {\bibinfo {author} {\bibfnamefont {J.}~\bibnamefont
  {Veitch}}, \bibinfo {author} {\bibfnamefont {W.~D.}\ \bibnamefont {Pozzo}},
  \bibinfo {author} {\bibfnamefont {A.}~\bibnamefont {Lyttle}}, \bibinfo
  {author} {\bibfnamefont {M.~J.}\ \bibnamefont {Williams}}, \bibinfo {author}
  {\bibfnamefont {C.}~\bibnamefont {Talbot}}, \bibinfo {author} {\bibfnamefont
  {M.}~\bibnamefont {Pitkin}}, \bibinfo {author} {\bibfnamefont
  {G.}~\bibnamefont {Ashton}}, \bibinfo {author} {\bibnamefont {Cody}},
  \bibinfo {author} {\bibfnamefont {M.}~\bibnamefont {Hübner}}, \bibinfo
  {author} {\bibfnamefont {A.}~\bibnamefont {Nitz}}, \bibinfo {author}
  {\bibfnamefont {D.}~\bibnamefont {Mihaylov}}, \bibinfo {author}
  {\bibfnamefont {D.}~\bibnamefont {Macleod}}, \bibinfo {author} {\bibfnamefont
  {G.}~\bibnamefont {Carullo}}, \bibinfo {author} {\bibfnamefont
  {G.}~\bibnamefont {Davies}}, \ and\ \bibinfo {author} {\bibnamefont {ttw}},\
  }\href {\doibase 10.5281/zenodo.7385221} {\enquote {\bibinfo {title}
  {johnveitch/cpnest: v0.11.5},}\ } (\bibinfo {year} {2022})\BibitemShut
  {NoStop}%
\bibitem [{\citenamefont {Del~Pozzo}\ \emph
  {et~al.}(2018{\natexlab{b}})\citenamefont {Del~Pozzo}, \citenamefont {Berry},
  \citenamefont {Ghosh}, \citenamefont {Haines}, \citenamefont {Singer},\ and\
  \citenamefont {Vecchio}}]{DelPozzo2018}%
  \BibitemOpen
  \bibfield  {author} {\bibinfo {author} {\bibfnamefont {W.}~\bibnamefont
  {Del~Pozzo}}, \bibinfo {author} {\bibfnamefont {C.~P.~L.}\ \bibnamefont
  {Berry}}, \bibinfo {author} {\bibfnamefont {A.}~\bibnamefont {Ghosh}},
  \bibinfo {author} {\bibfnamefont {T.~S.}\ \bibnamefont {Haines}}, \bibinfo
  {author} {\bibfnamefont {L.}~\bibnamefont {Singer}}, \ and\ \bibinfo {author}
  {\bibfnamefont {A.}~\bibnamefont {Vecchio}},\ }\href
  {https://doi.org/10.1093/mnras/sty1485} {\bibfield  {journal} {\bibinfo
  {journal} {Monthly Notices of the Royal Astronomical Society}\ }\textbf
  {\bibinfo {volume} {479}},\ \bibinfo {pages} {601} (\bibinfo {year}
  {2018}{\natexlab{b}})}\BibitemShut {NoStop}%
\bibitem [{\citenamefont {Blei}\ and\ \citenamefont {Jordan}(2006)}]{Blei2006}%
  \BibitemOpen
  \bibfield  {author} {\bibinfo {author} {\bibfnamefont {D.~M.}\ \bibnamefont
  {Blei}}\ and\ \bibinfo {author} {\bibfnamefont {M.~I.}\ \bibnamefont
  {Jordan}},\ }\href {\doibase 10.1214/06-BA104} {\bibfield  {journal}
  {\bibinfo  {journal} {Bayesian Analysis}\ }\textbf {\bibinfo {volume} {1}},\
  \bibinfo {pages} {121 } (\bibinfo {year} {2006})}\BibitemShut {NoStop}%
\bibitem [{\citenamefont {Haines}(10  )}]{haines_dpgmm}%
  \BibitemOpen
  \bibfield  {author} {\bibinfo {author} {\bibfnamefont {T.~S.~F.}\
  \bibnamefont {Haines}},\ }\href {https://github.com/thaines/helit} {\enquote
  {\bibinfo {title} {thaines/helit/dpgmm},}\ } (\bibinfo {year}
  {2010--})\BibitemShut {NoStop}%
\bibitem [{\citenamefont {Huang}\ \emph {et~al.}(2022)\citenamefont {Huang},
  \citenamefont {Chen}, \citenamefont {Haster}, \citenamefont {Sun},
  \citenamefont {Vitale},\ and\ \citenamefont {Kissel}}]{Huang:2022rdg}%
  \BibitemOpen
  \bibfield  {author} {\bibinfo {author} {\bibfnamefont {Y.}~\bibnamefont
  {Huang}}, \bibinfo {author} {\bibfnamefont {H.-Y.}\ \bibnamefont {Chen}},
  \bibinfo {author} {\bibfnamefont {C.-J.}\ \bibnamefont {Haster}}, \bibinfo
  {author} {\bibfnamefont {L.}~\bibnamefont {Sun}}, \bibinfo {author}
  {\bibfnamefont {S.}~\bibnamefont {Vitale}}, \ and\ \bibinfo {author}
  {\bibfnamefont {J.}~\bibnamefont {Kissel}},\ }\href@noop {} {\  (\bibinfo
  {year} {2022})},\ \Eprint {http://arxiv.org/abs/2204.03614} {arXiv:2204.03614
  [gr-qc]} \BibitemShut {NoStop}%
\bibitem [{\citenamefont {P\"urrer}\ and\ \citenamefont
  {Haster}(2020)}]{Purrer:2019jcp}%
  \BibitemOpen
  \bibfield  {author} {\bibinfo {author} {\bibfnamefont {M.}~\bibnamefont
  {P\"urrer}}\ and\ \bibinfo {author} {\bibfnamefont {C.-J.}\ \bibnamefont
  {Haster}},\ }\href {\doibase 10.1103/PhysRevResearch.2.023151} {\bibfield
  {journal} {\bibinfo  {journal} {Phys. Rev. Res.}\ }\textbf {\bibinfo {volume}
  {2}},\ \bibinfo {pages} {023151} (\bibinfo {year} {2020})},\ \Eprint
  {http://arxiv.org/abs/1912.10055} {arXiv:1912.10055 [gr-qc]} \BibitemShut
  {NoStop}%
\bibitem [{\citenamefont {Hu}\ and\ \citenamefont {Veitch}(2022)}]{Hu:2022rjq}%
  \BibitemOpen
  \bibfield  {author} {\bibinfo {author} {\bibfnamefont {Q.}~\bibnamefont
  {Hu}}\ and\ \bibinfo {author} {\bibfnamefont {J.}~\bibnamefont {Veitch}},\
  }\href {\doibase 10.1103/PhysRevD.106.044042} {\bibfield  {journal} {\bibinfo
   {journal} {Phys. Rev. D}\ }\textbf {\bibinfo {volume} {106}},\ \bibinfo
  {pages} {044042} (\bibinfo {year} {2022})},\ \Eprint
  {http://arxiv.org/abs/2205.08448} {arXiv:2205.08448 [gr-qc]} \BibitemShut
  {NoStop}%
\bibitem [{\citenamefont {Essick}(2022)}]{Essick:2022vzl}%
  \BibitemOpen
  \bibfield  {author} {\bibinfo {author} {\bibfnamefont {R.}~\bibnamefont
  {Essick}},\ }\href {\doibase 10.1103/PhysRevD.105.082002} {\bibfield
  {journal} {\bibinfo  {journal} {Phys. Rev. D}\ }\textbf {\bibinfo {volume}
  {105}},\ \bibinfo {pages} {082002} (\bibinfo {year} {2022})},\ \Eprint
  {http://arxiv.org/abs/2202.00823} {arXiv:2202.00823 [astro-ph.IM]}
  \BibitemShut {NoStop}%
\bibitem [{\citenamefont {Payne}\ \emph {et~al.}(2020)\citenamefont {Payne},
  \citenamefont {Talbot}, \citenamefont {Lasky}, \citenamefont {Thrane},\ and\
  \citenamefont {Kissel}}]{Payne:2020myg}%
  \BibitemOpen
  \bibfield  {author} {\bibinfo {author} {\bibfnamefont {E.}~\bibnamefont
  {Payne}}, \bibinfo {author} {\bibfnamefont {C.}~\bibnamefont {Talbot}},
  \bibinfo {author} {\bibfnamefont {P.~D.}\ \bibnamefont {Lasky}}, \bibinfo
  {author} {\bibfnamefont {E.}~\bibnamefont {Thrane}}, \ and\ \bibinfo {author}
  {\bibfnamefont {J.~S.}\ \bibnamefont {Kissel}},\ }\href {\doibase
  10.1103/PhysRevD.102.122004} {\bibfield  {journal} {\bibinfo  {journal}
  {Phys. Rev. D}\ }\textbf {\bibinfo {volume} {102}},\ \bibinfo {pages}
  {122004} (\bibinfo {year} {2020})},\ \Eprint
  {http://arxiv.org/abs/2009.10193} {arXiv:2009.10193 [astro-ph.IM]}
  \BibitemShut {NoStop}%
\bibitem [{\citenamefont {Pratten}\ \emph {et~al.}(2021)\citenamefont
  {Pratten}, \citenamefont {Garc{\'\i}a-Quir{\'o}s}, \citenamefont {Colleoni},
  \citenamefont {Ramos-Buades}, \citenamefont {Estell{\'e}s}, \citenamefont
  {Mateu-Lucena}, \citenamefont {Jaume}, \citenamefont {Haney}, \citenamefont
  {Keitel}, \citenamefont {Thompson} \emph
  {et~al.}}]{pratten2021computationally}%
  \BibitemOpen
  \bibfield  {author} {\bibinfo {author} {\bibfnamefont {G.}~\bibnamefont
  {Pratten}}, \bibinfo {author} {\bibfnamefont {C.}~\bibnamefont
  {Garc{\'\i}a-Quir{\'o}s}}, \bibinfo {author} {\bibfnamefont {M.}~\bibnamefont
  {Colleoni}}, \bibinfo {author} {\bibfnamefont {A.}~\bibnamefont
  {Ramos-Buades}}, \bibinfo {author} {\bibfnamefont {H.}~\bibnamefont
  {Estell{\'e}s}}, \bibinfo {author} {\bibfnamefont {M.}~\bibnamefont
  {Mateu-Lucena}}, \bibinfo {author} {\bibfnamefont {R.}~\bibnamefont {Jaume}},
  \bibinfo {author} {\bibfnamefont {M.}~\bibnamefont {Haney}}, \bibinfo
  {author} {\bibfnamefont {D.}~\bibnamefont {Keitel}}, \bibinfo {author}
  {\bibfnamefont {J.~E.}\ \bibnamefont {Thompson}},  \emph {et~al.},\ }\href
  {\doibase 10.1103/PhysRevD.103.104056} {\bibfield  {journal} {\bibinfo
  {journal} {Physical Review D}\ }\textbf {\bibinfo {volume} {103}},\ \bibinfo
  {pages} {104056} (\bibinfo {year} {2021})}\BibitemShut {NoStop}%
\end{thebibliography}%

\end{document}